\RequirePackage{snapshot}
\documentclass[a4paper,12pt]{article}
\usepackage{jcappub}
\usepackage{amsmath}
\usepackage{amssymb}
\usepackage[normalem]{ulem}
\usepackage[lmargin=2.5cm,rmargin=2.5cm,tmargin=2cm]{geometry}
\renewcommand{\vec}[1]{\mathbf{#1}}

\newcommand{\unit}[1]{\hat{\vec{#1}}}
\usepackage{mleftright}
\usepackage{diagbox}
\usepackage{graphicx}
\usepackage{enumitem}

\newcommand{\bn}{{\mathbf n}}

\newcommand{\bk}{{\mathbf k}}

\newcommand{\br}{{\mathbf r}}

\newcommand{\pa}{{\parallel}}

\newcommand{\HH}{{\cal H}}

\newcommand{\al}{\alpha}

\newcommand{\de}{\delta}
\newcommand{\De}{\Delta}

\newcommand{\la}{\lambda}
\newcommand{\Om}{\Omega}

\newcommand{\be}{\begin{equation}}
\newcommand{\ee}{\end{equation}}
\newcommand{\bea}{\begin{eqnarray}}
\newcommand{\eea}{\end{eqnarray}}
\newcommand{\bean}{\begin{eqnarray*}}
\newcommand{\eean}{\end{eqnarray*}}
\newcommand{\beal}{\begin{align}}
\newcommand{\enal}{\end{align}}

\newcommand{\dd}{\text{d}}

\usepackage{float}
\usepackage{booktabs}

\newcommand{\factorial}[2]{\begin{pmatrix} #1\\ #2 \end{pmatrix}}

\usepackage{caption}
\usepackage{subcaption}

\usepackage{hyperref}
\newcommand{\kcos}{\nu}

\usepackage[dvipsnames]{xcolor}
\usepackage{tikz}
\usepackage{subcaption}
\usepackage{xfp}

\newcommand\observerx{0}
\newcommand\observery{0}
\newcommand\zxa{-1}
\newcommand\zya{3}

\newcommand\zxb{\fpeval{-\zxa}}
\newcommand\zyb{5}
\newcommand\zymean{\fpeval{(\zya + \zyb) / 2.}}

\newcommand\randomCoordA[1]{\fpeval{#1 * \zxa}}
\newcommand\randomCoordB[1]{\fpeval{#1 * \zxb}}

\newcommand\arcDelta{30}

\newcommand\zaLabel{$z_1$}
\newcommand\zbLabel{$z_2$}
\newcommand\zmeanLabel{$\bar{z}$}

\usepackage{ntheorem}

\newtheorem*{theorem-non}{Theorem}

\title{The flat-sky approximation to galaxy number counts - redshift space correlation function}

\author{{Goran Jelic-Cizmek}}

\emailAdd{goran.jelic-cizmek@unige.ch}

\affiliation{
Universit\'e de Gen\`eve, D\'epartement de Physique Th\'eorique and CAP,
24 quai Ernest-Ansermet, CH-1211 Gen\`eve 4, Switzerland
}

\abstract{
We study the flat-sky approximation for galaxy number counts including relativistic effects, and numerically assess its performance and accuracy with respect to the full-sky result.
We find an agreement of up to 5\% for the local and lensing contributions to the 2-point correlation function and its multipoles at $z > 0.5$, and up to 1\% for the multipoles alone at $z > 1$ and separations $\lesssim 250$ Mpc/$h$, with a speed-up of over a factor of 1000.
Using a semi-analytic method, which has been implemented in a new version of the code COFFE\footnote{available at \url{https://github.com/JCGoran/coffe}}, along with the Limber approximation for the integrated contributions, we further increase the performance, allowing the computation of the flat-sky multipoles to be done over 10000 times faster than in the full-sky calculation, which could be used to greatly speed-up Markov chain Monte Carlo sampling for cosmological parameter estimation.%
}

\begin{document}

\maketitle

\section{Introduction}
\label{s:intro}

Cosmology in the 21st century is becoming a data driven science, and future galaxy surveys such as Euclid~\cite{Laureijs:2011gra}, SKA2~\cite{Maartens:2015mra}, and DESI~\cite{Aghamousa:2016zmz} will provide unprecedented amounts of data by probing the largest volumes and highest redshifts yet.
In order to make reliable predictions, we require fast and accurate modelling of various cosmological quantities.

In this paper, we assess the accuracy and performance of the 2-point correlation function (2PCF) of galaxy number counts and its multipoles using the flat-sky approximation, including the standard density and redshift-space distortions (RSD), as well as effects such as gravitational lensing and Doppler, and their respective cross-correlations.

When we count galaxies, we observe them in a given direction and at a given redshift.
The expression from linear perturbation theory for the over-density of galaxies at a redshift $z$ and in direction $\bn$ is given by~\cite{2009PhRvD..80h3514Y,Bonvin:2011bg,Challinor:2011bk}:
\begin{equation}
\begin{aligned}
\Delta(z, \mathbf{n})
&
=
b\cdot \delta
+
\frac{1}{\mathcal{H}}\partial_\chi^2V
+
\frac{5s - 2}{2 \chi}
\int_0^{\chi}
\dd \lambda\,
\frac{\chi-\lambda}{\lambda}\Delta_\Omega(\Phi+\Psi)\\
&
-
\partial_\chi V
-
\frac{1}{\mathcal{H}}\partial_\chi\dot{V}
+
\frac{1}{\mathcal{H}}\partial_\chi\Psi\\
&
-\left(
-5s-\frac{\dot{\mathcal{H}}}{\mathcal{H}^2}
+
\frac{5s-2}{\chi \mathcal{H}}
+
f_{\rm evo}
\right)
\partial_\chi V
\\
&+\frac{2-5s}{\chi}\int_0^{\chi} \dd \lambda(\Phi+\Psi)+(f_{\rm evo}-3)\mathcal{H}V+\Psi+(5s-2)\Phi\\
&+\frac{1}{\mathcal{H}}\dot{\Phi}+\left(\frac{\dot{\mathcal{H}}}{\mathcal{H}^2}+\frac{2-5s}{\chi \mathcal{H}}+5s -f_{\rm evo} \right)\left[\Psi+\int_0^{\chi} \dd \lambda\, (\dot{\Phi}+\dot{\Psi})\right]\, ,
\end{aligned}
\label{eq:number_counts}
\end{equation}
where $\chi=\chi(z)$ is the comoving distance to redshift $z$, and $V$ is the potential of the peculiar velocity in the longitudinal gauge, such that $\vec V = - \nabla V$.
The functions $b(z)$, $s(z)$ and $f_{\rm evo}(z)$ are the galaxy bias, the magnification bias and the galaxy evolution bias respectively.
They depend on the specifications of the catalog (which types of galaxies have been included) and on the instrument (what is the flux limit of the instrument in which frequency band).
The three terms on the first line of eq.~\eqref{eq:number_counts} denote the contributions from density, RSD, and lensing, the third line contains the Doppler term (which we will denote d1), and the fourth and fifth line denote various relativistic effects which are only important on the largest scales.
Note that in eq.~\eqref{eq:number_counts} we did not yet make use of the Euler equation for presureless nonrelativistic matter, given by:
\begin{equation}
\dot{\vec V} \cdot \vec n + \mathcal{H} \vec V \cdot \vec n + \partial_\chi \Psi = 0
\label{eq:euler}
\end{equation}
which causes the second line in eq.~\eqref{eq:number_counts} to vanish.
Throughout this paper, we will assume that the Euler equation holds, i.e. we shall assume our theory of gravity to be general relativity, and therefore neglect the second line in eq.~\eqref{eq:number_counts}.
Note that eq.~\eqref{eq:number_counts} contains two kinds of terms: 1) local terms, which just depend on the position $\bn$ and redshift $z$ of a galaxy, and 2) non-local, or integrated terms, which depend on the entire trajectory of a photon between the source galaxy and the observer.

The 2-point correlation function (2PCF) of the number counts, eq.~\eqref{eq:number_counts}, can be obtained from the expression:
\begin{equation}
\xi(z_1, z_2, \bn_1, \bn_2)
\equiv
\langle
\Delta(z_1, \mathbf{n}_1)
\Delta(z_2, \mathbf{n}_2)
\rangle
\label{eq:def_2pcf}
\end{equation}
The brackets in eq.~\eqref{eq:def_2pcf} are the theoretical ensemble average, but, if ergodicity holds (as it does for the case of a statistically homogeneous and isotropic Gaussian random field), they can be replaced by a spatial average in observations.
Due to isotropy, the 2PCF is a function of only three variables, usually taken to be $\{z_1, z_2, \cos \theta = \bn_1 \cdot \bn_2\}$, but of course, it can also be expressed in other variables; in this paper, we will use a mean redshift $\bar z$, a separation $r$ between the two points in comoving coordinates, and an angle $\mu$.
They are connected to $\{z_1, z_2, \cos \theta\}$ via:
\begin{align}
\bar z &= \frac{z_1 + z_2}{2}
\\
r &= \sqrt{\chi^2(z_1) + \chi^2(z_2) - 2\chi(z_1)\, \chi(z_2)\, \cos\theta}
\\
\mu &= \frac{\chi(z_1) - \chi(z_2)}{r}
\label{eq:relation_angles}
\end{align}
In the above and in the rest of this paper, we assume that the spatial curvature of the Universe is zero, $\Omega_K = 0$.
For $z_1 = z_2$, the 2PCF is just a function of two variables, and we will call this the angular correlation function.

The derivation of the full-sky 2PCF can be found in~\cite{Bonvin:2011bg,Challinor:2011bk,Tansella:2017rpi}; schematically, the contributions to the 2PCF from eq.~\eqref{eq:number_counts} are of the form, using the notation of~\cite{Tansella:2018sld} (the full expressions for the relevant terms are given in appendix~\ref{s:fullsky_expressions}):
\bea
\xi^{\text{L.-L.}}_{AB}(z_1, z_2, r)
\propto&
\hspace{-0.2cm}
&f_{AB}(z_1, z_2) \times I_\ell^n(r)
\label{eq:local_local}
\\
\xi^{\text{L.-N.L.}}_{AB}(z_1, z_2, r)
\propto&
\hspace{-0.2cm}
&\int_0^{\chi(z_2)}
\hspace{-0.3cm}
\dd \lambda\,
g_{AB}[z_1, z(\lambda)] \times I_\ell^n[r(\lambda)]
\label{eq:local_nonlocal}
\\
\xi^{\text{N.L.-N.L.}}_{AB}(z_1, z_2, r)
\propto&
\hspace{-0.2cm}
&\int_0^{\chi(z_1)}
\hspace{-0.3cm}
\dd \lambda_1\,
\int_0^{\chi(z_2)}
\hspace{-0.3cm}
\dd \lambda_2\,
h_{AB}[z(\lambda_1), z(\lambda_2)] \times I_\ell^n[r(\lambda_1, \lambda_2)]
\label{eq:nonlocal_nonlocal}
\eea
for arbitrary contributions $A$ and $B$, where L. and N.L. denote local and non-local (integrated) terms, respectively, and $f, g, h$ are functions that only depend on the two redshifts.
When it is clear from context, we will use the notation $A_i = A(z_i)$ for a redshift-dependent quantity $A$.
Additionally, we define the dimensionless integrals:
\begin{equation}
I_\ell^n(r) \equiv \frac{1}{2 \pi^2} \int_0^\infty \dd k\, k^2\, P(k)
\frac{
j_\ell (k r)
}
{
(k r)^n
}
\label{eq:fftlog}
\end{equation}
where $j_\ell$ denote the spherical Bessel functions of order $\ell$, and $P(k)$ is the linear matter power spectrum at present time.
The quantity~\eqref{eq:fftlog} can be computed quickly and accurately using a method based on Fourier transforms, commonly known as the FFTlog method~\cite{Hamilton_2000, Grasshorn_Gebhardt_2018}.
From now on, unless stated otherwise, we will write $P(k) \equiv P_\mathrm{linear} (k, z = 0)$ for brevity.

We define the multipoles of the 2PCF as:
\begin{equation}
\xi_\ell (\bar z, r)
\equiv
\frac{2 \ell + 1}{2}
\int_{-1}^{1}
\dd \mu\,
\xi(\bar z, r, \mu)\,
P_\ell(\mu)
\label{eq:def_multipoles}
\end{equation}
where $P_\ell$ denote the Legendre polynomials of degree $\ell$.

Therefore, to obtain the multipoles of the 2PCF with contributions from eqs.~\eqref{eq:local_local}--\eqref{eq:nonlocal_nonlocal}, in addition to the integration over $k$, which can be done using the FFTlog method, we need to compute either
\begin{enumerate}[label=\alph*)]
\item one integral over the angle $\mu$ (for local-local terms, such as density and RSD)
\item one integral along one of the lines of sight and one integral over the angle $\mu$ (for local - non-local terms, such as density-lensing),
\item two integrals along the two lines of sight and one integral over the angle $\mu$ (for non-local - non-local terms, such as lensing-lensing).
\end{enumerate}
The integrals over the line of sight are very computationally expensive operations, and we might hope to do better using various approximations.
For the angular power spectrum of various estimators (number counts, intensity mapping, etc.), the most well-known ones are the Limber approximation~\cite{kaiser_limber,LoVerde:2008re,Lemos_2017}, and the flat-sky approximation~\cite{Bernardeau_2011}.
The latter has recently been studied in greater detail for galaxy number counts~\cite{matthewson2020flatsky}, and here we study it for the redshift-space 2-point correlation function.

In section~\ref{s:flatsky} we go over the flat-sky approximation for the relevant non-integrated and integrated terms, and derive a semi-analytic result for the multipoles of the 2PCF.
In section~\ref{s:results} we show the performance and the accuracy of the flat-sky approximation, and in section~\ref{s:conclusions} we explain our results and discuss possible applications.

\section{The flat-sky approximation}
\label{s:flatsky}

The full-sky results given by eqs.~\eqref{eq:local_local}--\eqref{eq:nonlocal_nonlocal} are exact in linear theory, and have previously been implemented in the code COFFE~\cite{Tansella:2018sld}.
The largest contributions to the 2PCF and its multipoles come from density, RSD, and lensing terms, but, as mentioned above, the latter is very challenging to evaluate numerically.
In the flat-sky approximation for the 2PCF, we replace the two lines of sight to the galaxies, $\vec{n}_1$ and $\vec{n}_2$, with a single line of sight, which we will denote $\vec{n}_*$, and a small deviation, $\Delta \vec n$, such that $\vec{n}_1 = \vec{n}_* + \Delta \vec{n} / 2$ and $\vec{n}_2 = \vec{n}_* - \Delta \vec{n} / 2$.
We assume the distant observer approximation.
As we shall see, all of this greatly simplifies computations.
The flat-sky approximation has been studied in great detail in the angular power spectrum representation in~\cite{Bernardeau_2011,matthewson2020flatsky} for the CMB and the galaxy number counts, respectively, where it has also been compared to the well-known Limber approximation, in which we integrate out the radial Fourier modes, and can in principle be applied separately.
Notably, in the latter reference, the term ``flat-sky'' for the integrated terms could more accurately be referred to as either ``flat-sky Limber'' (in case of their eq. (2.26)), or as ``full-sky Limber'' (in case of their eq. (3.5)).

\subsection{Generalities - non-integrated terms}
\label{s:gen_nonint}

Here we give a short overview of the non-integrated contributions to the 2PCF and its multipoles in the flat-sky approximation.
We neglect terms from the last two lines of eq.~\eqref{eq:number_counts} in our discussion since they are only relevant at wide angles and large comoving separations~\cite{jeliccizmek2020importance}.

For the non-integrated terms, we may use expressions from appendix B of~\cite{Tansella:2017rpi}, which amount to a Taylor expansion of local-local terms, eq.~\eqref{eq:local_local}, in the variable $\mu r / \bar \chi$, from which we obtain the following expressions, written in generality for two populations of galaxies%
\footnote{The flat-sky limits (B.7)--(B.11) in~\cite{Tansella:2017rpi} contain some typos, and here we give the corrected expressions.}:
\begingroup
\allowdisplaybreaks
\begin{align}
\xi_\mathrm{flat}^\text{den}(\bar z, r, \mu)
&=
D_1^2(\bar z)\,
b_1\,
b_2\,
I_0^0(r)
\label{eq:den_flatsky}
\\
\xi_\mathrm{flat}^\text{RSD}(\bar z, r, \mu)
&=
D_1^2(\bar z)
\bigg[
\frac{1}{5} f^2 I_0^0 (r)
-
\frac{4}{7} f^2 I_2^0 (r)
P_2(\mu)
+
\frac{8}{35} f^2 I_4^0 (r)
P_4(\mu)
\bigg]
\label{eq:rsd_flatsky}
\\
\xi_\mathrm{flat}^\text{den-RSD}(\bar z, r, \mu)
&=
D_1^2(\bar z)\,
\bigg[
\frac{1}{3}
\big(
b_1\,
f
+
b_2\,
f
\big)
I_0^0(r)
-
\frac{2}{3}
\big(
b_1\,
f
+
b_2\,
f
\big)
I_2^0(r)\,
P_2(\mu)
\bigg]
\label{eq:den_rsd_flatsky}
\\
\xi_\mathrm{flat}^\text{d1}(\bar z, r, \mu)
&=
D_1^2(\bar z)
\bigg[
\frac{1}{3}
\mathcal{H}^2
f^2\,
G_1\,
G_2\,
r^2\,
I_0^2(r)
+
\frac{2}{3}
\mathcal{H}^2\,
f^2\,
G_1\,
G_2\,
r^2\,
I_2^2(r)\,
P_2 (\mu)
\bigg]
\label{eq:d1_flatsky}
\\
\xi_\mathrm{flat}^\text{den-d1}(\bar z, r, \mu)
&=
D_1^2(\bar z)\,
(b_1\, G_2 - b_2\, G_1)
f\,
\mathcal{H}\,
r\,
I_1^1 (r)\,
P_1 (\mu)
\label{eq:den_d1_flatsky}
\\
\xi_\mathrm{flat}^\text{RSD-d1}(\bar z, r, \mu)
&=
D_1^2(\bar z)\,
f^2\,\mathcal{H}\, r\,
(G_1 - G_2)
\bigg[
-\frac{3}{5}\,
I_1^1 (r)\,
P_1 (\mu)
+
\frac{2}{5}
I_3^1 (r)\,
P_3 (\mu)
\bigg]
\label{eq:rsd_d1_flatsky}
\end{align}
\endgroup
where $b_i = b_i (\bar z)$, $i = 1,2$, $D_1(\bar z)$ is the linear matter growth function, $f = f(\bar z) = \dd \log D_1 / \dd \log a$ is the growth rate, and we defined:
\begin{equation}
G(z)= \frac{\dot \HH}{\HH^2}+\frac{2-5s}{\chi \HH}+5 s - f_\text{evo}
\end{equation}
Note that here we use the notation $\xi^A = \langle \Delta^A \Delta ^A \rangle$ for the auto-, and $\xi^{AB} = \langle \Delta^{A} \Delta^{B} \rangle + \langle \Delta^{B} \Delta^{A} \rangle$ for the cross-correlation terms.
These expressions do not use the Limber approximation, since that would yield a result which is a linear combination of $\delta(\chi_1 - \chi_2)$ and $\Theta(\chi_1 - \chi_2)$ (see appendix~\ref{s:limber_nonint}).

Eqs.~\eqref{eq:den_flatsky},~\eqref{eq:rsd_flatsky}, and~\eqref{eq:den_rsd_flatsky} are the familiar density, RSD, and their cross-correlations, respectively, which are commonly known as the ``standard'' terms, while eqs.~\eqref{eq:d1_flatsky},~\eqref{eq:den_d1_flatsky}, and~\eqref{eq:rsd_d1_flatsky} are the Doppler auto-correlation term and its cross-correlations with density and RSD, respectively.

From the functional form of the above, we may conclude the following for the flat-sky approximation of the non-integrated terms:

\begin{enumerate}[label=\alph*)]
\item
the standard terms only generate the $\ell = \{0, 2, 4\}$ multipoles; additionally, for the case of constant galaxy bias $b$, the density auto-correlation term exactly corresponds to the full-sky result.
\item
The Doppler-Doppler term only contributes to the monopole ($\ell = 0$) and the quadrupole ($\ell = 2$).
\item
The Doppler cross-correlation with density is always zero, unless we consider multiple populations of galaxies with different galaxy, magnification, or evolution bias, i.e. the case $b_1 \neq b_2$ or $G_1 \neq G_2$, where they generate only the dipole ($\ell = 1$)~\cite{Bonvin_2016}.
\item
The Doppler cross-correlation with RSD is always zero, unless we consider multiple populations of galaxies with different magnification or evolution bias, i.e. the case $G_1 \neq G_2$, where they generate the dipole ($\ell = 1$) and the octupole ($\ell = 3$).
\end{enumerate}

For completeness, we could also consider the other non-integrated terms in the flat-sky approximation: the terms with $I_\ell^n$ in full-sky induce a multipole of order $\ell$ with a $(\mathcal{H} r)^n$ dependence on separation; they are, however, very small compared to the other terms listed above.\footnote{These terms also require counterterms to be well-behaved, see section 2.1 of~\cite{Tansella:2017rpi} for additional details.}

Due to the simple dependence on the angle $\mu$, the multipoles of expressions~\eqref{eq:den_flatsky}--\eqref{eq:rsd_d1_flatsky} are easily obtained analytically using eq.~\eqref{eq:def_multipoles}, and, owing to the FFTlog transformation, can be computed in a fast and accurate manner.

The key take-away from this section is that, in the flat-sky approximation, unlike in the full-sky case, we can compute the multipoles analytically, which is what we will exploit later for the integrated terms as well.

\subsection{Generalities - integrated terms}
\label{s:gen_int}

The flat-sky approximation of the 2PCF for the terms integrated along the line of sight, i.e. of the form of eqs.~\eqref{eq:local_nonlocal} and~\eqref{eq:nonlocal_nonlocal}, is somewhat less well known, but has been studied in~\cite{Tansella:2017rpi}, and here we just note the results.
We will primarily focus on the contributions from density-lensing and lensing-lensing, since, in full-sky, the RSD-lensing and Doppler-lensing contributions are usually much smaller than density-lensing, and can be shown to be zero in the flat-sky approximation.
Furthermore, for all of the integrated terms considered, we will use the Limber approximation, in which we integrate out the radial Fourier modes, which greatly simplify the final result.
In principle, we could go beyond the Limber approximation by performing the expansion outlined in~\cite{LoVerde:2008re}, however, we leave this ``extended Limber'' approximation of the integrated terms for future work.
The full derivation of the relevant equations is available in appendix~\ref{s:flat_sky_derivation}.

\subsubsection{Density-lensing 2PCF}
\label{s:den_len_flatsky}

\begin{figure}
\centering

\begin{subfigure}[t]{0.49\textwidth}
\centering
\begin{tikzpicture}
\coordinate (observer) at (\observerx, \observery);
\coordinate (z1) at (\zxa, \zya);
\coordinate (z2) at (\zxb, \zyb);
\coordinate (zmean) at (\observerx, \zymean);
\coordinate (randomRedshiftA1) at (\randomCoordA{1}, \fpeval{(\zya - \observery) / (\zxa - \observerx) * (\randomCoordA{1} - \zxa) + \zya});
\coordinate (randomRedshiftB1) at (\randomCoordB{0.5}, \fpeval{(\zyb - \observery) / (\zxb - \observerx) * (\randomCoordB{0.5} - \zxb) + \zyb});

\coordinate (randomRedshiftA2) at (\randomCoordA{1}, \fpeval{(\zya - \observery) / (\zxa - \observerx) * (\randomCoordA{1} - \zxa) + \zya});
\coordinate (randomRedshiftB2) at (\randomCoordB{0.8}, \fpeval{(\zyb - \observery) / (\zxb - \observerx) * (\randomCoordB{0.8} - \zxb) + \zyb});

\coordinate (randomRedshiftA3) at (\randomCoordA{1}, \fpeval{(\zya - \observery) / (\zxa - \observerx) * (\randomCoordA{1} - \zxa) + \zya});
\coordinate (randomRedshiftB3) at (\randomCoordB{0.9}, \fpeval{(\zyb - \observery) / (\zxb - \observerx) * (\randomCoordB{0.9} - \zxb) + \zyb});

\draw[thick] (z1) -- (observer) -- (z2);
\draw[dotted, thick] (\zxa, \zya) arc (\fpeval{90 + \arcDelta}:\fpeval{90 - \arcDelta}:\fpeval{\zya * 0.5});
\draw[dotted, thick] (\zxb, \zyb) arc (\fpeval{90 - \arcDelta}:\fpeval{90 + \arcDelta}:\fpeval{\zyb * 0.5});
\filldraw[black] (observer) circle (2pt) node[anchor=north] {Observer};
\draw[dashed, red] (randomRedshiftA1) -- (randomRedshiftB1);
\draw[dashed, blue] (randomRedshiftA2) -- (randomRedshiftB2);
\draw[dashed, ForestGreen] (randomRedshiftA3) -- (randomRedshiftB3);
\filldraw[red] (randomRedshiftA1) circle (1pt);
\filldraw[red] (randomRedshiftB1) circle (1pt);
\filldraw[blue] (randomRedshiftA2) circle (1pt);
\filldraw[blue] (randomRedshiftB2) circle (1pt);
\filldraw[ForestGreen] (randomRedshiftA3) circle (1pt);
\filldraw[ForestGreen] (randomRedshiftB3) circle (1pt);
\filldraw[black] (z1) circle (2pt) node[anchor=east] {\zaLabel};
\filldraw[black] (z2) circle (2pt) node[anchor=west] {\zbLabel};
\filldraw[black] (zmean) circle (2pt) node[anchor=east] {\zmeanLabel};
\end{tikzpicture}
\end{subfigure}
\begin{subfigure}[t]{0.49\textwidth}
\centering
\begin{tikzpicture}
\coordinate (observer) at (\observerx, \observery);
\coordinate (zmean1) at (\zxa, \zymean);
\coordinate (zmean2) at (\zxb, \zymean);
\draw[thick] (zmean1) -- (observer) -- (zmean2);

\coordinate (randomRedshiftA1) at (\randomCoordA{1}, \fpeval{(\zymean - \observery) / (\zxa - \observerx) * (\randomCoordA{1} - \zxa) + \zymean});
\coordinate (randomRedshiftB1) at (\randomCoordB{1}, \fpeval{(\zymean - \observery) / (\zxb - \observerx) * (\randomCoordB{1} - \zxb) + \zymean});

\draw[dashed, red] (randomRedshiftA1) -- (randomRedshiftB1);

\filldraw[red] (randomRedshiftA1) circle (1pt);
\filldraw[red] (randomRedshiftB1) circle (1pt);

\draw[dotted, thick] (\zxb, \zymean) arc (\fpeval{90 - \arcDelta}:\fpeval{90 + \arcDelta}:\fpeval{\zymean * 0.5});
\filldraw[black] (observer) circle (2pt) node[anchor=north] {Observer};
\filldraw[black] (zmean1) circle (2pt) node[anchor=east] {\zmeanLabel};
\filldraw[black] (zmean2) circle (2pt) node[anchor=west] {\zmeanLabel};
\end{tikzpicture}
\end{subfigure}

\caption{\textit{Left}: full-sky geometry, \textit{right}: flat-sky geometry, for density-lensing. The dashed colored lines indicate one of the infinitely many paths which contribute to the result.
In the flat-sky case, only the point at the mean redshift has a non-zero contribution, and only if $\chi_\text{den} < \chi_\text{len}$.
Note that this is meant to be merely a visualization aid; from the diagram above, we would erroneously conclude that $\chi_1 = \chi_2$, from which it follows that $\mu = 0$, and from eq.~\eqref{eq:den_len_flat_2pcf}, this implies $\xi^\text{den-len}_\text{flat} = 0$ everywhere, which is incorrect.}
\label{fig:fullsky_vs_flatsky_den_len_tikz}
\end{figure}
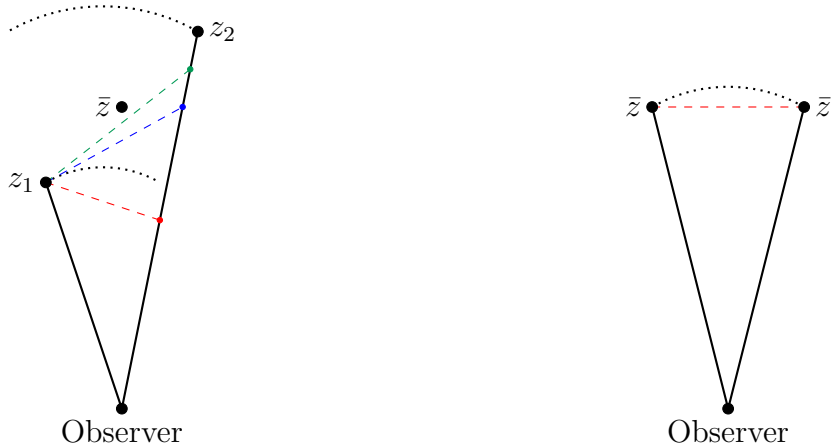

The density-lensing contribution to the 2PCF in the flat-sky Limber approximation reads, for two populations of galaxies\footnote{Note that there is a missing factor of $D_1(\bar z)$ in eq. (E.8) of reference~\cite{Tansella:2017rpi} with respect to the result written here.}:
\begin{align}
\xi^\text{den-len}_\mathrm{flat}
(\bar z, r, \mu)
=&
-\frac{3}{8\pi}
\Om_mH_0^2\,
D_1^2(\bar z)\,
(1 + \bar z)\,
r
\nonumber
\\
&
\times
\bigg\{
\left[
(2-5s_1(\bar z))b_2(\bar z)
-
(2-5s_2(\bar z))b_1(\bar z)
\right]
\mu
\nonumber
\\
&+
\left[
(2-5s_1(\bar z))b_2(\bar z)
+
(2-5s_2(\bar z))b_1(\bar z)
\right]
|\mu|
\bigg\}
\nonumber
\\
&
\times
i(r\sqrt{1-\mu^2})
\label{eq:den_len_flat_2pcf}
\end{align}
where we defined:
\begin{equation}
i(x)
\equiv
\int_0^\infty
\dd k\,
k\,
P(k)\,
J_0(k x)
\label{eq:def_i}
\end{equation}
For simplicity, in what follows, we will restrict ourselves to one population of galaxies.
The geometry of full-sky and flat-sky for density-lensing is shown schematically in figure~\ref{fig:fullsky_vs_flatsky_den_len_tikz}.
The schematic is meant to be a visualization aid, and does not represent the actual geometry of the sky.

Note that the quantity in eq.~\eqref{eq:def_i} can be written as:
\begin{equation}
\hspace{-0.3cm}
i(x)
=
x
\int^\infty_0
\hspace{-0.2cm}
\dd k\,
k^2\,
P(k)\,
\frac{J_0(k x)}{k x}
=
\sqrt{\frac{2}{\pi}}x
\int^\infty_0
\hspace{-0.2cm}
\dd k\,
k^2\,
P(k)\,
\frac{j_{-\frac{1}{2}}(k x)}{(k x)^\frac{1}{2}}
=
(2 \pi)^\frac{3}{2}\,
x\,
I_{-1/2}^{1/2}(x)
\label{eq:i_fftlog}
\end{equation}
where $I$ is defined by eq.~\eqref{eq:fftlog}, and can be computed with the same FFTlog method mentioned earlier.

We note several features of the 2PCF for density-lensing in the flat-sky Limber approximation:

\begin{enumerate}[label=\alph*)]
\item
For $r \rightarrow 0$, or alternatively, $\chi_1 \rightarrow \chi_2$, $\theta \rightarrow 0$, we have that $\xi_\mathrm{flat}^\text{den-len} \rightarrow 0$, which is \textit{not} the case for $\xi_\mathrm{full}^\text{den-len}$; we can see this by directly computing the limit using the full-sky expression, eq.~\eqref{eq:den-len}:
\allowdisplaybreaks
\begin{flalign*}
&\lim\limits_{r \rightarrow 0} \xi^\text{den-len}_\mathrm{full}(\bar z, r, \mu)
=
- 3\Omega_m b_1 \HH_0^2 (2-5s_2) D_1(z)
\int_0^{\chi}
\hspace{-0.1cm}
\dd\la\,
(\chi-\la)\,
\frac{D_1(\la)}{ a(\la)}\,
I^1_1[r(\lambda)]&
\end{flalign*}
where $r (\lambda)$ inside the integral is given by $r(\lambda) = \chi - \lambda$.
We can numerically show that $I_1^1(r)$ is positive for all $0 \leq r \leq \chi$, hence the integrand is positive in the entire range of integration, and consequently we have that:
\begin{equation}
\lim\limits_{r \rightarrow 0} \xi^\text{den-len}_\mathrm{full}(\bar z, r, \mu)
\neq
0
\end{equation}
\item
$\xi_\mathrm{flat}^\text{den-len}(\bar z, r, 0) = 0$, which is not the case for $\xi_\mathrm{full}^\text{den-len}$.
In other words, the angular correlation function of density-lensing in the flat-sky approximation vanishes, while the full-sky result does not.
\end{enumerate}

From the above, we expect that the flat-sky approximation for density-lensing actually becomes worse at small separations, and for small values of the angle $\mu$.

\subsubsection{Density-lensing multipoles}

The multipoles of density-lensing can be computed by plugging eq.~\eqref{eq:den_len_flat_2pcf} into eq.~\eqref{eq:def_multipoles}.
We may swap the order of the integrals in $k$ and $\mu$, so that the result is written as:
\begin{align}
\xi^\text{den-len}_{\ell,\text{flat}}
(\bar z, r)
=&
-\frac{2\ell + 1}{2}
\frac{3}{8\pi}
\Om_mH_0^2
D_1^2(\bar z)
(1 + \bar z)
r
\nonumber
\\
&\times
\left[(2-5s_1(\bar z))b_2(\bar z) +(2-5s_2(\bar z))b_1(\bar z)\right]
\nonumber
\\
&\times
\int_0^\infty
\dd k\,
k\,
P(k)\,
\mathcal{J}_\ell(k r)
\end{align}
where we defined:
\begin{equation}
\mathcal{J}_\ell(\alpha)
\equiv
\int_{-1}^1
\dd \mu\,
|\mu|\,
P_\ell(\mu)\,
J_0(\alpha \sqrt{1 - \mu^2})
\label{eq:flatsky_density_lensing}
\end{equation}
This integral has a closed form solution, and is given by (see appendix~\ref{s:den_len_appendix}):
\begin{equation}
\mathcal{J}_\ell(\alpha)
=
\frac{
[(-1)^\ell + 1]
}
{
2^\frac{\ell}{2}
}
\sum\limits_{k = 0}^{\left \lfloor \frac{\ell}{2} \right \rfloor}
\frac{(-1)^k}{2^k}
\begin{pmatrix} \ell \\ k \end{pmatrix}
\begin{pmatrix} 2\ell - 2k \\ \ell \end{pmatrix}
\Gamma\left[\frac{\ell}{2} - k + 1\right]
\frac{
J_{\frac{\ell}{2} - k + 1} (\alpha)
}
{
\alpha^{\frac{\ell}{2} - k + 1}
}
\label{eq:density_lensing_multipoles}
\end{equation}
where $J_\nu$ denote the Bessel functions of the first kind of order $\nu$.
From the above, we can see that the odd multipoles vanish, while the even ones are given in terms of a linear combination of integrals of Bessel functions of integer order.
Note that this is not true for two populations of galaxies, which, due to the dependence of eq.~\eqref{eq:den_len_flat_2pcf} on $\mu = P_1(\mu)$, also generate odd multipoles.

For completeness, we write down the entire expression for the density-lensing multipoles in terms of $I_\ell^n$ below:
\begin{align}
\xi^\text{den-len}_{\ell,\text{flat}}
(\bar z, r)
=&
-\frac{2\ell + 1}{2}
\frac{3}{8\pi}
\Om_mH_0^2
D_1^2(\bar z)
(1 + \bar z)
r^2
\nonumber
\\
&\times
\left[(2-5s_1(\bar z))b_2(\bar z) +(2-5s_2(\bar z))b_1(\bar z)\right]
\nonumber
\\
&\times
\pi^\frac{3}{2}\
\frac{
2^\frac{5}{2}
}
{
2^\frac{\ell}{2}
}
\sum\limits_{k = 0}^{\left \lfloor \frac{\ell}{2} \right \rfloor}
\frac{(-1)^k}{2^k}
\begin{pmatrix} \ell \\ k \end{pmatrix}
\begin{pmatrix} 2\ell - 2k \\ \ell \end{pmatrix}
\left[\frac{\ell}{2} - k\right]!\,
I^{\ell / 2 - k + 3 / 2}_{\ell / 2 - k + 1 / 2} (r)
\label{eq:den_len_analytic}
\end{align}
where we've assumed that $\ell$ is even.
Since the above is just a linear combination of $I^n_\ell$, it can be quickly computed by applying the FFTlog method on each term.

\subsubsection{Lensing-lensing 2PCF}

The flat-sky Limber lensing-lensing 2PCF can be shown to be equal to (see appendix~\ref{s:flat_sky_derivation}):
\begin{align}
\xi^\mathrm{len}_\text{flat}
(\bar z, r, \mu)
=&
\frac{(3\Omega_m H_0^2)^2(2-5s_1(\bar z))(2-5s_2(\bar z))}{8 \pi \bar\chi^2}
\nonumber
\\
&\times
\int_0^{\bar\chi}
\dd \lambda\,
\int_0^\infty
\dd k_\perp\,
k_\perp\,
P (k_\perp \bar\chi / \lambda)
J_0\left(k_\perp r\sqrt{1-\mu^2}\right)
\nonumber
\\
&\times
\left(\frac{\lambda}{\bar\chi}\right)^2
\left[\frac{(\bar\chi-\lambda)\bar\chi^2}{\lambda}\right]^2\,
D_1^2(z(\lambda))\,
(1+z(\lambda))^2
\label{eq:lens_lens_2pcf_1}
\end{align}
where $J_0$ is again the Bessel function of order 0.

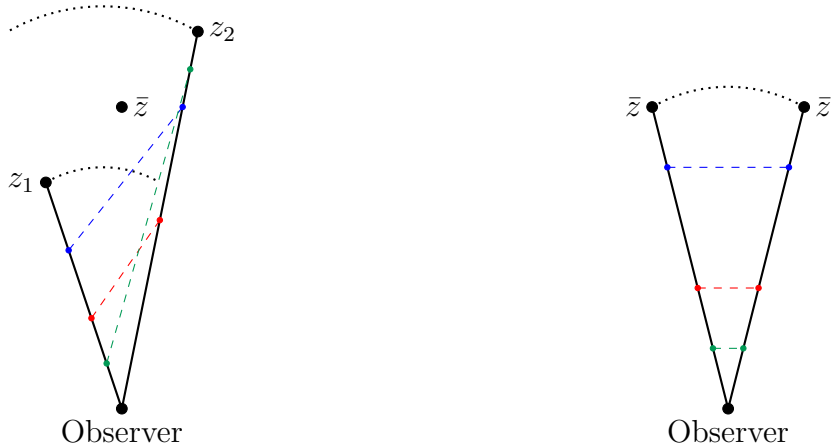
\begin{figure}
\centering

\begin{subfigure}[t]{0.49\textwidth}
\centering
\begin{tikzpicture}
\coordinate (observer) at (\observerx, \observery);
\coordinate (z1) at (\zxa, \zya);
\coordinate (z2) at (\zxb, \zyb);
\coordinate (zmean) at (\observerx, \zymean);
\coordinate (randomRedshiftA1) at (\randomCoordA{0.4}, \fpeval{(\zya - \observery) / (\zxa - \observerx) * (\randomCoordA{0.4} - \zxa) + \zya});
\coordinate (randomRedshiftB1) at (\randomCoordB{0.5}, \fpeval{(\zyb - \observery) / (\zxb - \observerx) * (\randomCoordB{0.5} - \zxb) + \zyb});

\coordinate (randomRedshiftA2) at (\randomCoordA{0.7}, \fpeval{(\zya - \observery) / (\zxa - \observerx) * (\randomCoordA{0.7} - \zxa) + \zya});
\coordinate (randomRedshiftB2) at (\randomCoordB{0.8}, \fpeval{(\zyb - \observery) / (\zxb - \observerx) * (\randomCoordB{0.8} - \zxb) + \zyb});

\coordinate (randomRedshiftA3) at (\randomCoordA{0.2}, \fpeval{(\zya - \observery) / (\zxa - \observerx) * (\randomCoordA{0.2} - \zxa) + \zya});
\coordinate (randomRedshiftB3) at (\randomCoordB{0.9}, \fpeval{(\zyb - \observery) / (\zxb - \observerx) * (\randomCoordB{0.9} - \zxb) + \zyb});

\draw[thick] (z1) -- (observer) -- (z2);
\draw[dotted, thick] (\zxa, \zya) arc (\fpeval{90 + \arcDelta}:\fpeval{90 - \arcDelta}:\fpeval{\zya * 0.5});
\draw[dotted, thick] (\zxb, \zyb) arc (\fpeval{90 - \arcDelta}:\fpeval{90 + \arcDelta}:\fpeval{\zyb * 0.5});
\filldraw[black] (observer) circle (2pt) node[anchor=north] {Observer};
\filldraw[black] (z1) circle (2pt) node[anchor=east] {\zaLabel};
\draw[dashed, red] (randomRedshiftA1) -- (randomRedshiftB1);
\draw[dashed, blue] (randomRedshiftA2) -- (randomRedshiftB2);
\draw[dashed, ForestGreen] (randomRedshiftA3) -- (randomRedshiftB3);
\filldraw[red] (randomRedshiftA1) circle (1pt) node[anchor=east]{};
\filldraw[red] (randomRedshiftB1) circle (1pt) node[anchor=east]{};
\filldraw[blue] (randomRedshiftA2) circle (1pt) node[anchor=east]{};
\filldraw[blue] (randomRedshiftB2) circle (1pt) node[anchor=east]{};
\filldraw[ForestGreen] (randomRedshiftA3) circle (1pt) node[anchor=east]{};
\filldraw[ForestGreen] (randomRedshiftB3) circle (1pt) node[anchor=east]{};
\filldraw[black] (z2) circle (2pt) node[anchor=west] {\zbLabel};
\filldraw[black] (zmean) circle (2pt) node[anchor=west] {\zmeanLabel};
\end{tikzpicture}
\end{subfigure}
\begin{subfigure}[t]{0.49\textwidth}
\centering
\begin{tikzpicture}
\coordinate (observer) at (\observerx, \observery);
\coordinate (zmean1) at (\zxa, \zymean);
\coordinate (zmean2) at (\zxb, \zymean);
\draw[thick] (zmean1) -- (observer) -- (zmean2);

\coordinate (randomRedshiftA1) at (\randomCoordA{0.4}, \fpeval{(\zymean - \observery) / (\zxa - \observerx) * (\randomCoordA{0.4} - \zxa) + \zymean});
\coordinate (randomRedshiftB1) at (\randomCoordB{0.4}, \fpeval{(\zymean - \observery) / (\zxb - \observerx) * (\randomCoordB{0.4} - \zxb) + \zymean});

\coordinate (randomRedshiftA2) at (\randomCoordA{0.8}, \fpeval{(\zymean - \observery) / (\zxa - \observerx) * (\randomCoordA{0.8} - \zxa) + \zymean});
\coordinate (randomRedshiftB2) at (\randomCoordB{0.8}, \fpeval{(\zymean - \observery) / (\zxb - \observerx) * (\randomCoordB{0.8} - \zxb) + \zymean});

\coordinate (randomRedshiftA3) at (\randomCoordA{0.2}, \fpeval{(\zymean - \observery) / (\zxa - \observerx) * (\randomCoordA{0.2} - \zxa) + \zymean});
\coordinate (randomRedshiftB3) at (\randomCoordB{0.2}, \fpeval{(\zymean - \observery) / (\zxb - \observerx) * (\randomCoordB{0.2} - \zxb) + \zymean});

\draw[dashed, red] (randomRedshiftA1) -- (randomRedshiftB1);
\draw[dashed, blue] (randomRedshiftA2) -- (randomRedshiftB2);
\draw[dashed, ForestGreen] (randomRedshiftA3) -- (randomRedshiftB3);

\filldraw[red] (randomRedshiftA1) circle (1pt);
\filldraw[red] (randomRedshiftB1) circle (1pt);
\filldraw[blue] (randomRedshiftA2) circle (1pt);
\filldraw[blue] (randomRedshiftB2) circle (1pt);
\filldraw[ForestGreen] (randomRedshiftA3) circle (1pt);
\filldraw[ForestGreen] (randomRedshiftB3) circle (1pt);

\draw[dotted, thick] (\zxb, \zymean) arc (\fpeval{90 - \arcDelta}:\fpeval{90 + \arcDelta}:\fpeval{\zymean * 0.5});
\filldraw[black] (observer) circle (2pt) node[anchor=north] {Observer};
\filldraw[black] (zmean1) circle (2pt) node[anchor=east] {\zmeanLabel};
\filldraw[black] (zmean2) circle (2pt) node[anchor=west] {\zmeanLabel};
\end{tikzpicture}
\end{subfigure}

\caption{\textit{Left}: full-sky geometry, \textit{right}: flat-sky geometry, for lensing-lensing. The dashed colored lines indicate one of the infinitely many paths which contribute to the result.
In the flat-sky case, only points at equal redshifts have a non-zero contribution.}
\label{fig:fullsky_vs_flatsky_tikz}
\end{figure}

Note that the flat-sky Limber approximation for lensing-lensing is such that only the correlations at equal redshifts contribute to the final result, which can be seen in figure~\ref{fig:fullsky_vs_flatsky_tikz}.

After performing a change of variables $k'=k_\perp \bar\chi / \lambda$, we obtain:
\begin{align}
\xi^\mathrm{len}_\text{flat}
(\bar z, r, \mu)
=&
\frac{(3\Omega_m H_0^2)^2(2-5s_1(\bar z))(2-5s_2(\bar z))}{8 \pi \bar\chi^2}
\nonumber
\\
&\times
\int_0^{\bar\chi}
\dd \lambda\,
\int_0^\infty
\dd k'\,
k'\,
P (k')
J_0\left(k' \frac{\lambda}{\bar\chi} r\sqrt{1-\mu^2}\right)
\nonumber
\\
&\times
\left(
\frac{\lambda}{\bar\chi}
\right)^4
\left[
\frac{(\bar\chi-\lambda)\bar\chi^2}{\lambda}
\right]^2\,
D_1^2(z(\lambda))\,
(1+z(\lambda))^2
\end{align}
or, somewhat more compactly:
\begin{align}
\xi^\mathrm{len}_\text{flat}
(\bar z, r, \mu)
=&
\frac{(3\Omega_m H_0^2)^2(2-5s_1(\bar z))(2-5s_2(\bar z))}{8 \pi \bar\chi^2}
\times
\int_0^{\bar\chi}
\dd \lambda\,
i\left(\frac{\lambda}{\bar\chi} r\sqrt{1-\mu^2}\right)
\nonumber
\\
&\times
\left[
\lambda(\bar\chi-\lambda)
\right]^2\,
D_1^2(z(\lambda))\,
(1+z(\lambda))^2
\label{eq:lens_lens_2pcf}
\end{align}
where $i(x)$ is given by eq.~\eqref{eq:def_i}.

\subsubsection{Lensing-lensing multipoles}
\label{s:analytic}

The multipoles of lensing-lensing in the flat-sky approximation are given by:
\begin{align}
\xi^\mathrm{len}_{\ell,\text{flat}}
(\bar z, r)
=&
\frac{2\ell + 1}{2}
\frac{(3\Omega_m H_0^2)^2(2-5s_1(\bar z))(2-5s_2(\bar z))}{8 \pi \bar\chi^2}
\nonumber
\\
&\times
\int_0^{\bar\chi}
\dd \lambda\,
\left[
\lambda(\bar\chi-\lambda)
\right]^2\,
D_1^2(z(\lambda))\,
(1+z(\lambda))^2
\nonumber
\\
&\times
\int_0^\infty
\dd k\,
k\,
P(k)\,
\mathcal{I}_\ell\left(k \frac{\lambda}{\bar\chi} r\right)
\end{align}
where now we define:
\begin{equation}
\mathcal{I}_\ell(\alpha)
\equiv
\int_{-1}^{1}
\dd \mu\,
P_\ell(\mu)\,
J_0(\alpha \sqrt{1 - \mu^2})
\label{eq:len_len_integral_multipoles}
\end{equation}
After a straightforward calculation (see appendix~\ref{s:len_len_appendix}), we obtain a simple analytic result:
\begin{equation}
\mathcal{I}_\ell(\alpha)
=
\begin{cases}
\displaystyle
C(\ell) j_\ell(\alpha),\quad \ell\;\mathrm{even}
\\
0,\quad \ell\;\mathrm{odd}
\end{cases}
\label{eq:integral_bessel}
\end{equation}
where the coefficients $C(\ell)$ are defined as:
\begin{equation}
C(\ell)
\equiv
\frac{\ell!}{2^{\ell - 1} [(\ell/2)!]^2}
\label{eq:coefficients}
\end{equation}
The end result is very similar to density-lensing, in that we just need to find integrals of spherical Bessel functions of the form:
\begin{equation}
\mathcal{K}_\ell(\alpha)
=
\alpha
\int_0^\infty
\dd k\,
k^2\,
P(k)\,
\frac{j_\ell(k \alpha)}{k \alpha}
=
2 \pi^2 \alpha I_\ell^1(\alpha)
\label{eq:lensing_lensing_analytic}
\end{equation}
which can again be computed using the FFTlog method.
The analytic expression for flat-sky lensing-lensing multipoles is thus given by:
\begin{align}
\xi^\mathrm{len}_{\ell,\text{flat}}
(\bar z, r)
=&
C(\ell)
\frac{2\ell + 1}{2}
\frac{(3\Omega_m H_0^2)^2(2-5s_1(\bar z))(2-5s_2(\bar z))}{8 \pi \bar\chi^2}
\nonumber
\\
&\times
\int_0^{\bar\chi}
\dd \lambda\,
\left[
\lambda(\bar\chi-\lambda)
\right]^2\,
D_1^2(z(\lambda))\,
(1+z(\lambda))^2\,
\mathcal{K}_\ell\left(\frac{\lambda}{\bar\chi} r\right)
\label{eq:lens_lens_analytic}
\end{align}
Expressions~\eqref{eq:den_len_analytic} and~\eqref{eq:lens_lens_analytic} are the main analytical results of this paper.
To obtain the multipoles in the flat-sky approximation, we only need to evaluate the result at a single point (for density-lensing), or compute a single integral along the line of sight (for lensing-lensing).
On the other hand, the multipoles of the full-sky results, eqs.~\eqref{eq:local_nonlocal} and~\eqref{eq:nonlocal_nonlocal}, require the numerical computation of a double and triple integral, respectively.

Additionally, since we obtained eq.~\eqref{eq:lens_lens_analytic} analytically, it will not suffer from any numerical issues as the direct evaluation of the double integral over eq.~\eqref{eq:lens_lens_2pcf} (one over $\lambda$ and one over $\mu$).
For convenience, in table~\ref{t:fftlogs} we report the FFTlog integrals which we need to compute for density-lensing and lensing-lensing, which, at least in linear theory, need to be computed only at $z = 0$.

\begin{table}[H]
\caption{The various FFTlog integrals needed for the computation of the different contributions to the 2PCF or its multipoles in the flat-sky approximation.}
\centering
\begin{tabular}{c c}
\toprule
quantity & integral(s) needed\\
\midrule
$\xi^\text{den-len}(\bar z, r, \mu)$ & $I^{1/2}_{-1/2}$\\
$\xi^\text{den-len}_\ell(\bar z, r)$ & $I^{\ell/2 - k + 3/2}_{\ell/2 - k + 1/2}$, $k = 0, \ldots, \ell/2$\\
$\xi^\text{len-len}(\bar z, r, \mu)$ & $I^{1/2}_{-1/2}$\\
$\xi^\text{len-len}_\ell(\bar z, r)$ & $I^1_\ell$\\
\bottomrule
\end{tabular}
\label{t:fftlogs}
\end{table}

\subsection{Generalizations and extensions}
\label{s:generalizations}

Below we briefly discuss how to incorporate a redshift-dependent bias, nonlinearities, and certain phenomenological modifications of gravity in the flat-sky approximation.

\subsubsection{Redshift-dependent bias}
\label{s:bias}

When deriving the flat-sky results eq.~\eqref{eq:den_len_analytic} and~\eqref{eq:lens_lens_analytic}, we have implicitly assumed that the galaxy and magnification bias are redshift-independent within a given redshift bin, which is usually not the case for a real survey, and can present a problem for a rapidly-varying bias when using large redshift bins.
Note that at linear order in perturbation theory, the galaxy bias can be described by a constant plus a term inversely proportional to the growth rate $D_1$~\citep{Desjacques_2018}.%
However, in what follows, we will use $Q(z)$ to denote any redshift-dependent bias (galaxy, magnification, etc.).
The bias always enters as a prefactor for the flat-sky 2PCF, i.e. we have one of the following situations:
\begin{align}
\xi_\text{flat}(\bar z, r, \mu)
=&
Q_i(z_{1,2})
f(\bar z, r, \mu)
\\
\xi_\text{flat}(\bar z, r, \mu)
=&
Q_i(z_1) Q_j(z_2)
g(\bar z, r, \mu)
\end{align}
where $f, g$ are some functions in the flat-sky approximation that do not depend on $Q_i$ and $Q_j$.
The first case arises only in the case of certain cross-correlations (such as density-RSD), while the second can appear in both auto- and cross-correlation terms.
For concreteness, we focus on the second case; the calculation for the first proceeds in an analogous way.

To fix the problem of a redshift-dependent bias for the 2PCF, we can simply make the replacement $Q_{i}(\bar z) \rightarrow Q_{i}(z_{1,2})$, i.e. just evaluate the bias parameters at the actual redshifts $z_1$ and $z_2$, as is done for the full-sky 2PCF; unfortunately, then the multipoles cannot be obtained semi-analytically with the method used previously.

On the other hand, for a given cosmology, we can in principle expand any bias $Q_i(z)$ as a power series in the comoving distance:
\begin{equation}
Q_i(z)
=
\sum_n
a_n^{(i)}\,
\chi^n
\label{eq:bias_taylor}
\end{equation}
The above can be made arbitrarily accurate on a given redshift interval $[0, z_\text{max}]$ by increasing the number of terms in the sum.
Since from the geometry of the problem we have $\chi_{1,2} = \bar\chi \pm r\,\mu/2$, the multipoles of, for instance, lensing-lensing, can be shown to be proportional to (see appendix~\ref{s:bias_full} for the full computation):
\begin{align}
\xi_{\ell, \text{flat}}
\propto&
\int_{-1}^1
\dd \mu\,
P_\ell(\mu)\,
Q_i(z_1)\,
Q_j(z_2)\,
J_0(\alpha\, \sqrt{1 - \mu^2})
\nonumber
\\
=&
\sum_{n_1,n_2}
a_{n_1}^{(i)}\,
b_{n_2}^{(j)}\,
\sum_{k_1 = 0, k_2 = 0}^{n_1, n_2}
\factorial{n_1}{k_1}\,
\factorial{n_2}{k_2}\,
\bar\chi^{n_1 + n_2 - k_1 - k_2}
\left(
\frac{r}{2}
\right)^{k_1 + k_2}
(-1)^{k_1}
\nonumber
\\
&
\times
\left[
(-1)^{k_1 + k_2 + \ell} + 1
\right]
\Omega\left(k_1 + k_2, \ell; \alpha \right)
\label{eq:bias_analytic}
\end{align}
where $\Omega(n, \ell; \alpha)$ is given by eq.~\eqref{eq:omega_def}, and we've omitted the integrals over $\lambda$ and $k$ for brevity.

The last line contains exactly the integral~\eqref{eq:len_len_integral_multipoles}; therefore, we can account for a redshift-dependent bias at the expense of computing additional integrals of the form $I^n_\ell$.
An analogous computation with the same conclusion can be done for the case of density-lensing, as well as the non-integrated terms.

\subsubsection{Nonlinearities}
\label{s:nl}

To account for nonlinearities for either local or non-local terms in full-sky, we could naively make the substitution:
\begin{equation}
D_1(z_1) D_1(z_2) P(k) \rightarrow P_\mathrm{NL}(k, z_1, z_2)
\label{eq:nonlinear_substitution}
\end{equation}
where $P_\text{NL}$ denotes the nonlinear matter unequal-time cross-spectrum between redshifts $z_1$ and $z_2$; however, there are several subtleties related to the above.

For one, we need a way to model the cross-spectrum; this is usually accomplished by writing $P_\text{NL}(k, z_1, z_2)$ in terms of the more well studied equal-time correlators $P_\text{NL}(k, z)$, which can be computed using an analytical model such as halofit~\cite{Takahashi:2012em}, the augmented halo model~\cite{Mead_2015}, or, alternatively, using an emulator such as EuclidEmulator~\cite{euclidemu2019}.
To relate the two, we can either use the geometric mean approximation~\cite{Kitching_2017,Chisari_2019}, in which we assume that:
\begin{equation}
P^\mathrm{geom}_\mathrm{NL}(k, z_1, z_2)
\approx
\sqrt{P_\mathrm{NL}(k, z_1) P_\mathrm{NL}(k, z_2)}
\label{eq:geom_mean_approx}
\end{equation}
or, if we insist on higher accuracy, the recently studied Zel'dovich approximation~\cite{Chisari_2019}:
\begin{equation}
P^\mathrm{Zel}_\mathrm{NL}
\approx
\sqrt{P_\mathrm{NL}(k, z_1) P_\mathrm{NL}(k, z_2)}\,
\mathrm{e}^{-\left[ D_1(z_1) - D_1(z_2) \right]^2 \left( k / k_\mathrm{NL} \right)^2}
\label{eq:zel_approx}
\end{equation}
where $D_1(z)$ is the growth rate obtained from linear theory, and we defined:
\begin{equation}
k_\mathrm{NL}^{-2}
\equiv
\frac{1}{12 \pi^2}
\int_0^\infty
\dd k\,
P(k)
\label{eq:k_nl_def}
\end{equation}
where $P(k)$ is the linear matter power spectrum today.
Note that the latter approximation is simply a generalization of the former, so that the Zel'dovich approximation in linear theory becomes:
\begin{equation}
P^\text{Zel}_\text{NL}(k, z_1, z_2)
\xrightarrow{\text{lin. th.}}
D_1(z_1)\,
D_1(z_2)\,
P(k)\,
\mathrm{e}^{-\left[ D_1(z_1) - D_1(z_2) \right]^2 \left( k / k_\mathrm{NL} \right)^2}
\end{equation}
If we additionally set $k_\text{NL} \rightarrow \infty$, we recover the standard linear theory result.

On the other hand, we do not have the problem above in the flat-sky approximation, since only equal redshifts contribute to the final result, so we may simply use the nonlinear power spectrum $P_\text{NL}(k, z)$.

The second problem is that the contributions to eq.~\eqref{eq:number_counts} involve the velocity field $\vec{V}$, which can be written as:
\begin{equation}
\vec V
=
-\nabla V
+
\vec{v}_R
\label{eq:velocity_decomposition}
\end{equation}
where $V$ is again the velocity potential, and $\vec{v}_R$ is the rotational part of the velocity field, such that $\nabla \cdot \vec{v}_R = 0$ and $\boldsymbol \omega = \nabla \times \vec{v}_R$, where $\boldsymbol \omega$ is the vorticity field.

$\vec V$ is related to the overdensity $\delta$ via the continuity equation, and in linear theory we set $\vec{v}_R = 0$, but this no longer holds as we go beyond linear theory, and therefore we need to accurately model the various nonlinear spectra that could enter into eq.~\eqref{eq:number_counts}, namely $P_{\delta}$, $P_{V}$, $P_{\vec{v}_R}$, and their corresponding cross-spectra, which become increasingly important at low redshift~\cite{Jelic_Cizmek_2018}.

Finally, for terms involving lensing (such as density-lensing or lensing-lensing), we have the additional effect of second order lensing~\cite{Bonvin_2015}, which we have to consistently take into consideration.

With the above in mind, to obtain the naive nonlinear result in full-sky, we need to modify our eq.~\eqref{eq:fftlog} slightly, and instead consider:
\begin{equation}
I_\ell^{n,\text{(NL)}}(z_1, z_2, r)
=
\frac{1}{2\pi^2}
\int_0^\infty
\dd k\,
k^2\,
P_\text{NL}(k, z_1, z_2)
\frac{
j_\ell (k r)
}
{
(k r)^n
}
\label{eq:fftlog_nl_fullsky}
\end{equation}
The above needs to be calculated for each pair of redshifts $\{z_1, z_2\}$, which is very time consuming, especially if we are computing the correlation of non-local - non-local terms.

In the flat-sky case, due to the fact that only equal redshifts contribute, we just need to evaluate the above for $z_1 = z_2$, which, combined with the semi-analytic flat-sky results, eq.~\eqref{eq:den_len_analytic} and~\eqref{eq:lens_lens_analytic}, greatly reduces the total computation time.

\subsubsection{Modifications of gravity}
\label{s:mond}

As an example of modifications of gravity, we may parametrize deviations from general relativity (GR) with the functions $\Sigma(k, z)$ and $\gamma(k, z)$ via the following equations (see for instance~\cite{Zhao_2009}):
\begin{align}
-k^2\, \Phi (\vec k, z)
&=
4 \pi G\, \Sigma (k, z)\, a^2\, \rho\, \delta(\vec k, z)
\label{eq:def_sigma}
\\
\gamma (k, z)
&=
\frac{\Psi(\vec k, z)}{\Phi(\vec k, z)}
\label{eq:def_gamma}
\end{align}

Of course, for $\Sigma = 1$ and $\gamma = 1$ we recover the familiar GR values.

If we assume that $\Sigma$ and $\gamma$ are deterministic, and not random variables, this amounts to the substitution:
\[
P(k, z_1, z_2) \rightarrow \frac{1}{2}\left[ 1 + \gamma(k, z_1) \right] \Sigma(k, z_1) \frac{1}{2}\left[ 1 + \gamma(k, z_2) \right] \Sigma(k, z_2) P(k, z_1, z_2)
\]
in the case of lensing-lensing, and:
\[
P(k, z_1, z_2) \rightarrow \frac{1}{2}\left[ 1 + \gamma(k, z_2) \right] \Sigma(k, z_2) P(k, z_1, z_2)
\]
in the case of density-lensing.

\section{Results}
\label{s:results}

In the below we discuss the accuracy and the performance of the flat-sky approximation compared to the full-sky results as implemented in a new version of the code COFFE.

\subsection{Accuracy}
\label{s:accuracy}

For concreteness, we assume a flat $\Lambda$CDM cosmology, and the parameters we use are shown in table~\ref{t:params}.
We take the values $b = 1$, $s = 0$, and $f_\text{evo} = 0$ for the galaxy, magnification, and evolution bias, respectively.
The linear matter power spectrum has been generated with the code CLASS~\cite{Blas:2011rf}.
The maximum separation in the plots below has been set so that the size of each redshift bin is constant, with bin half-width $\Delta z = 0.1$.

In what follows, we show the comparison between the flat-sky approximation and full-sky, with contributions to the 2PCF from:

\begingroup
\begin{enumerate}[label=\alph*)]
\setlength\itemsep{0.1em}
\item
density + RSD + Doppler (auto- and cross-correlations)
\item
density + lensing (cross-correlations only)
\item
lensing (auto-correlation)
\item
density + RSD + Doppler + lensing (auto- and cross-correlations)
\end{enumerate}
\endgroup
\begin{table}[ht!]
\caption{
The fiducial $\Lambda$CDM parameters used for the calculation
}
\centering
\begin{tabular}{c | c c c c c}
\toprule
parameter & $\Omega_\mathrm{b}$ & $\Omega_\mathrm{cdm}$ & $h$ & $n_s$ & $\log 10^{10} A_s$
\\
\midrule
value & 0.05 & 0.25 & 0.67 & 0.96 & 3.06
\\
\bottomrule
\end{tabular}
\label{t:params}
\end{table}
\subsubsection{Non-integrated terms}
\label{s:nonintegrated}

\begin{figure}[H]
\begin{subfigure}[t]{0.45\textwidth}
\includegraphics[trim=20 0 20 2,clip,width=\textwidth]{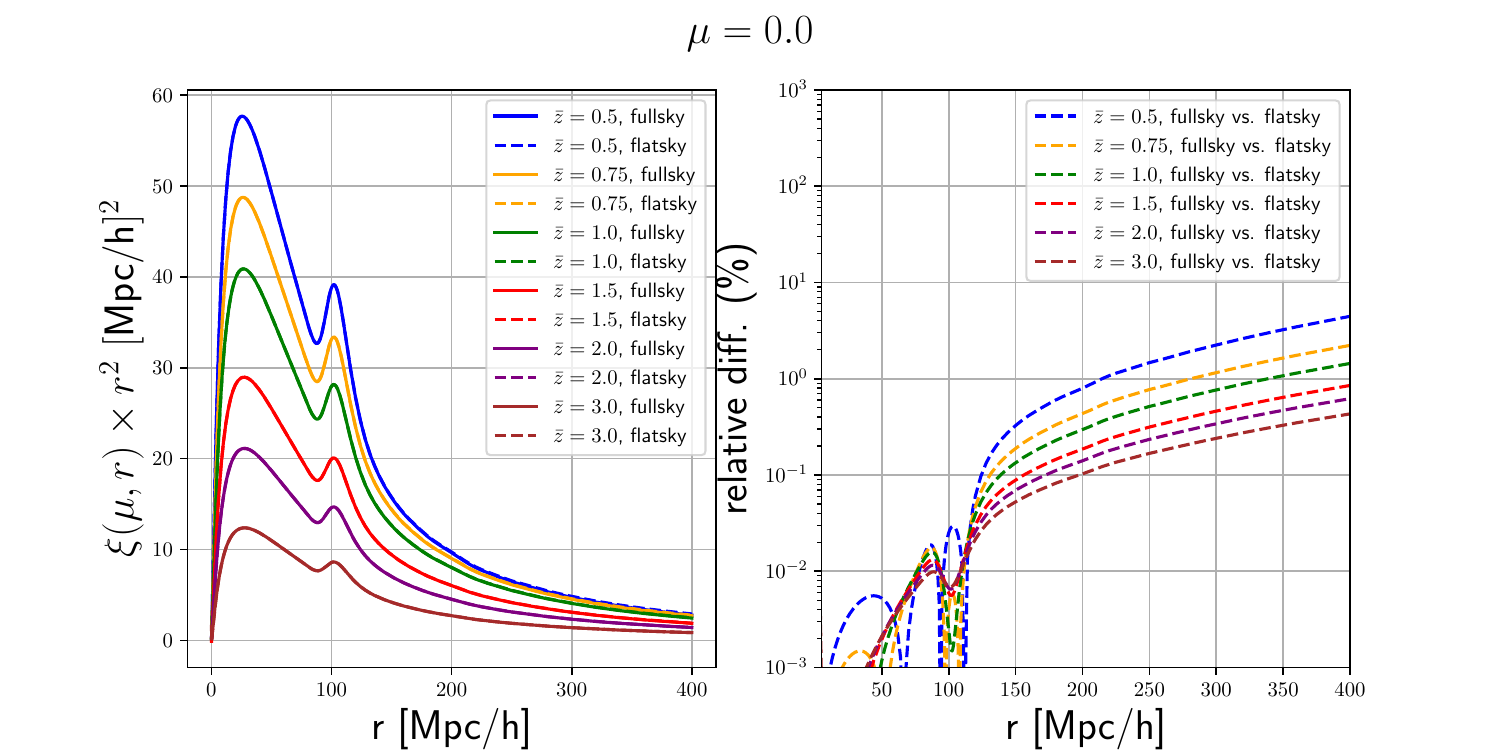}
\label{f:den_rsd_d1_mu0}
\end{subfigure}
\begin{subfigure}[t]{0.45\textwidth}
\includegraphics[trim=20 0 20 2,clip,width=\textwidth]{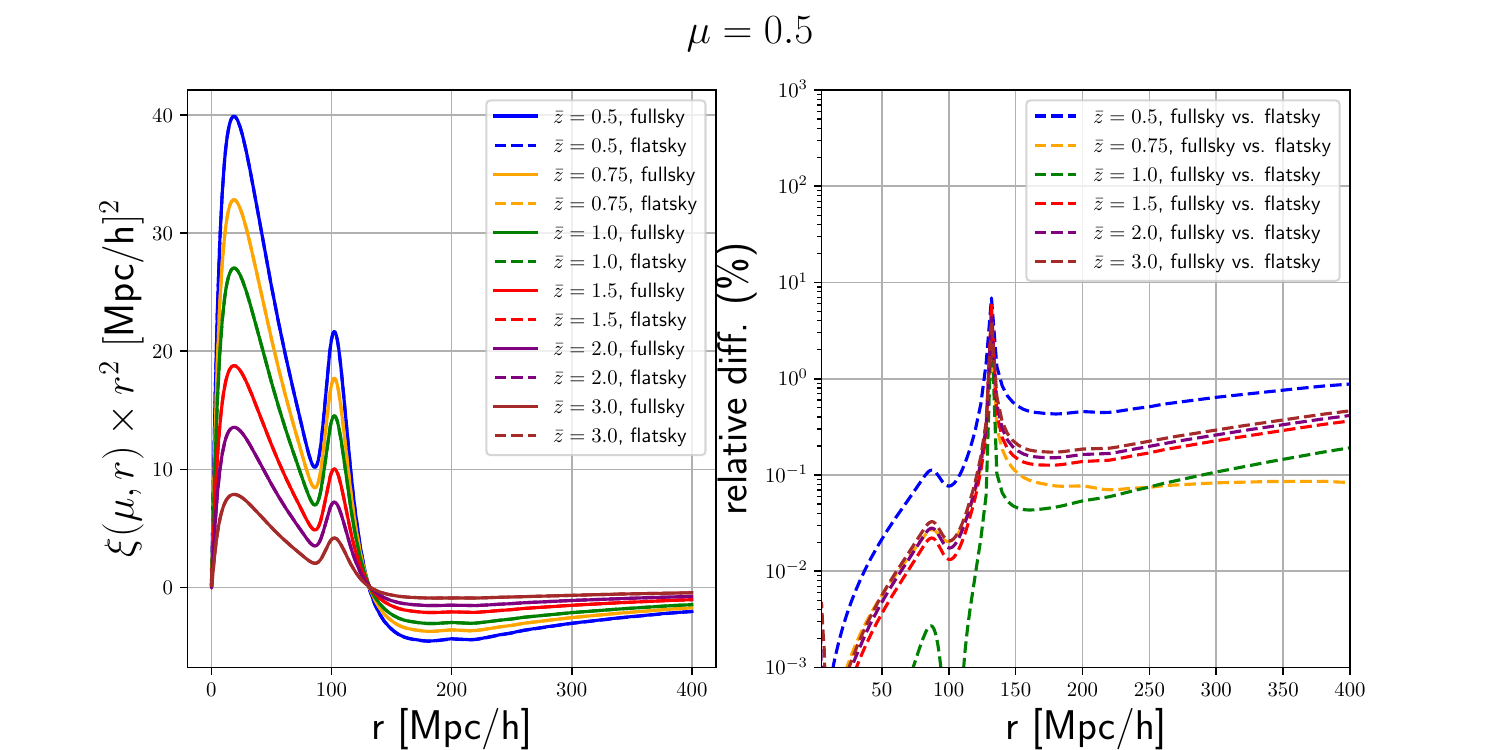}
\label{f:den_rsd_d1_mu1}
\end{subfigure}
\begin{subfigure}[t]{0.45\textwidth}
\includegraphics[trim=20 0 20 2,clip,width=\textwidth]{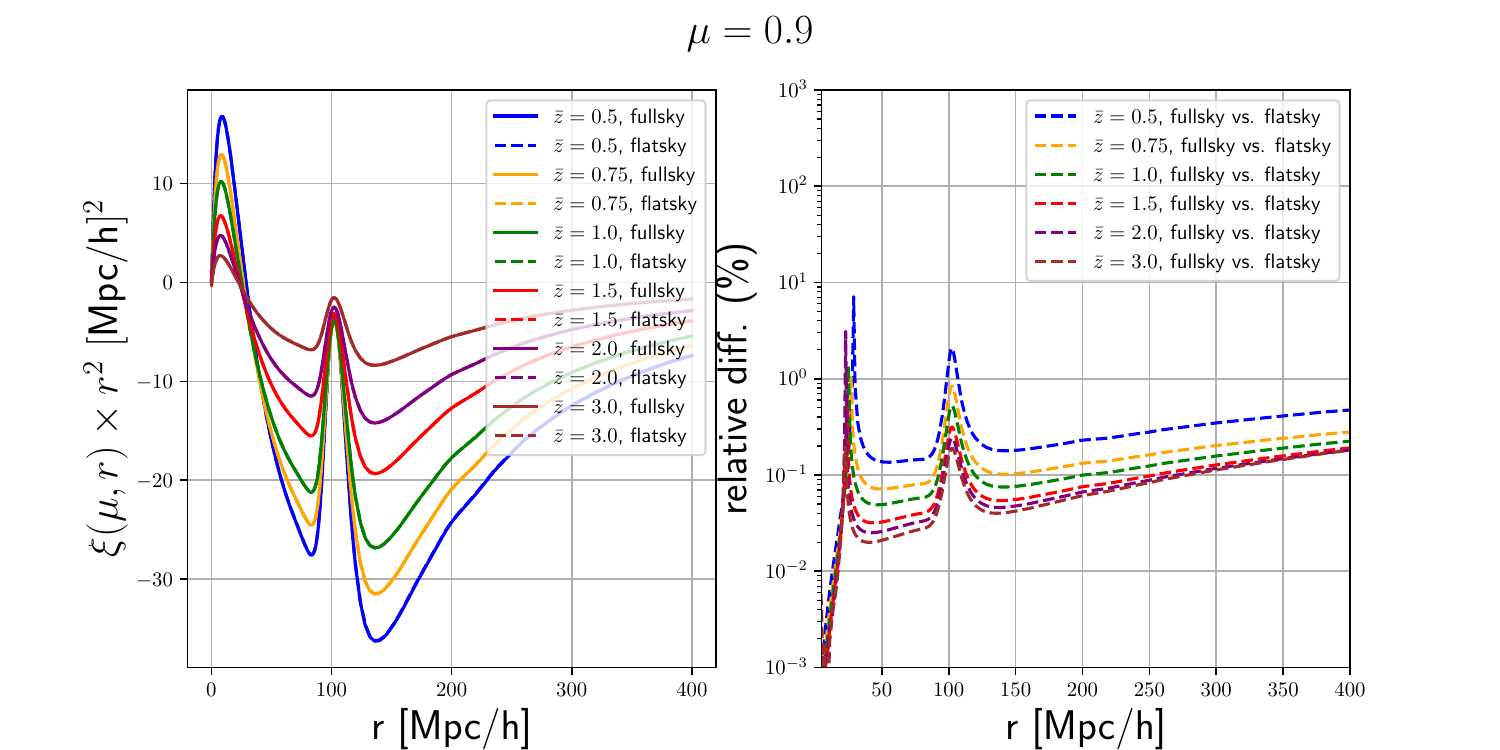}
\end{subfigure}
\label{f:den_rsd_d1_mu2}
\begin{subfigure}[t]{0.45\textwidth}
\includegraphics[trim=20 0 20 2,clip,width=\textwidth]{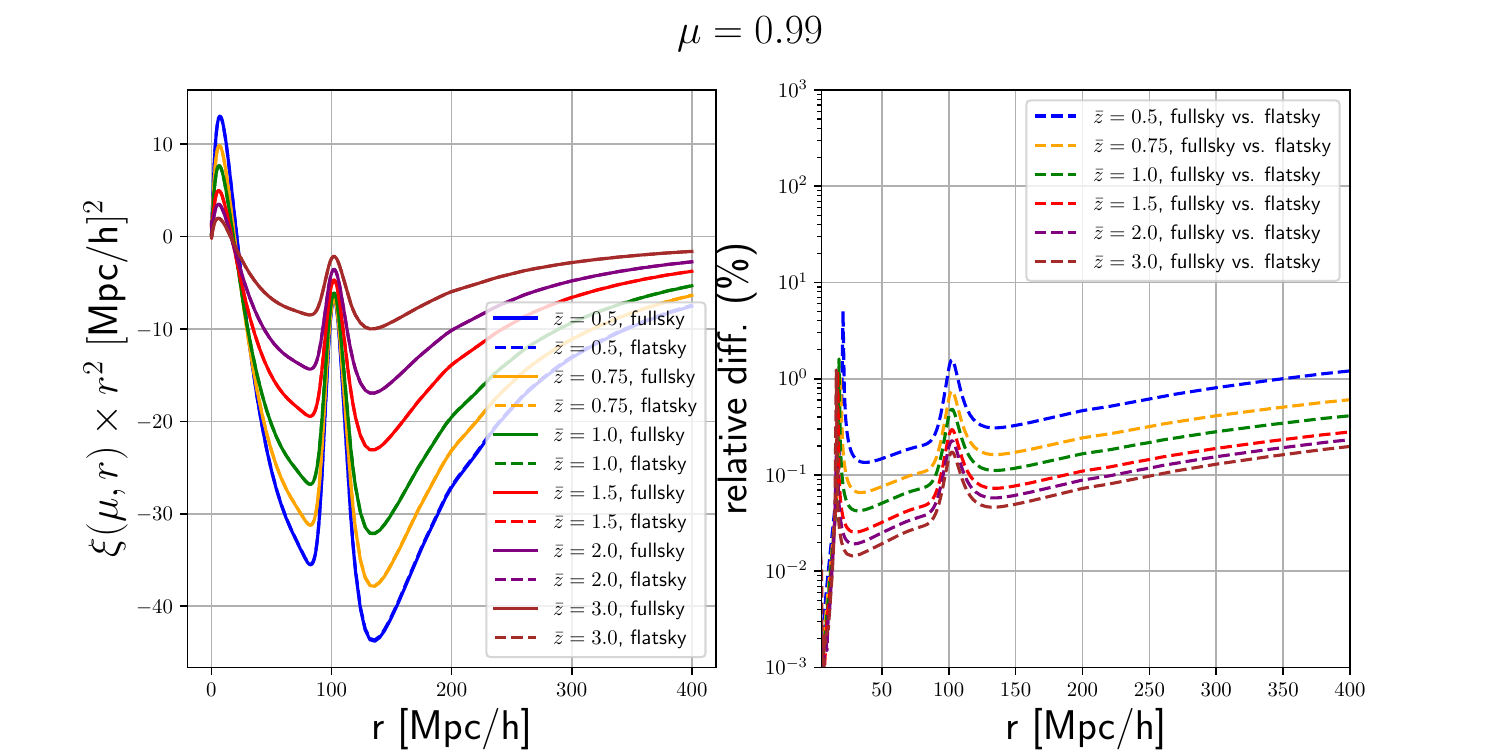}
\label{f:den_rsd_d1_mu3}
\end{subfigure}
\caption{The 2PCF for $\mu = 0$ (top left), $\mu = 0.5$ (top right), $\mu = 0.9$ (bottom left), and $\mu = 0.99$ (bottom right), with contributions from density, RSD, and Doppler, computed at various redshifts, in full-sky (solid) and flat-sky (dashed).
The relative difference (in percent) between full-sky and flat-sky is indicated on the right plot of each figure.
}
\label{f:den_rsd_d1_2pcf}
\end{figure}

\begin{figure}[H]
\centering
\begin{subfigure}[t]{0.49\textwidth}
\includegraphics[trim=20 0 20 2,clip,width=\textwidth]{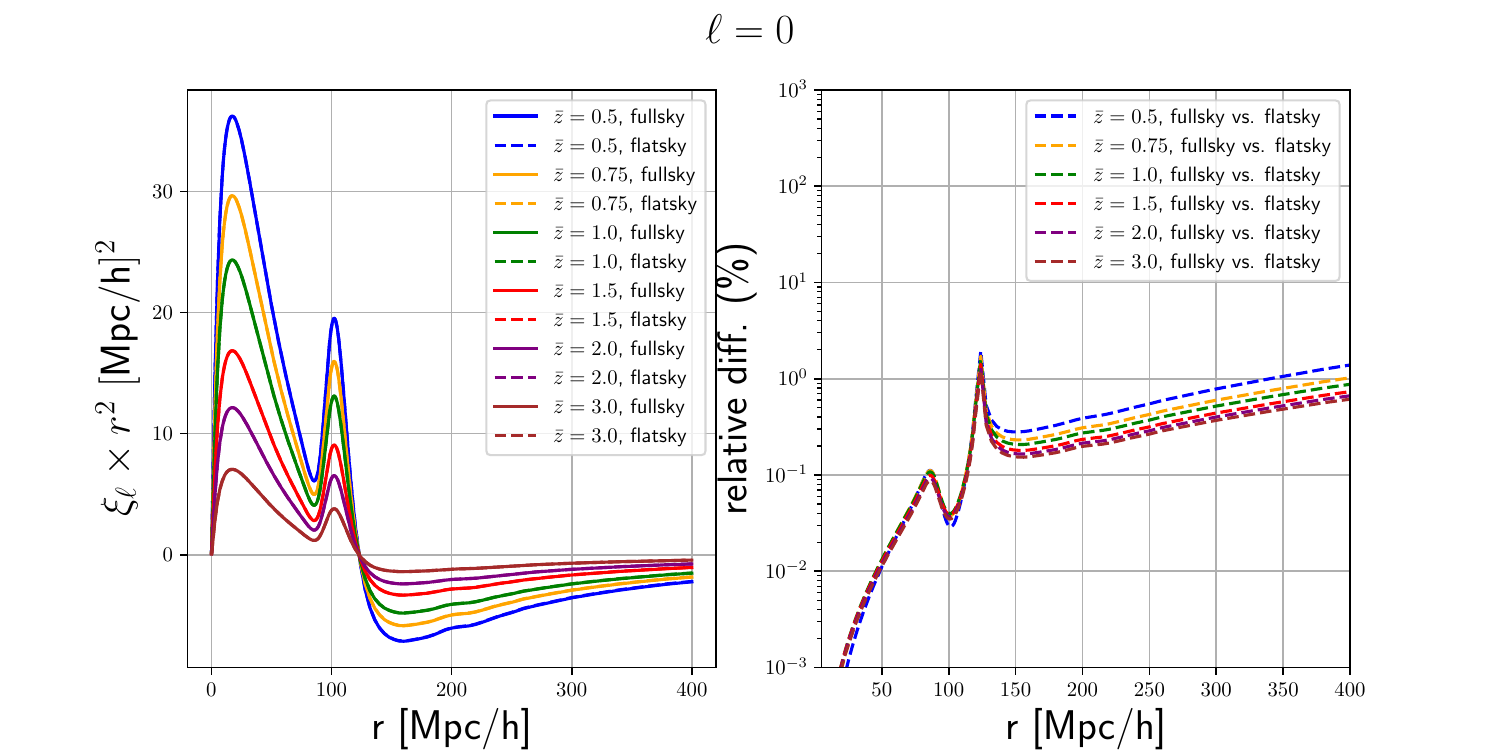}
\label{f:den_rsd_d10}
\end{subfigure}
\begin{subfigure}[t]{0.49\textwidth}
\includegraphics[trim=20 0 20 2,clip,width=\textwidth]{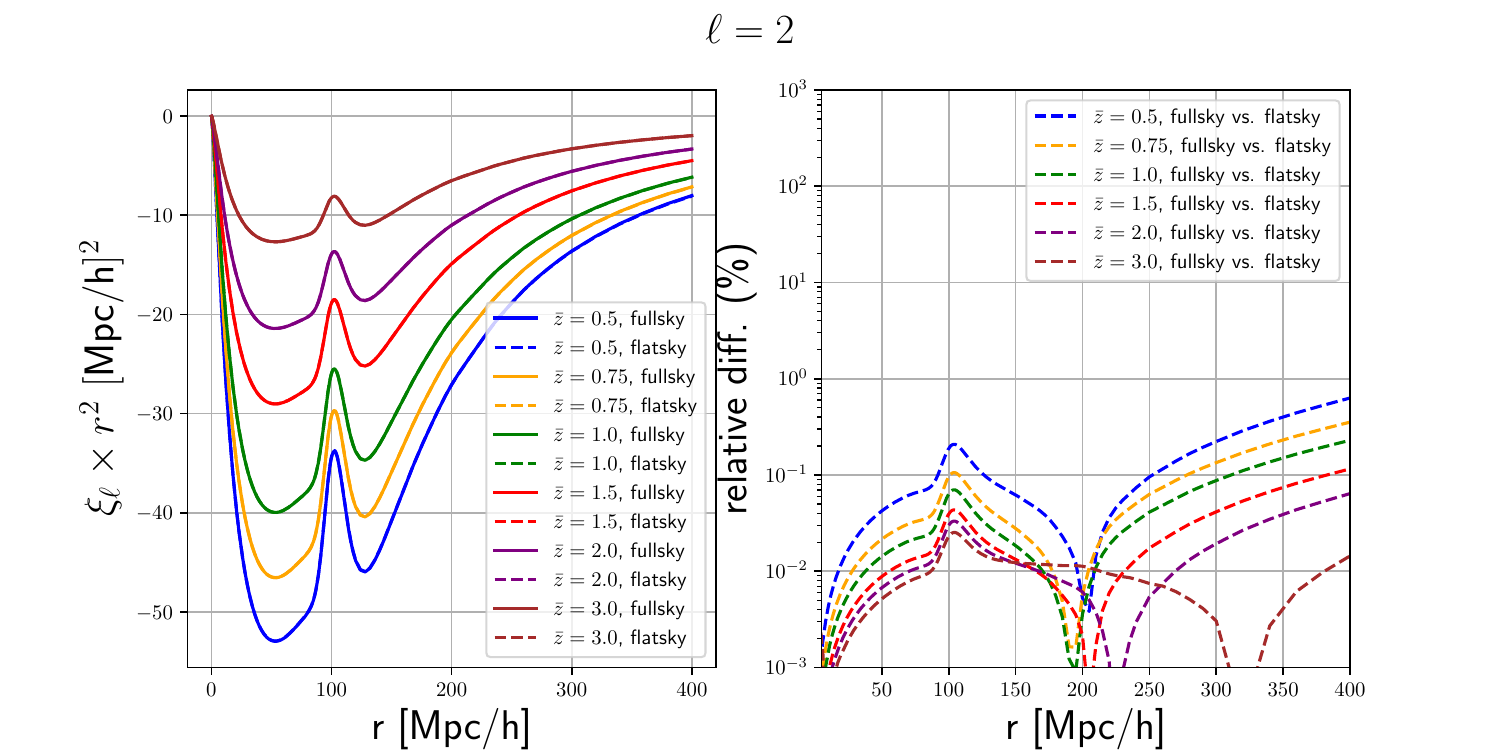}
\label{f:den_rsd_d12}
\end{subfigure}
\begin{subfigure}[]{0.6\textwidth}
\includegraphics[trim=20 0 20 2,clip,width=\textwidth]{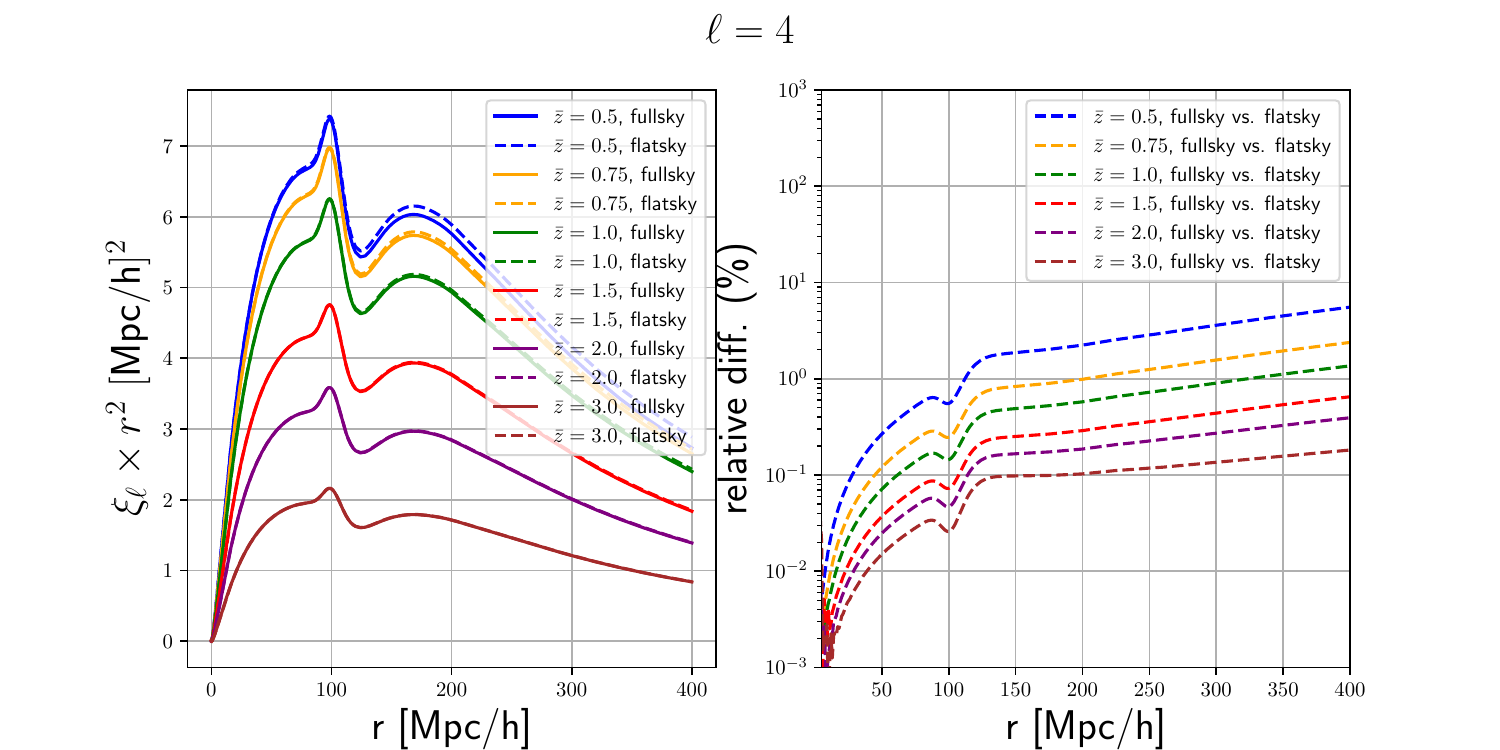}
\label{f:den_rsd_d14}
\end{subfigure}
\caption{The $\ell = 0$ (top left), $\ell = 2$ (top right), $\ell = 4$ (bottom) multipoles of the 2PCF, with contributions from density, RSD, and Doppler, computed at various redshifts, in full-sky (solid) and flat-sky (dashed).
The relative difference (in percent) between full-sky and flat-sky is indicated on the right plot of each figure.
The 'glitch' at $r \sim 120\; \mathrm{Mpc}/h$ comes from the monopole passing through zero.
}
\label{f:den_rsd_d1_multipoles}
\end{figure}

The results for the non-integrated terms, the sum of eqs.~\eqref{eq:den_flatsky}--\eqref{eq:rsd_d1_flatsky}, are shown in figures~\ref{f:den_rsd_d1_2pcf} and \ref{f:den_rsd_d1_multipoles}, for the 2PCF and its multipoles, respectively.

The 2PCF in flat-sky seems to be quite accurate; for all configurations considered, the relative error with respect to full-sky is less than 10\%.
The 'glitch' at $r \approx 120\; \text{Mpc}/h$ that appears on the right-hand side of the figures for certain values of the angle $\mu$ is caused by the correlation function passing through zero.
As we go to higher redshifts ($z \gtrsim 1$), the error decreases, and aside from the behavior in the vicinity of the 'glitch', is at most 1\% for all separations.
The difference between flat-sky and full-sky at $\mu = 0$, i.e. for the angular correlation function, is caused solely by the RSD and Doppler terms, because, in flat-sky, the contribution to the angular correlation function from the density auto-correlation term is exactly equal to the full-sky result.

Likewise, the accuracy of the flat-sky approximation for the monopole and the quadrupole is better than 1\% for redshifts $z \gtrsim 1$, and reaches at most $\approx 2 \%$ for lower redshifts.
The hexadecapole is somewhat worse, and at $z < 1$ the error becomes larger than 5\% for large separations, $r \gtrsim 300\,\textrm{Mpc}/h$.
This is a consequence of the fact that the density-RSD and Doppler terms do not contribute at all to the $\ell = 4$ multipole in flat-sky, while they have a non-negligible contribution in full-sky.

\subsubsection{Density-lensing}

\begin{figure}[H]
\centering
\begin{subfigure}[t]{0.49\textwidth}
\includegraphics[trim=20 0 20 2,clip,width=\textwidth]{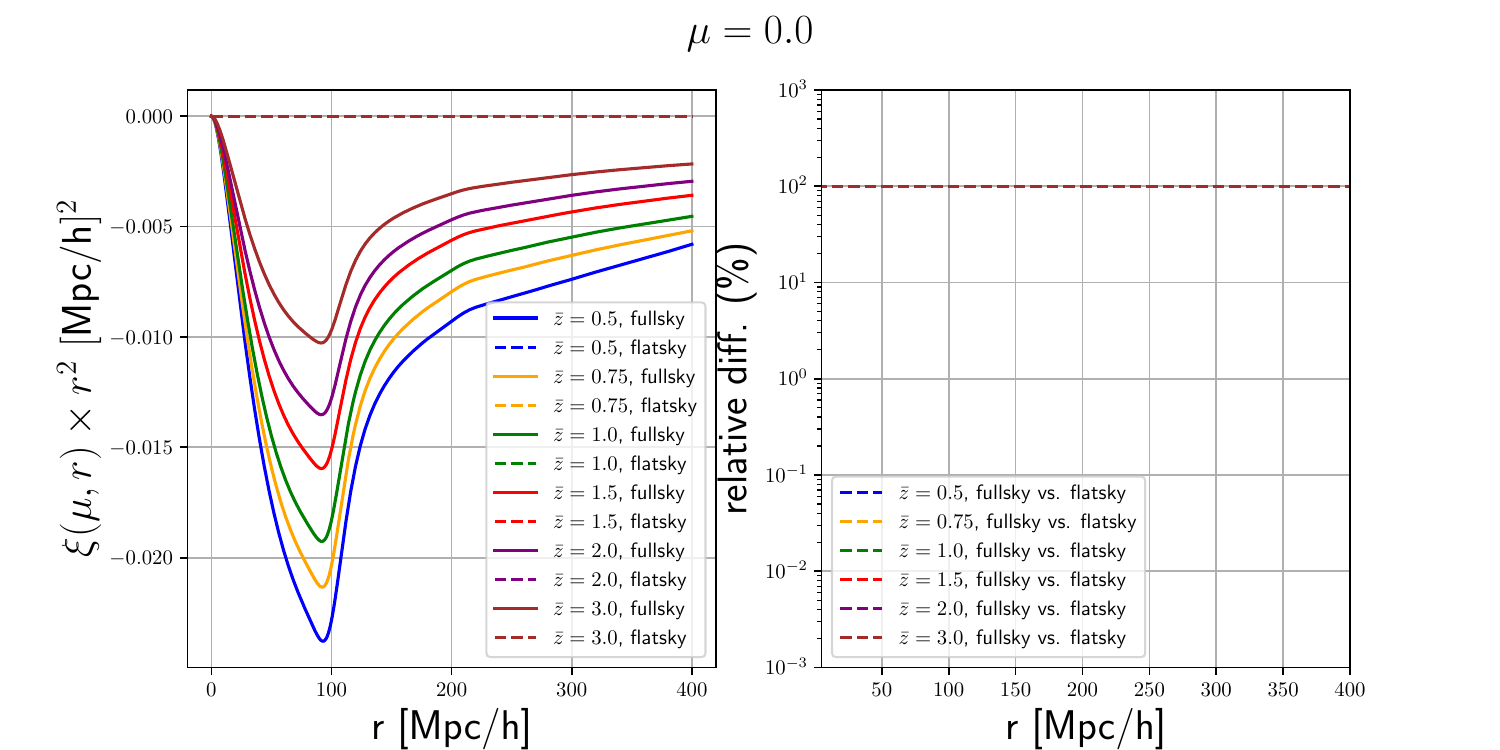}
\label{f:den_len_mu0}
\end{subfigure}
\begin{subfigure}[t]{0.49\textwidth}
\includegraphics[trim=20 0 20 2,clip,width=\textwidth]{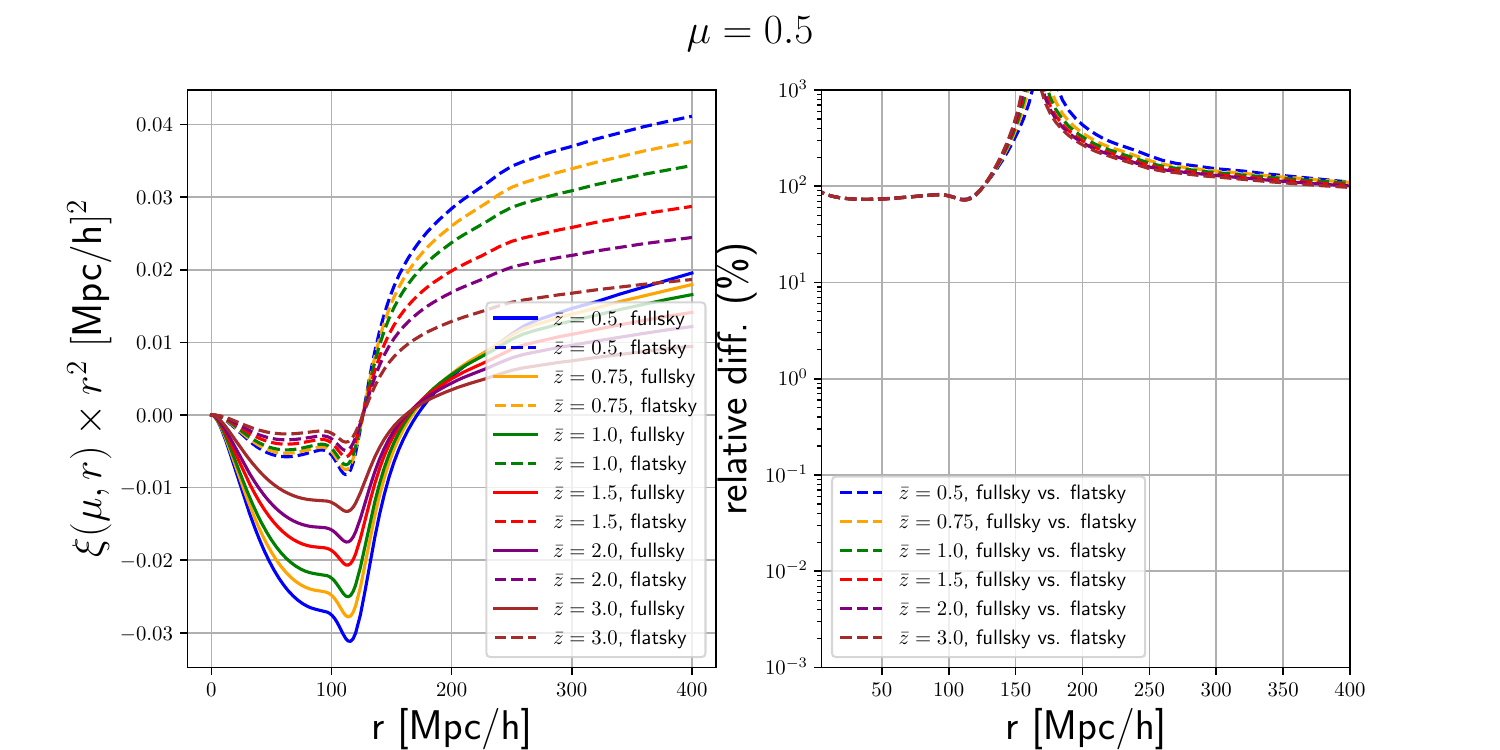}
\label{f:den_len_mu1}
\end{subfigure}
\begin{subfigure}[t]{0.49\textwidth}
\includegraphics[trim=20 0 20 2,clip,width=\textwidth]{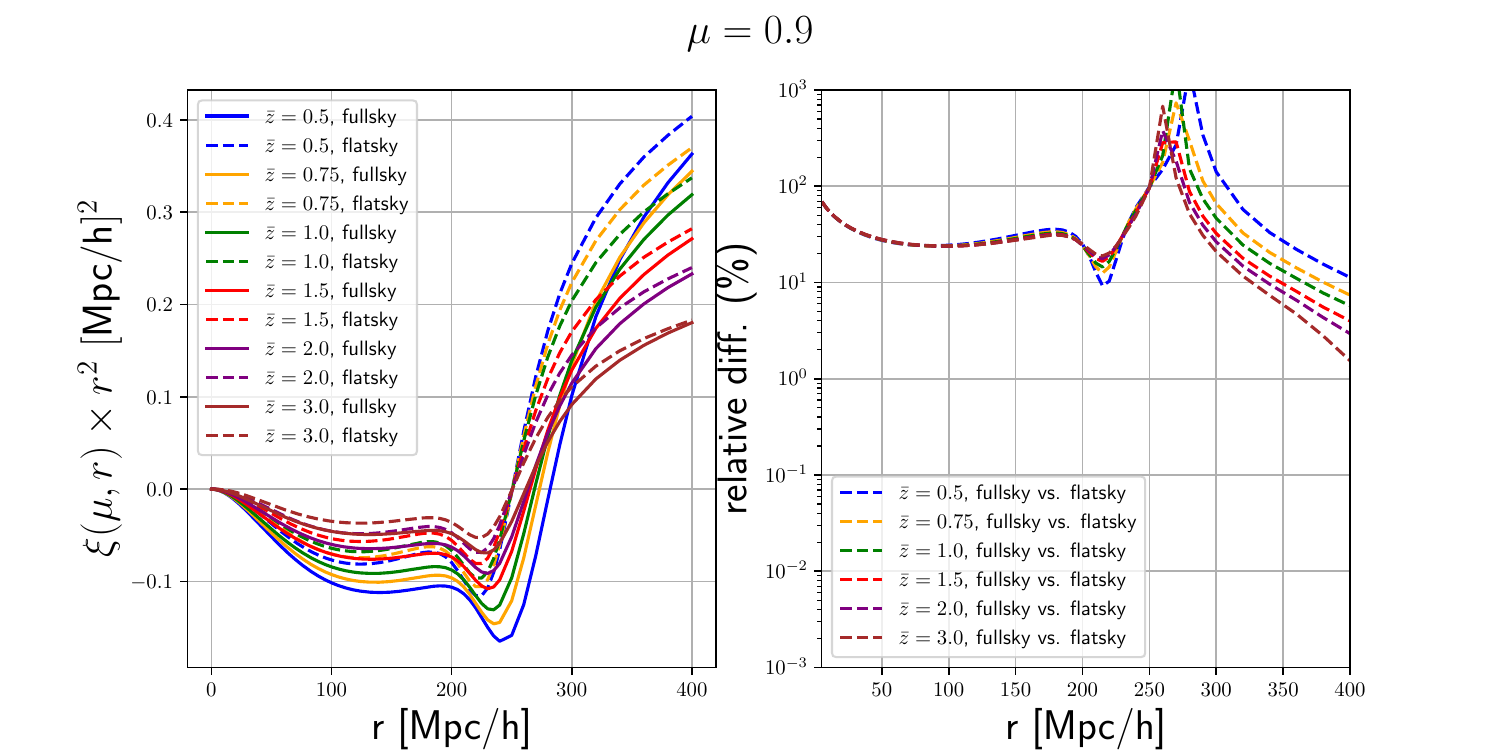}
\end{subfigure}
\label{f:den_len_mu2}
\begin{subfigure}[t]{0.49\textwidth}
\includegraphics[trim=20 0 20 2,clip,width=\textwidth]{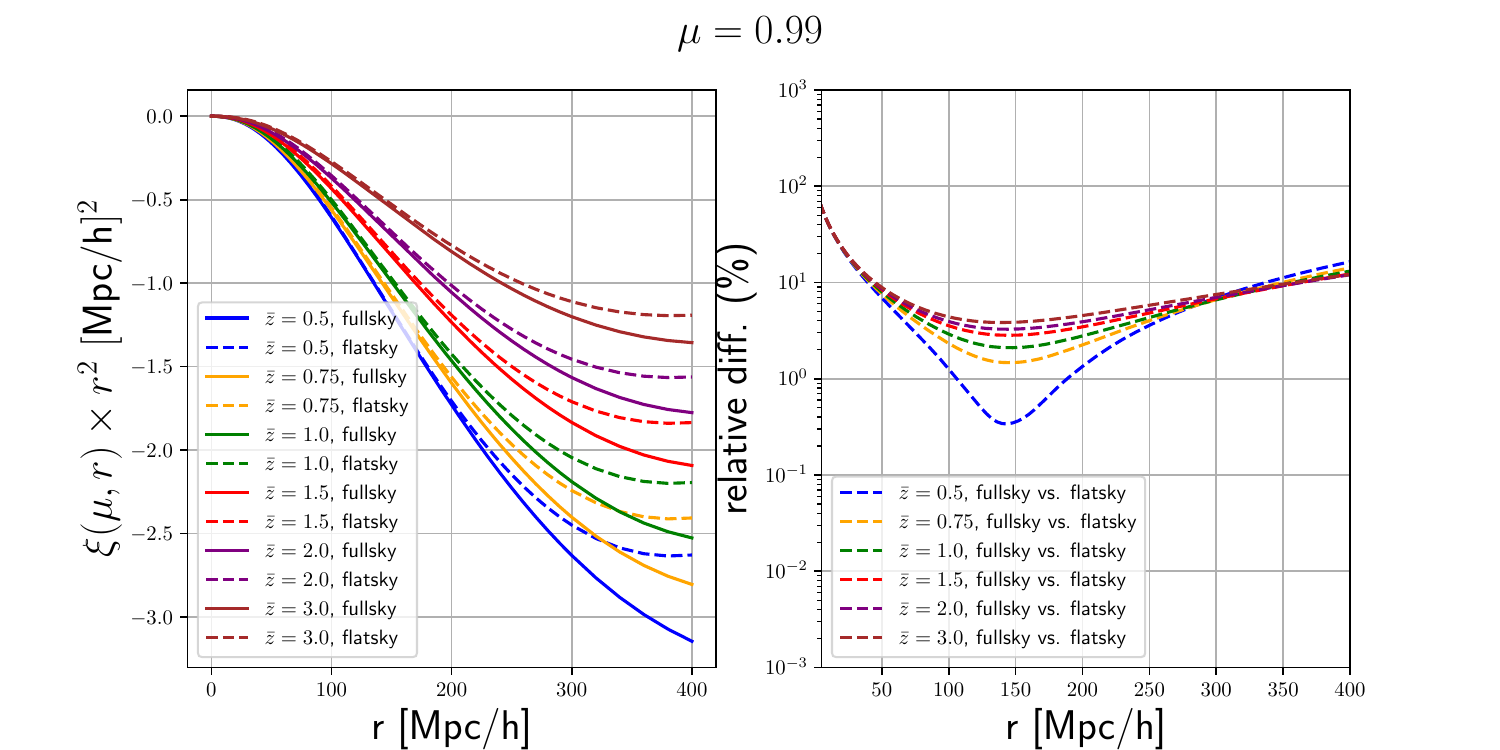}
\label{f:den_len_mu3}
\end{subfigure}
\begin{subfigure}[t]{0.49\textwidth}
\includegraphics[trim=20 0 20 2,clip,width=\textwidth]{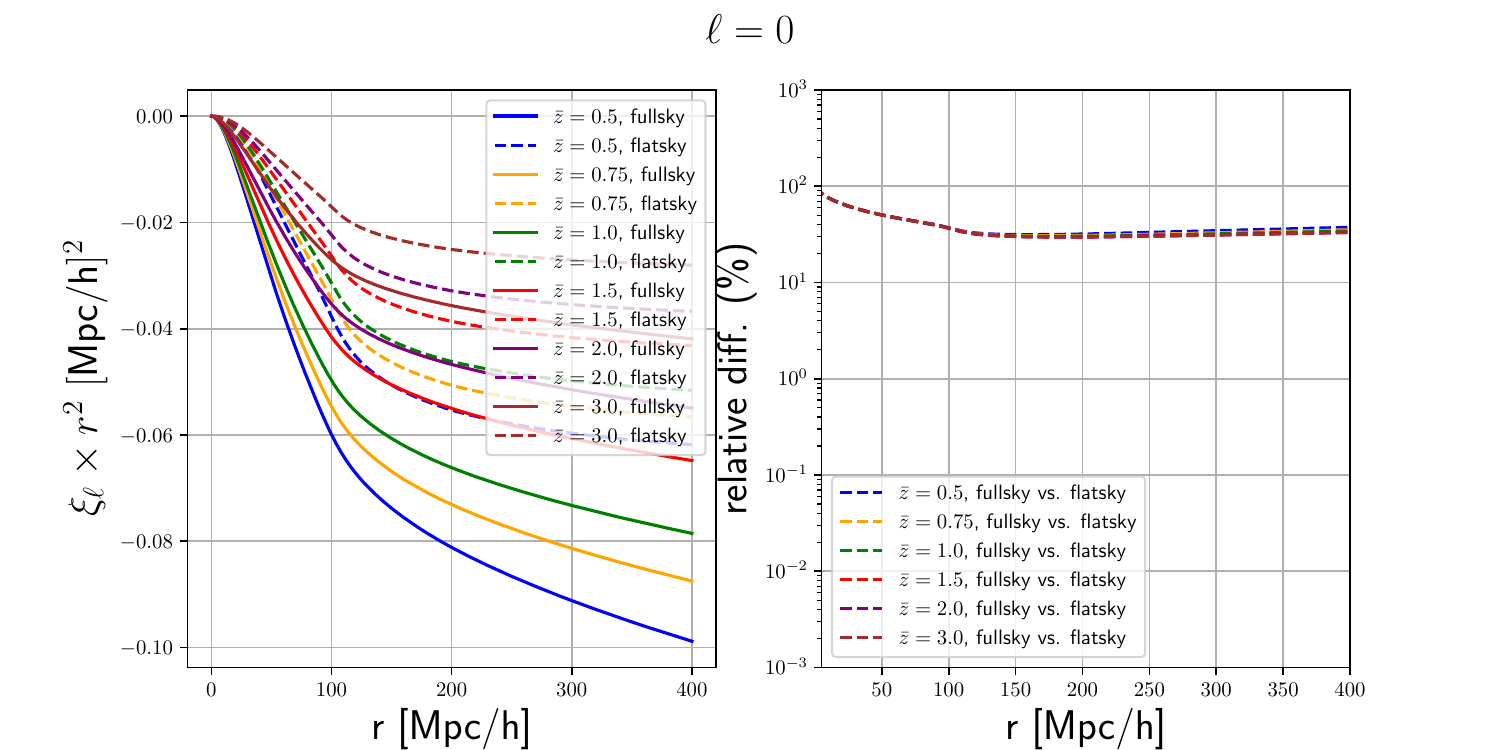}
\label{f:den_len0}
\end{subfigure}
\begin{subfigure}[t]{0.49\textwidth}
\includegraphics[trim=20 0 20 2,clip,width=\textwidth]{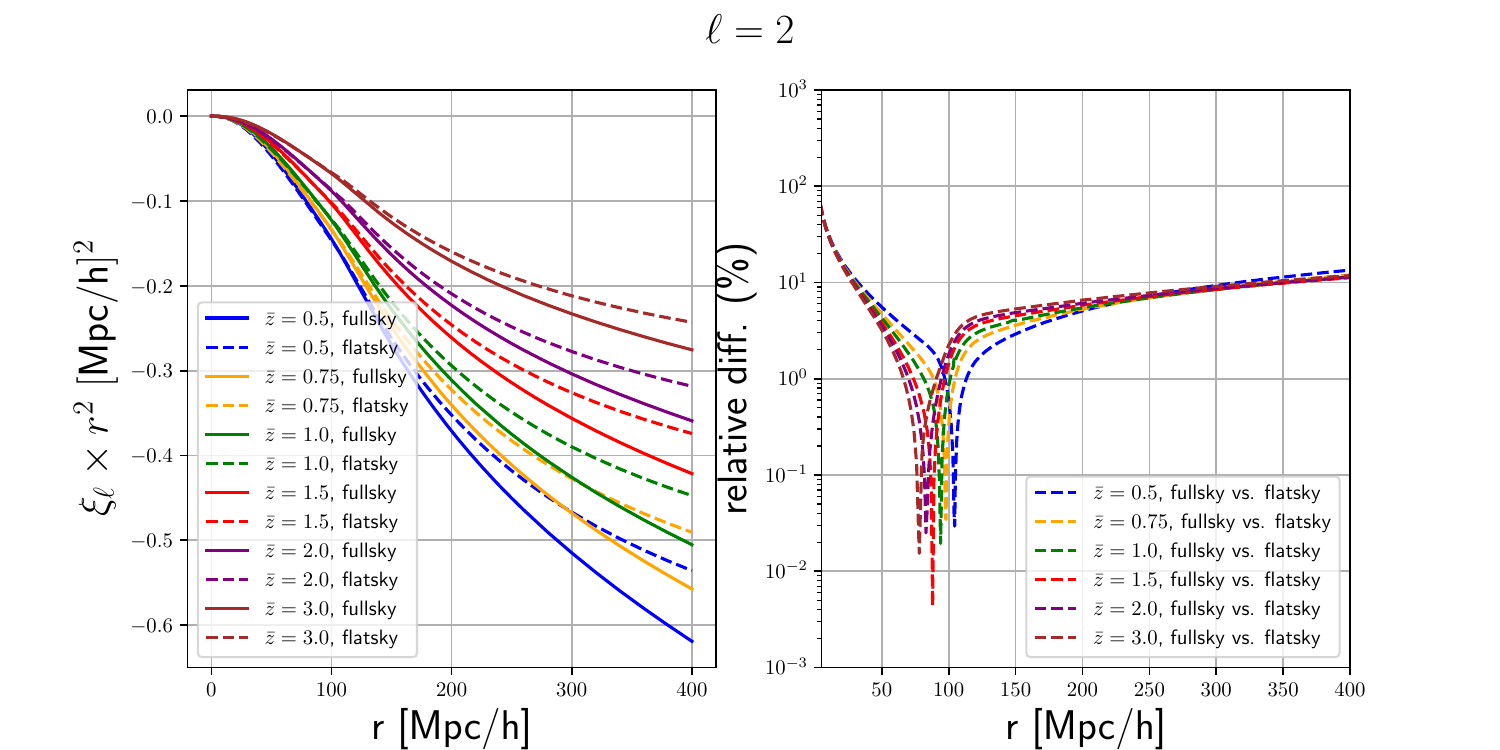}
\label{f:den_len2}
\end{subfigure}
\begin{subfigure}[]{0.6\textwidth}
\includegraphics[trim=20 0 20 2,clip,width=\textwidth]{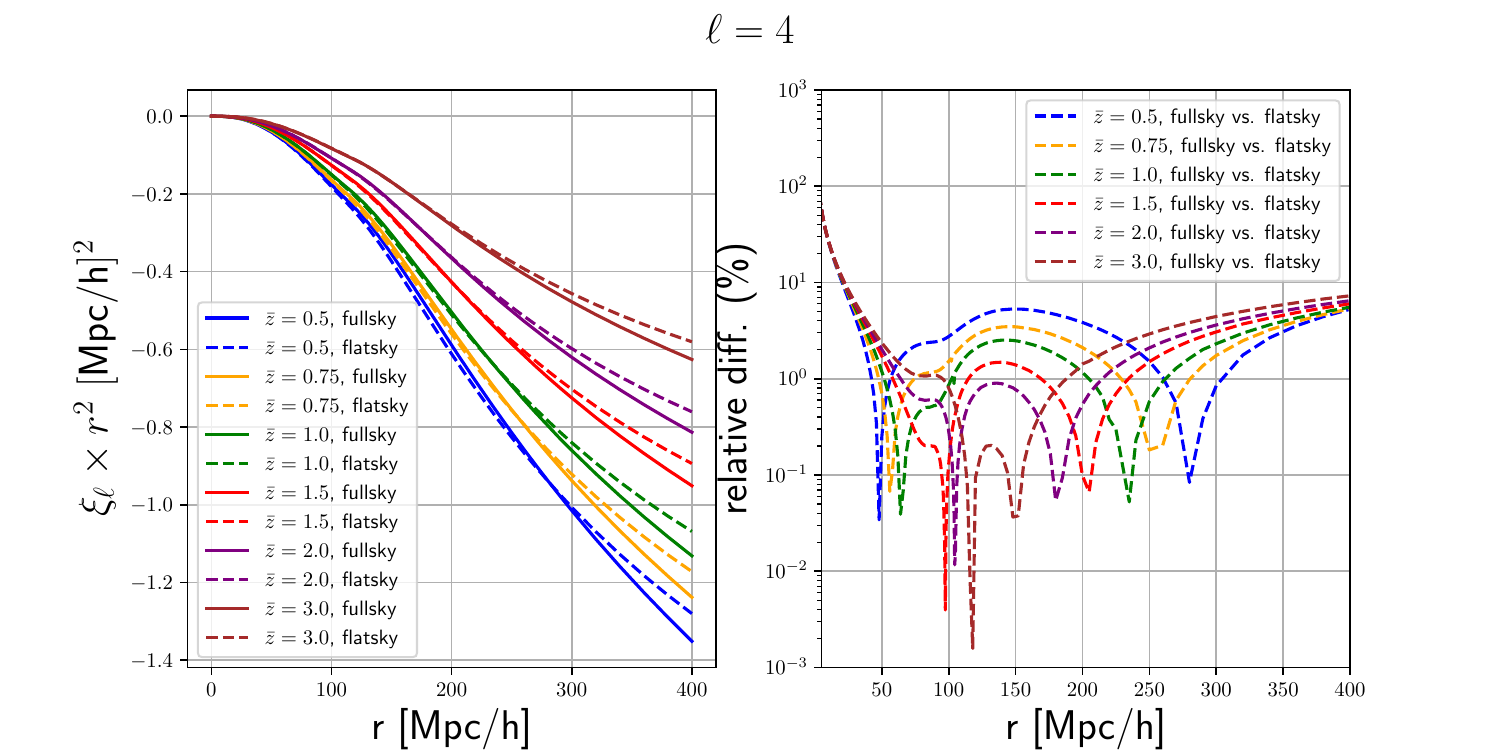}
\label{f:den_len4}
\end{subfigure}
\caption{\textit{Left to right, top to bottom}: the 2PCF for $\mu = \{0, 0.5, 0.9, 0.99\}$, and the $\ell = \{0, 2, 4\}$ multipoles, with only contributions from the cross-correlation of density-lensing, computed at various redshifts, in full-sky (solid) and flat-sky (dashed).
The relative difference (in percent) between full-sky and flat-sky is indicated on the right plot of each figure.
}
\label{f:den_len_multipoles}
\end{figure}

As is noticeable from the results in figure~\ref{f:den_len_multipoles}, the flat-sky Limber approximation for density-lensing is not very accurate.
As mentioned in section~\ref{s:den_len_flatsky}, the 2PCF at $\mu = 0$ in the flat-sky approximation is by construction zero everywhere, while the full-sky result gains contributions from the integration along the line of sight, so the discrepancy there is 100\%.
Furthermore, since $\lim\limits_{r \rightarrow 0} \xi_\mathrm{flat}^\text{den-len} = 0$ and $\lim\limits_{r \rightarrow 0} \xi_\mathrm{full}^\text{den-len} \neq 0$, we also have a discrepancy at small separations, at all values of $\mu$.
The approximation becomes more accurate for $\mu \rightarrow 1$ and separations in the range $[50, 300]\;\mathrm{Mpc}/h$, where the agreement is mostly better than 10\%, and for $\mu = 0.99$ is consistently better than 1\% for all redshifts at separations in the range $[50, 300]\;\text{Mpc}/h$.
The monopole has an error that is consistently larger than 10\% with respect to the full-sky result, while the quadrupole and the hexadecapole have an error below 10\% for all redshifts, but only for separations in the range $[50, 300]\; \mathrm{Mpc}/h$; outside of that range, the difference becomes much larger, due to the same reasons as the 2PCF.

Overall, the flat-sky Limber approximation for density-lensing seems to be of limited utility due to its poor accuracy for most configurations.

\subsubsection{Lensing-lensing}
\label{s:lensing_lensing}

\begin{figure}[H]
\centering
\begin{subfigure}[t]{0.49\textwidth}
\includegraphics[trim=20 0 20 2,clip,width=\textwidth]{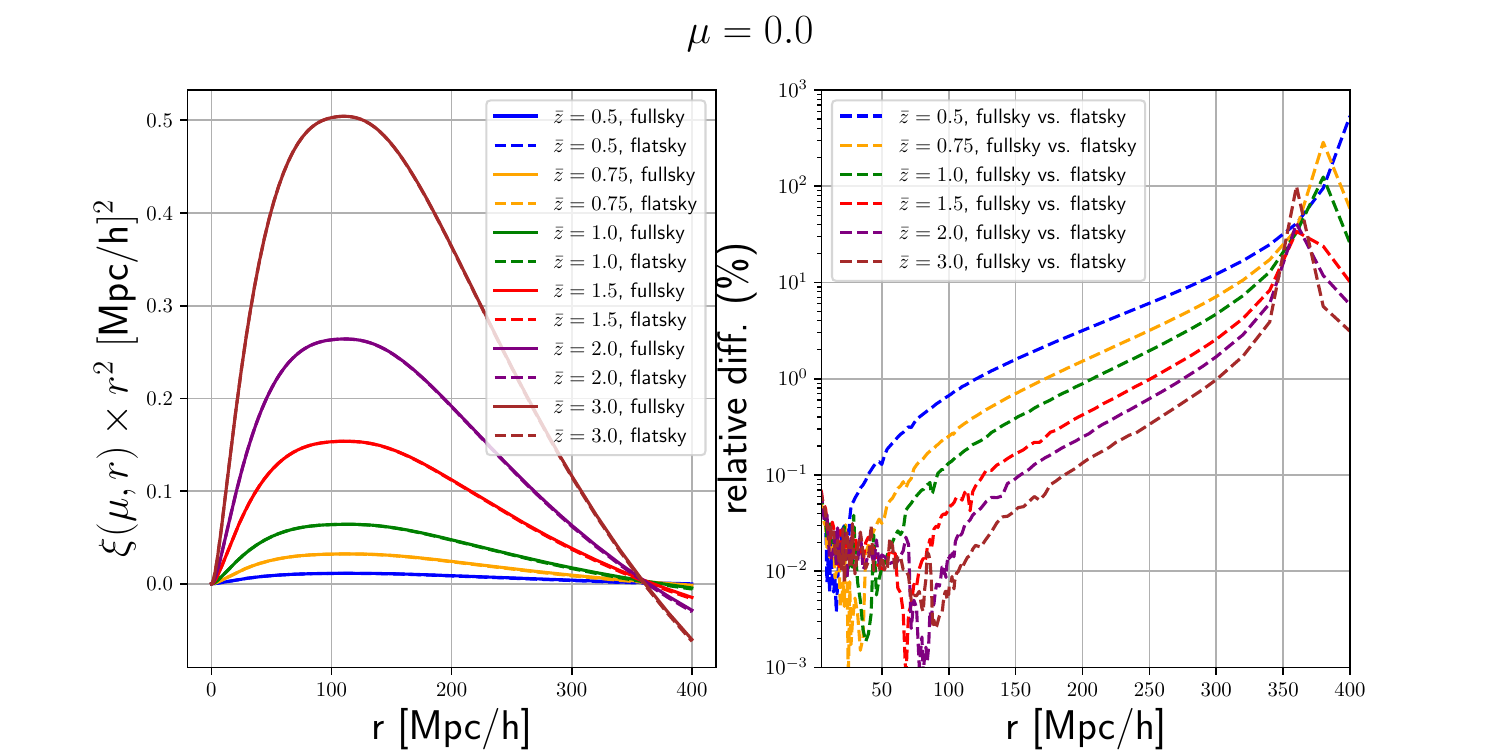}
\label{f:len_len_mu0}
\end{subfigure}
\begin{subfigure}[t]{0.49\textwidth}
\includegraphics[trim=20 0 20 2,clip,width=\textwidth]{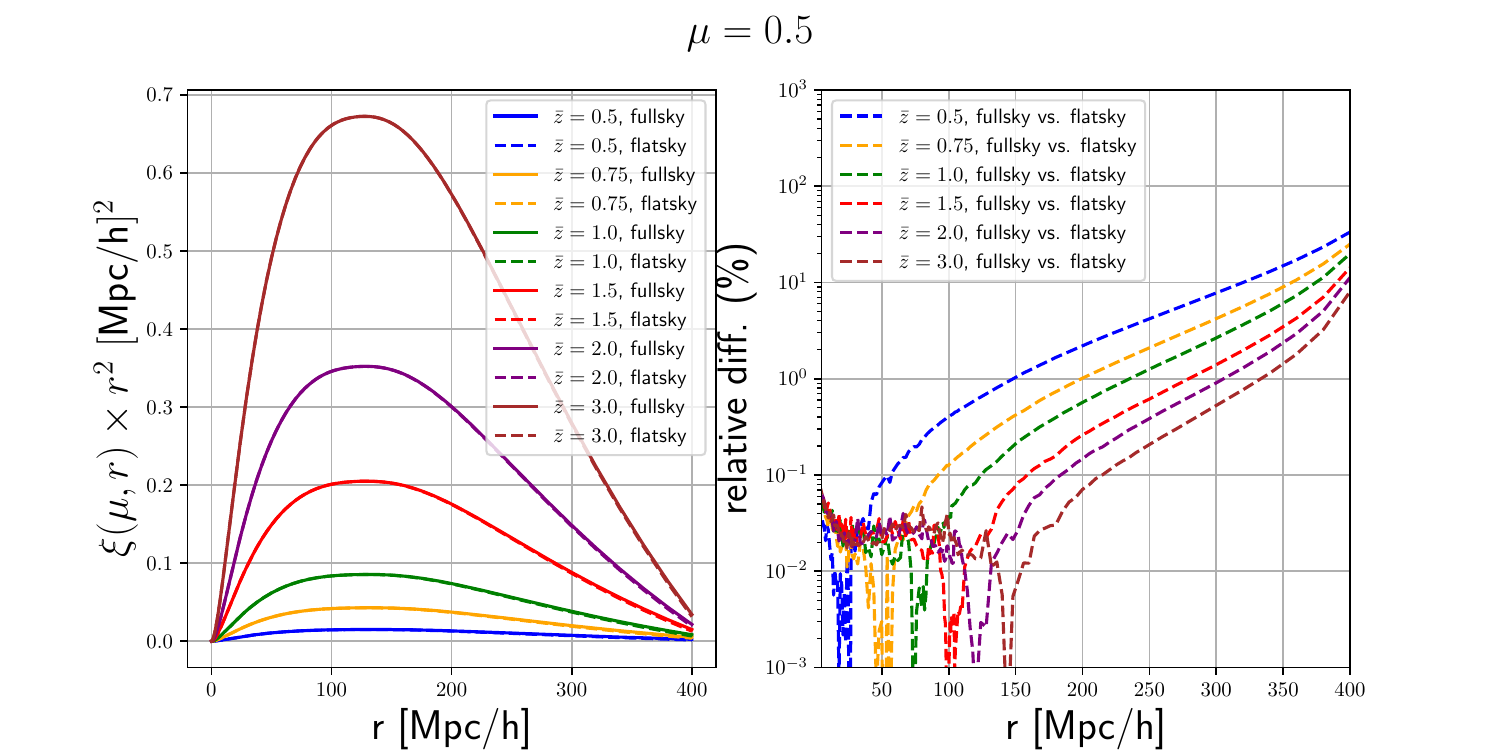}
\label{f:len_len_mu1}
\end{subfigure}
\begin{subfigure}[t]{0.49\textwidth}
\includegraphics[trim=20 0 20 2,clip,width=\textwidth]{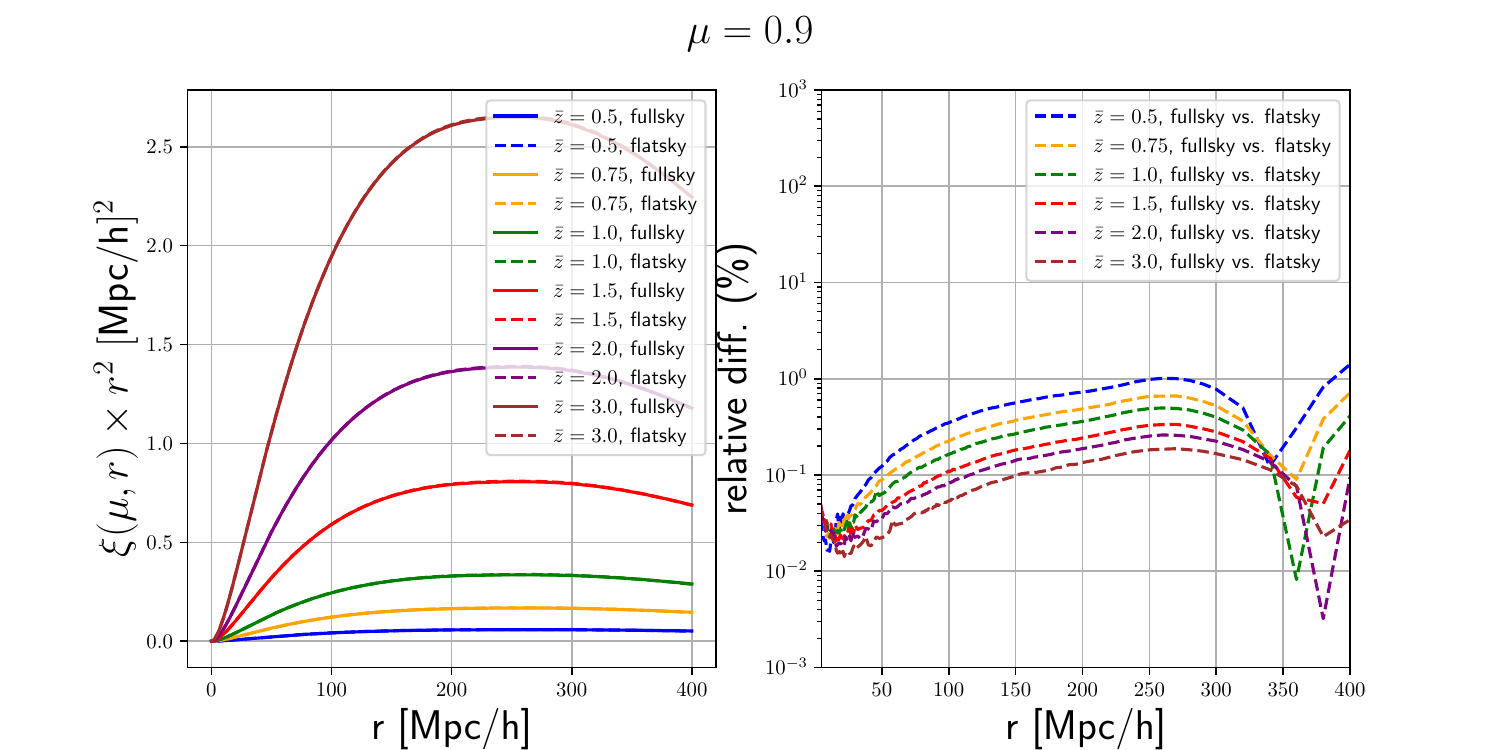}
\end{subfigure}
\label{f:len_len_mu2}
\begin{subfigure}[t]{0.49\textwidth}
\includegraphics[trim=20 0 20 2,clip,width=\textwidth]{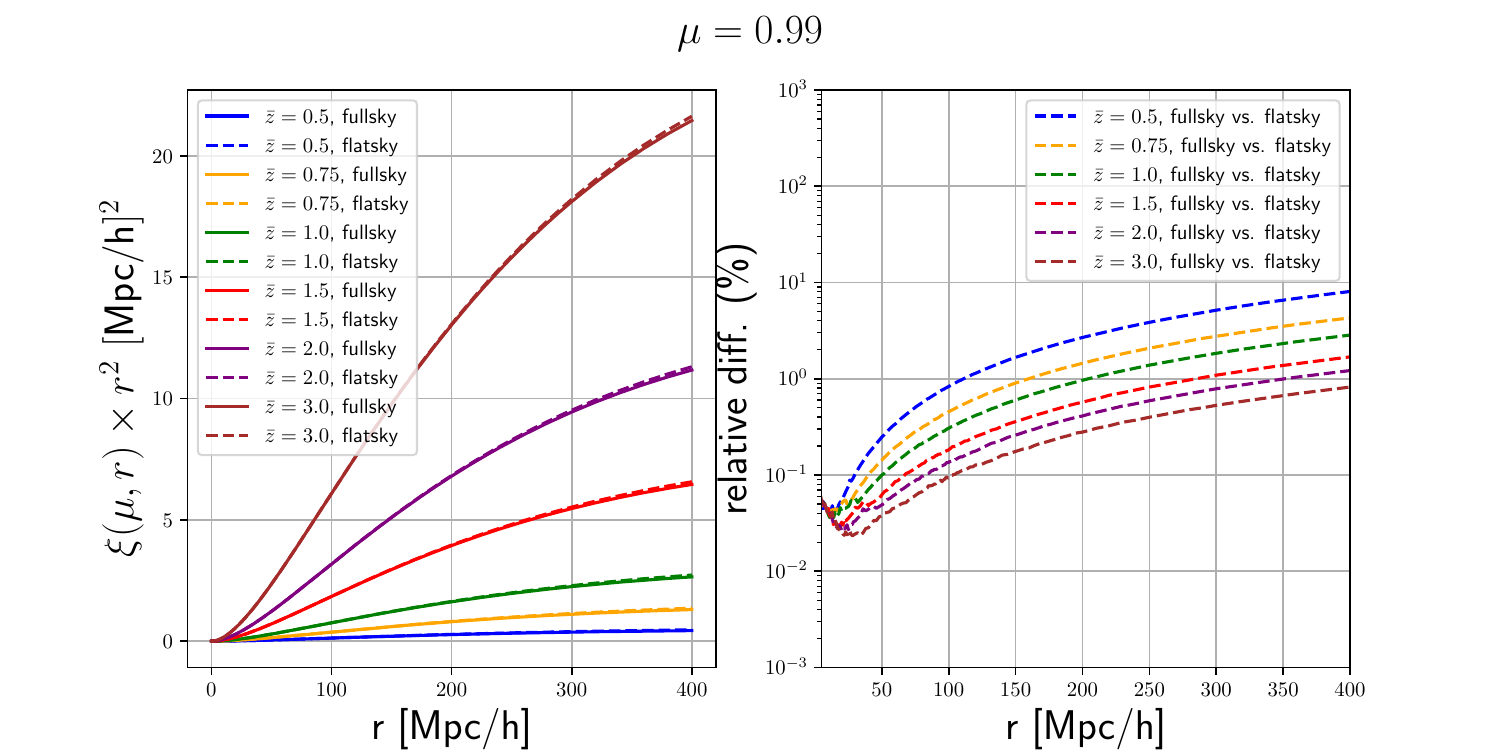}
\label{f:len_len_mu3}
\end{subfigure}
\begin{subfigure}[t]{0.49\textwidth}
\includegraphics[trim=20 0 20 2,clip,width=\textwidth]{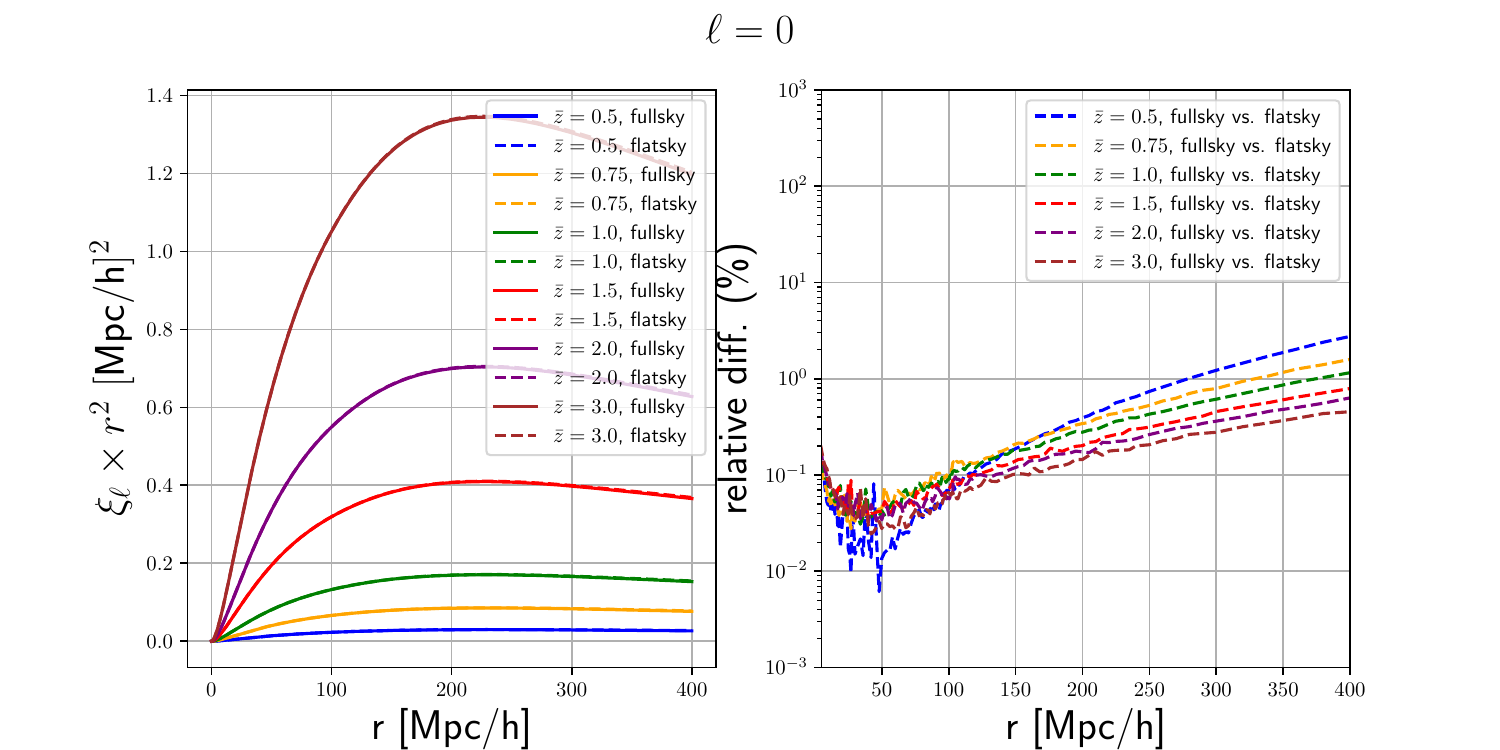}
\label{f:len_len0}
\end{subfigure}
\begin{subfigure}[t]{0.49\textwidth}
\includegraphics[trim=20 0 20 2,clip,width=\textwidth]{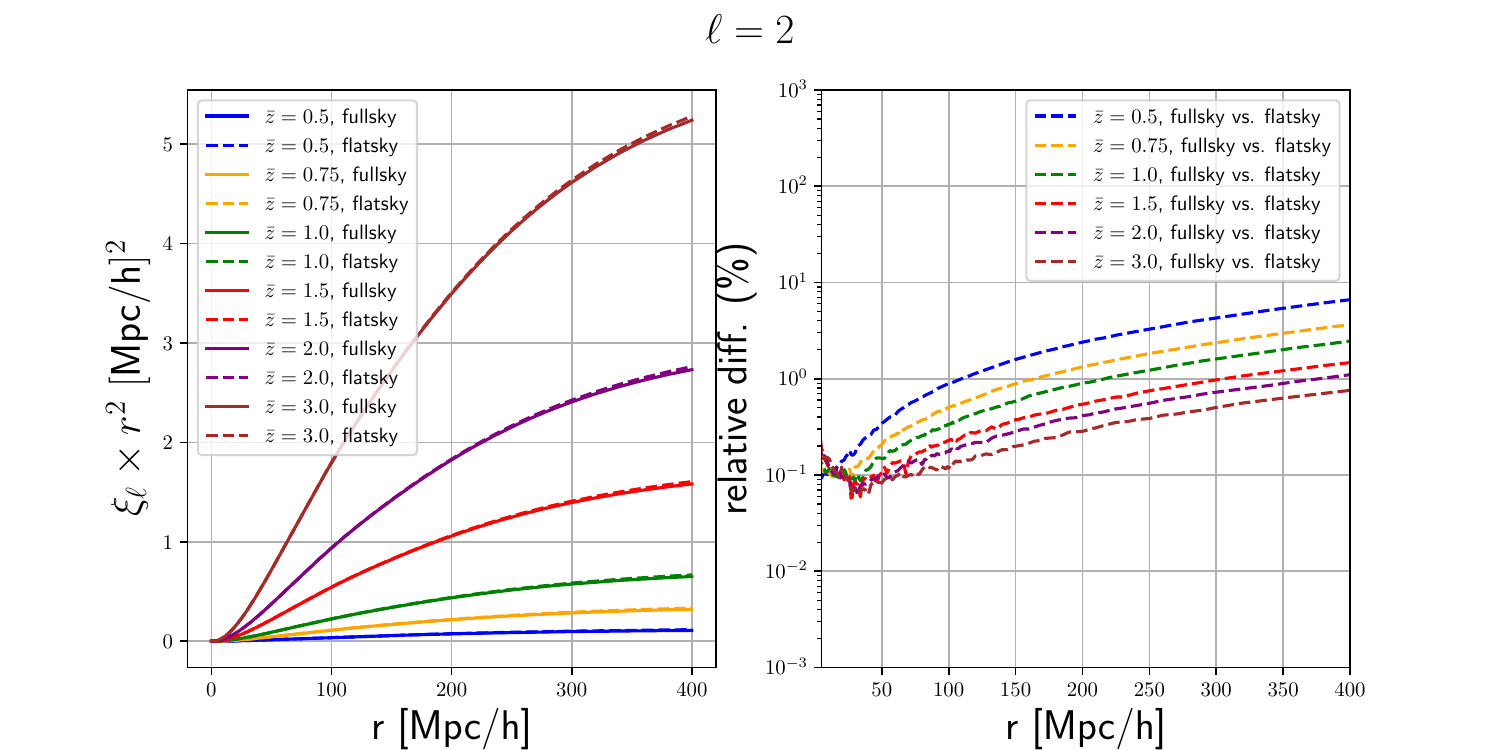}
\label{f:len_len2}
\end{subfigure}
\begin{subfigure}[]{0.6\textwidth}
\includegraphics[trim=20 0 20 2,clip,width=\textwidth]{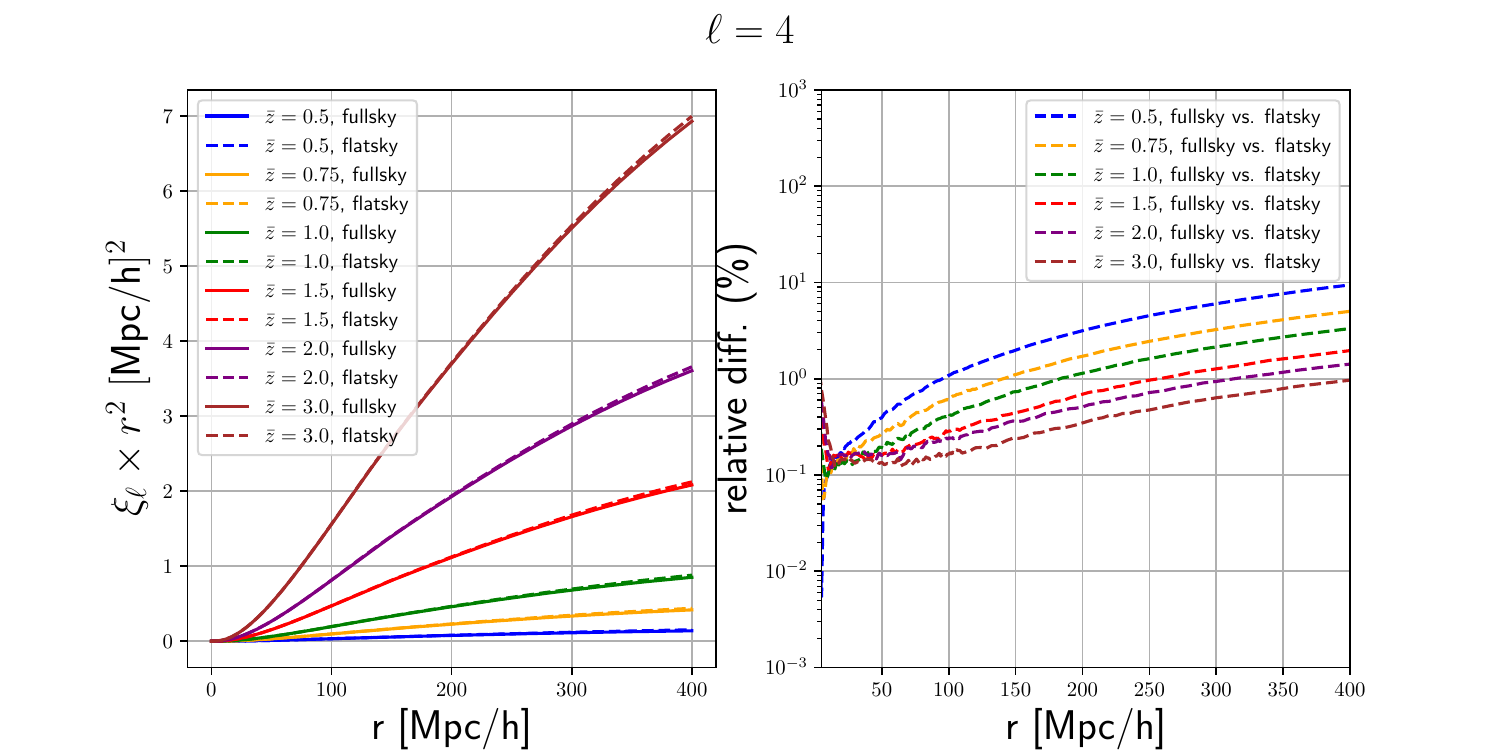}
\label{f:len_len4}
\end{subfigure}
\caption{\textit{Left to right, top to bottom}: the 2PCF for $\mu = \{0, 0.5, 0.9, 0.99\}$, and the $\ell = \{0, 2, 4\}$ multipoles, with only contributions from lensing-lensing, computed at various redshifts, in full-sky (solid) and flat-sky (dashed).
The relative difference (in percent) between full-sky and flat-sky is indicated on the right plot of each figure.
}
\label{f:len_len_multipoles}
\end{figure}

The flat-sky Limber approximation for lensing-lensing, shown in figure~\ref{f:len_len_multipoles} for the 2PCF and its multipoles, respectively, is surprisingly accurate; for the 2PCF, we see that for $\mu = 0$ we have sub-percent agreement for all redshifts at separations $r \lesssim 100\; \mathrm{Mpc}/h$, and for $\mu \rightarrow 1$ we consistently obtain sub-percent accuracy below $200\; \mathrm{Mpc}/h$ for redshifts $z \gtrsim 1$.

Likewise, for all redshifts considered, the results for the $\ell = 0$ multipole agree with the full-sky approximation to better than 1\%, while the $\ell = 2$ and $\ell = 4$ multipoles show an agreement of $\sim 1\%$ only at $z > 1$.
In general, for $z \gtrsim 1$, we get an agreement to better than $5 \%$ for all separations, for both the 2PCF and its multipoles.

To ensure that our particular choice of cosmological parameters doesn't affect the accuracy of the approximation, we have checked the dependence of the accuracy on the cosmological parameters by varying the value of each parameter from table~\ref{t:params} by $\pm20 \%$, while keeping the others fixed, and have found no noticeable improvement nor degradation with respect to the results discussed above.

Additionally, we've verified that using a redshift-dependent magnification bias does not significantly alter the results.
For concreteness, we assume an SKA2-like bias with the following parametrization~\cite{Bull_2016}:
\begin{align}
s(z) &= s_0 +s_1\, z + s_2\, z^2 + s_3\, z^3
\label{eq:bias}
\end{align}
with $s_0 = -0.106875$, $s_1 = 1.35999$, $s_2 = -0.620008$, and $s_3 = 0.188594$.
We find that the agreement with full-sky is somewhat worse that the case with constant magnification bias, but is nevertheless better than $5 \%$ for $r \lesssim 300\; \mathrm{Mpc}/h$ at redshifts $z \gtrsim 1$.
Of course, we could simply use the method described in section~\ref{s:bias} to fully account the effects of a redshift-dependent bias.

Overall, this result suggests that, at least in linear theory, the cross-correlations between different redshifts have an almost negligible impact on the lensing-lensing contribution to the 2PCF.

\subsubsection{All terms}

\begin{figure}[H]
\centering
\begin{subfigure}[t]{0.49\textwidth}
\includegraphics[trim=20 0 20 2,clip,width=\textwidth]{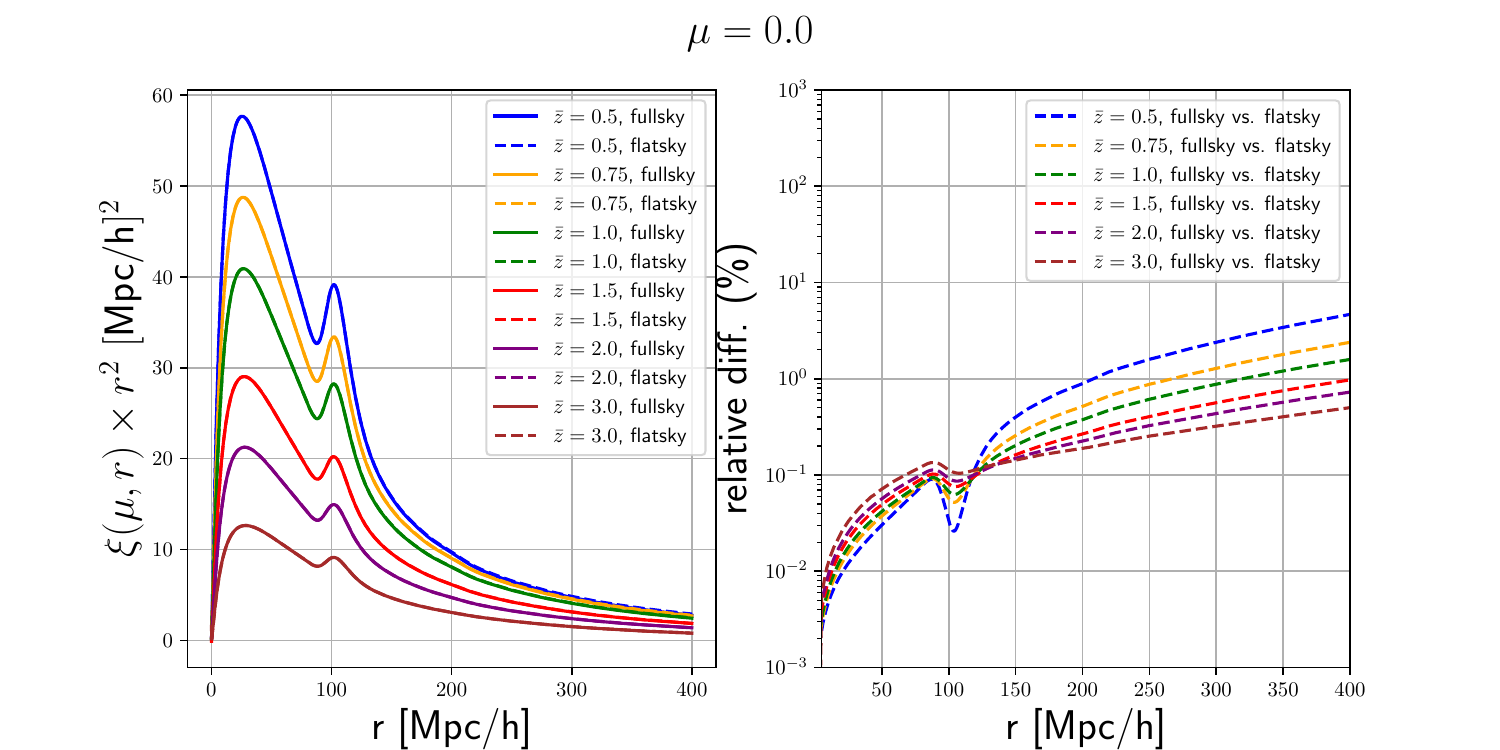}
\label{f:all_all_mu0}
\end{subfigure}
\begin{subfigure}[t]{0.49\textwidth}
\includegraphics[trim=20 0 20 2,clip,width=\textwidth]{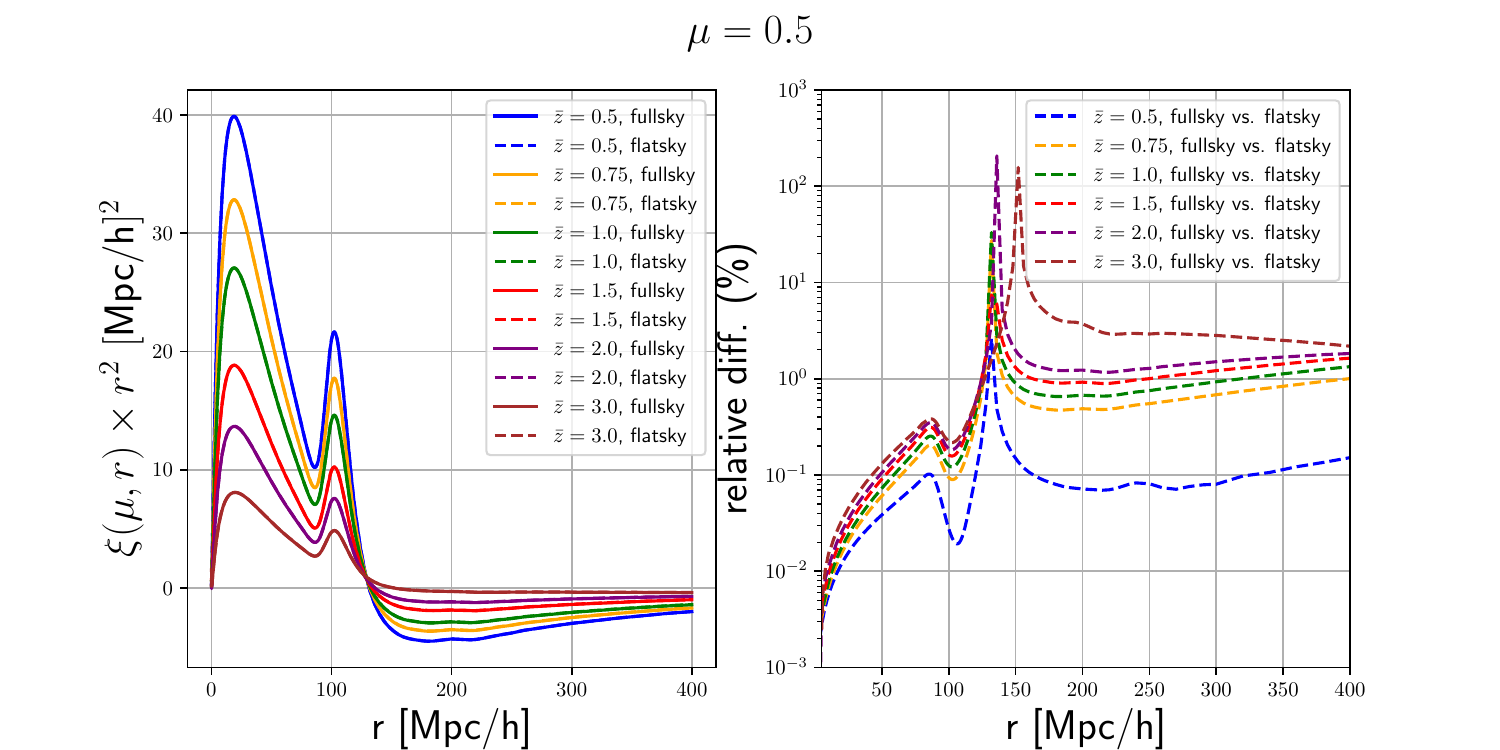}
\label{f:all_all_mu1}
\end{subfigure}
\begin{subfigure}[t]{0.49\textwidth}
\includegraphics[trim=20 0 20 2,clip,width=\textwidth]{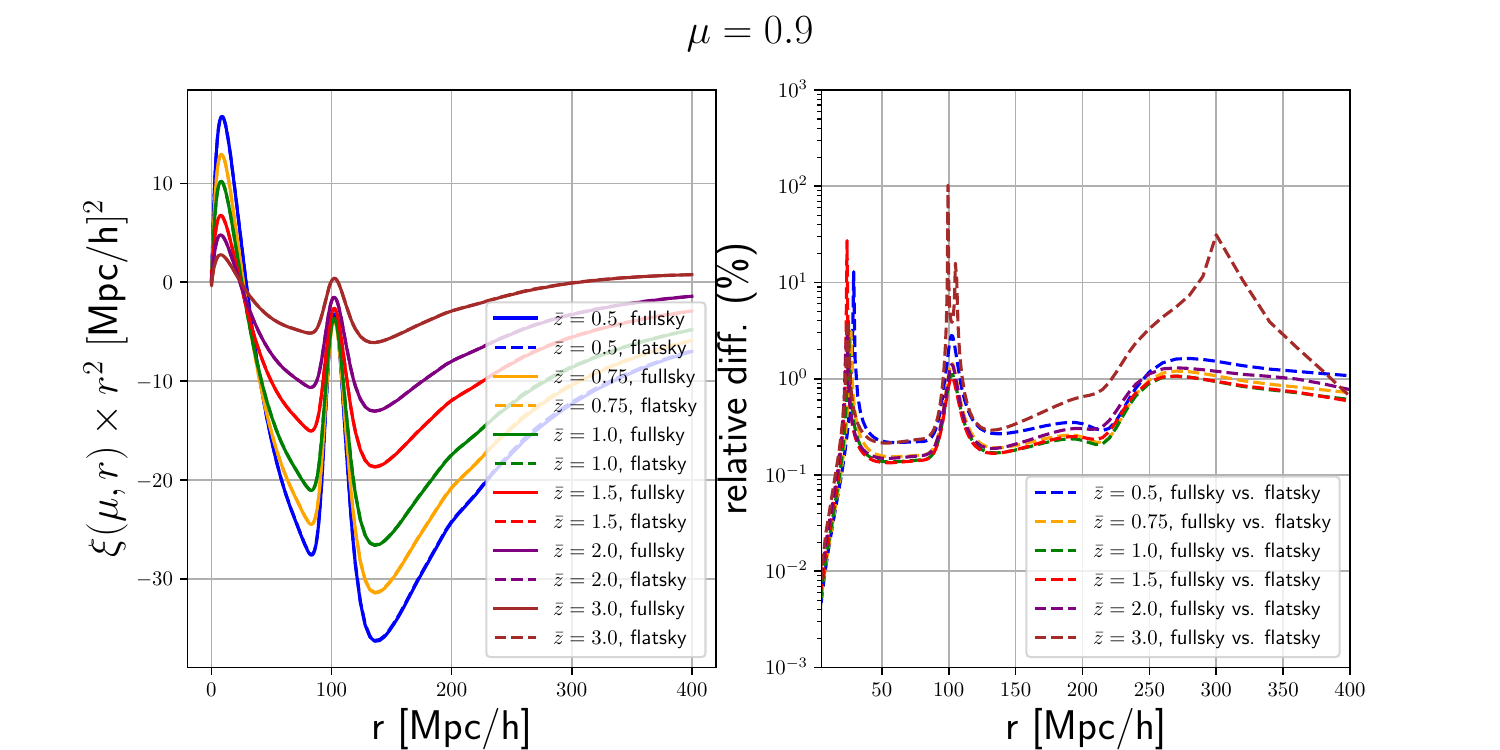}
\end{subfigure}
\label{f:all_all_mu2}
\begin{subfigure}[t]{0.49\textwidth}
\includegraphics[trim=20 0 20 2,clip,width=\textwidth]{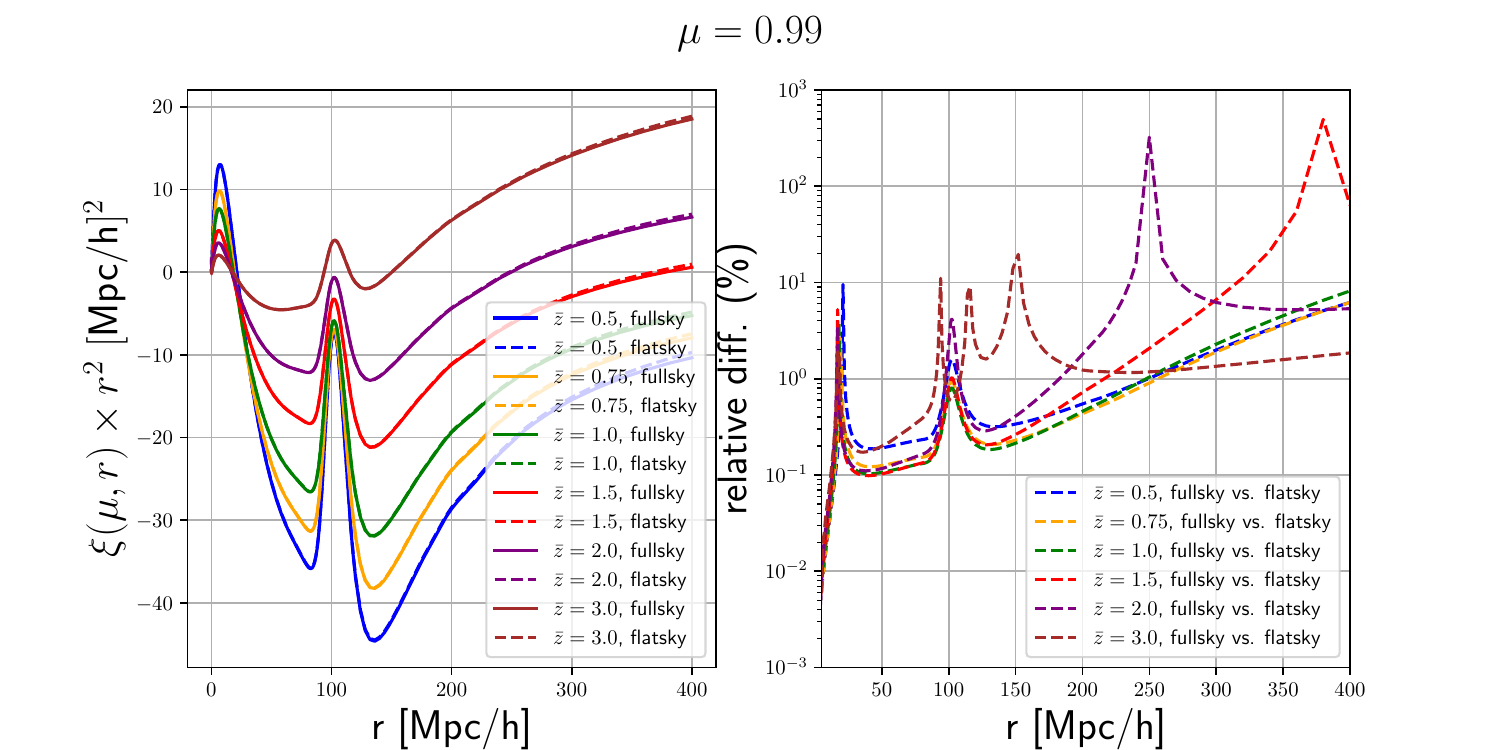}
\label{f:all_all_mu3}
\end{subfigure}
\begin{subfigure}[t]{0.49\textwidth}
\includegraphics[trim=20 0 20 2,clip,width=\textwidth]{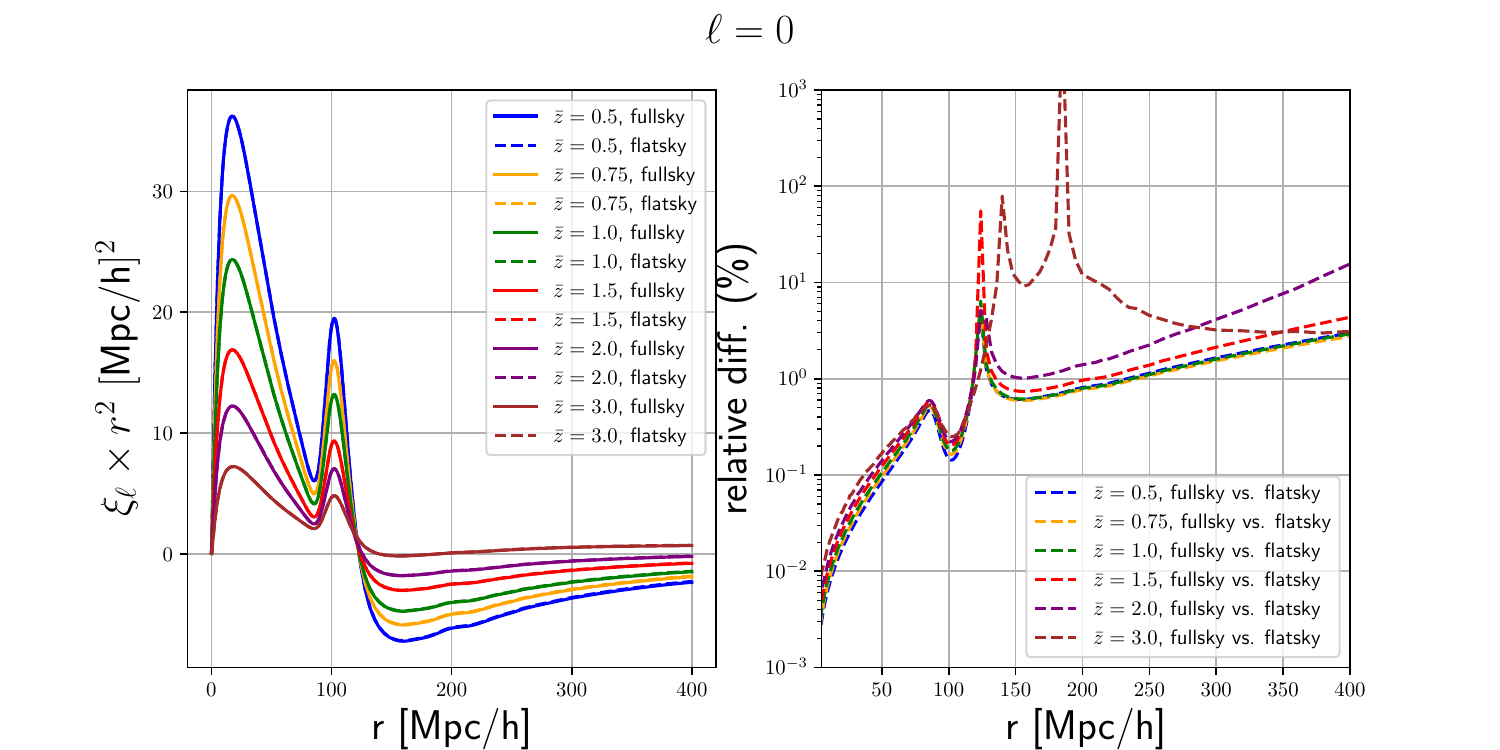}
\label{f:all_all0}
\end{subfigure}
\begin{subfigure}[t]{0.49\textwidth}
\includegraphics[trim=20 0 20 2,clip,width=\textwidth]{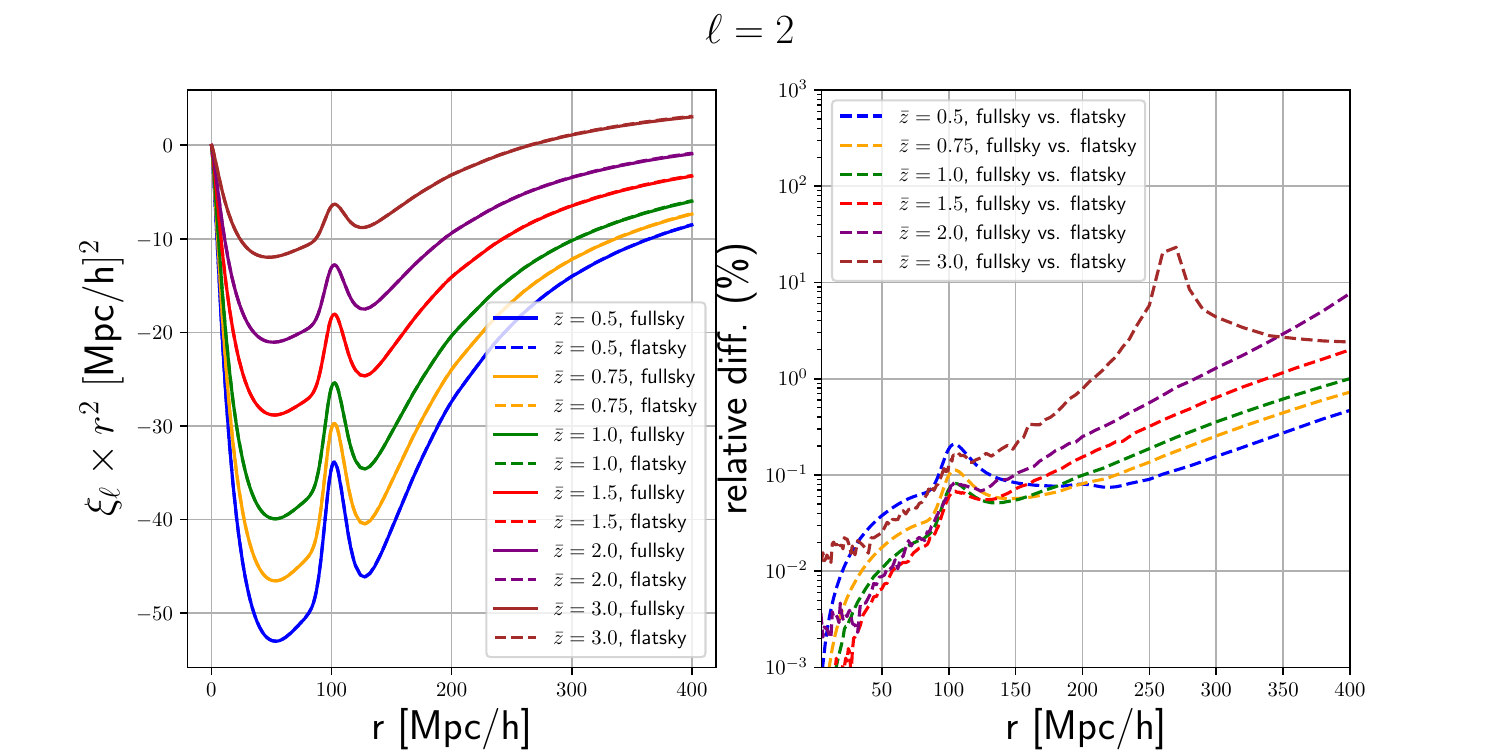}
\label{f:all_all2}
\end{subfigure}
\begin{subfigure}[]{0.6\textwidth}
\includegraphics[trim=20 0 20 2,clip,width=\textwidth]{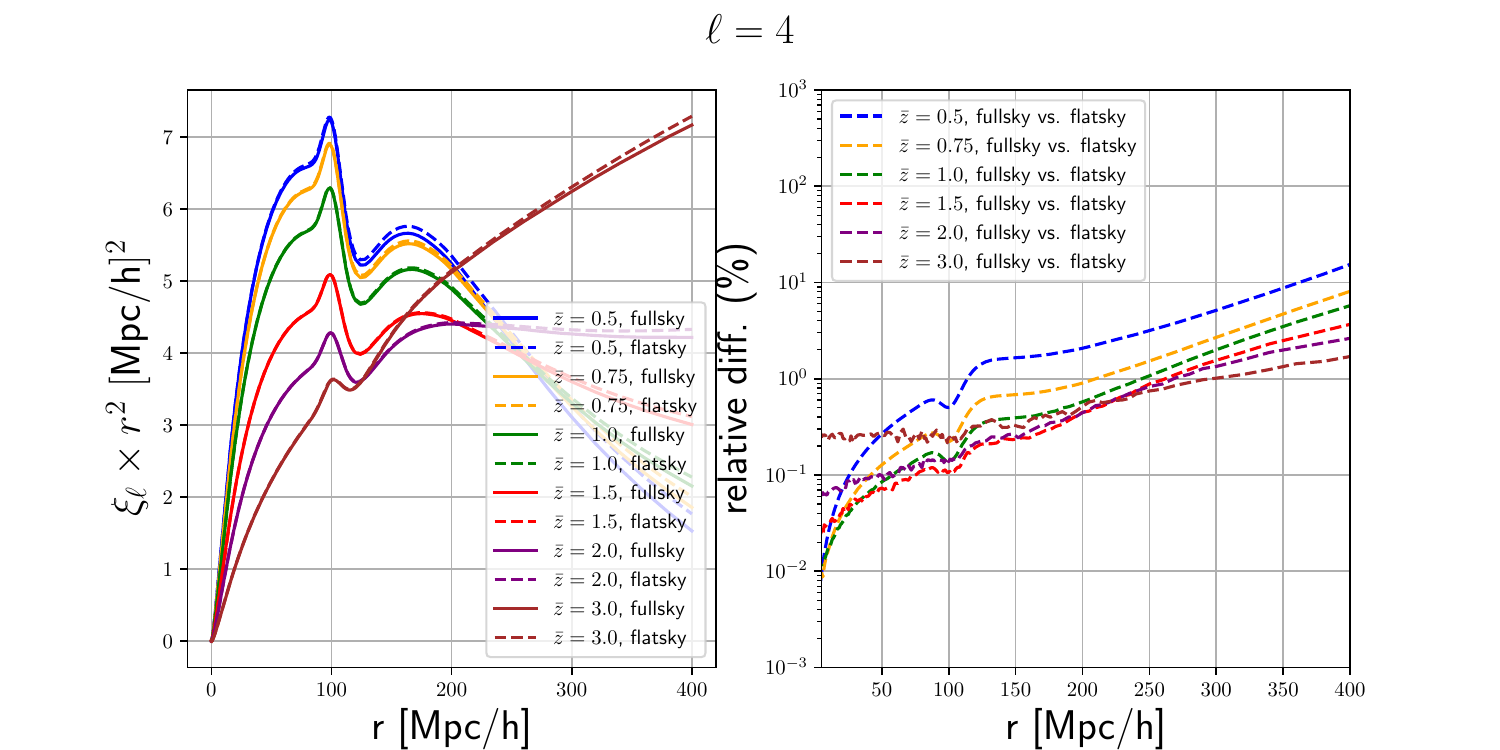}
\label{f:all_all4}
\end{subfigure}
\caption{\textit{Left to right, top to bottom}: the 2PCF for $\mu = \{0, 0.5, 0.9, 0.99\}$, and the $\ell = \{0, 2, 4\}$ multipoles, with contributions from density, RSD, Doppler, and lensing, computed at various redshifts, in full-sky (solid) and flat-sky (dashed).
The relative difference (in percent) between full-sky and flat-sky is indicated on the right plot of each figure.
}
\label{f:all_all_multipoles}
\end{figure}

In figure~\ref{f:all_all_multipoles} we show the comparison of the full-sky result with the flat-sky approximation, when using all of the terms discussed earlier (density, RSD, Doppler, and lensing, and their cross-correlations).
The results indicate that the flat-sky approximation agrees up to 10\% with the full-sky result for the monopole and the quadrupole for all separations and redshifts, while the hexadecapole is somewhat worse, and the error rises above 10\% with respect to the full-sky result for $r \gtrsim 200\,\text{Mpc}/h$ at all redshifts.

\begin{figure}[H]
\centering
\begin{subfigure}[t]{0.49\textwidth}
\includegraphics[trim=20 0 20 2,clip,width=\textwidth]{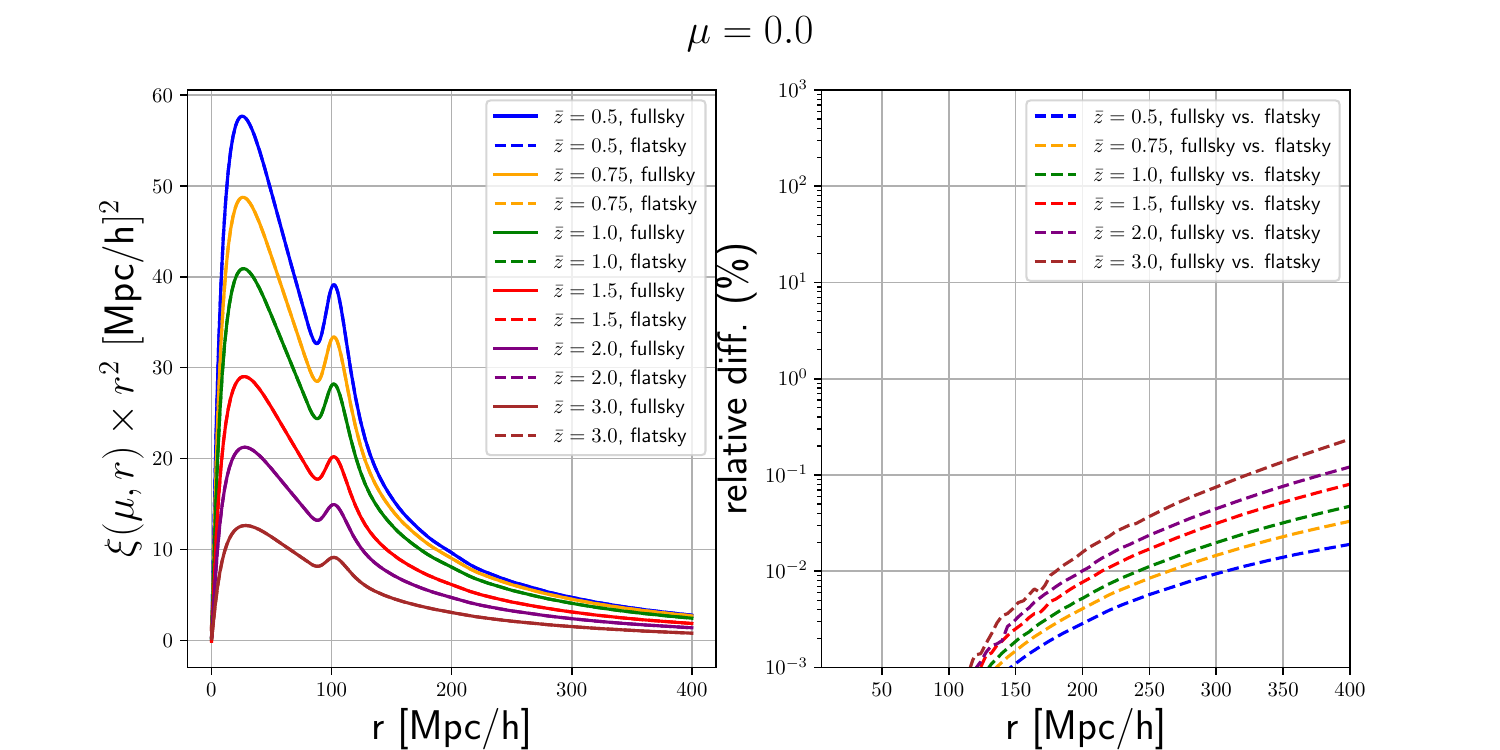}
\label{f:all_all_len_flat_mu0}
\end{subfigure}
\begin{subfigure}[t]{0.49\textwidth}
\includegraphics[trim=20 0 20 2,clip,width=\textwidth]{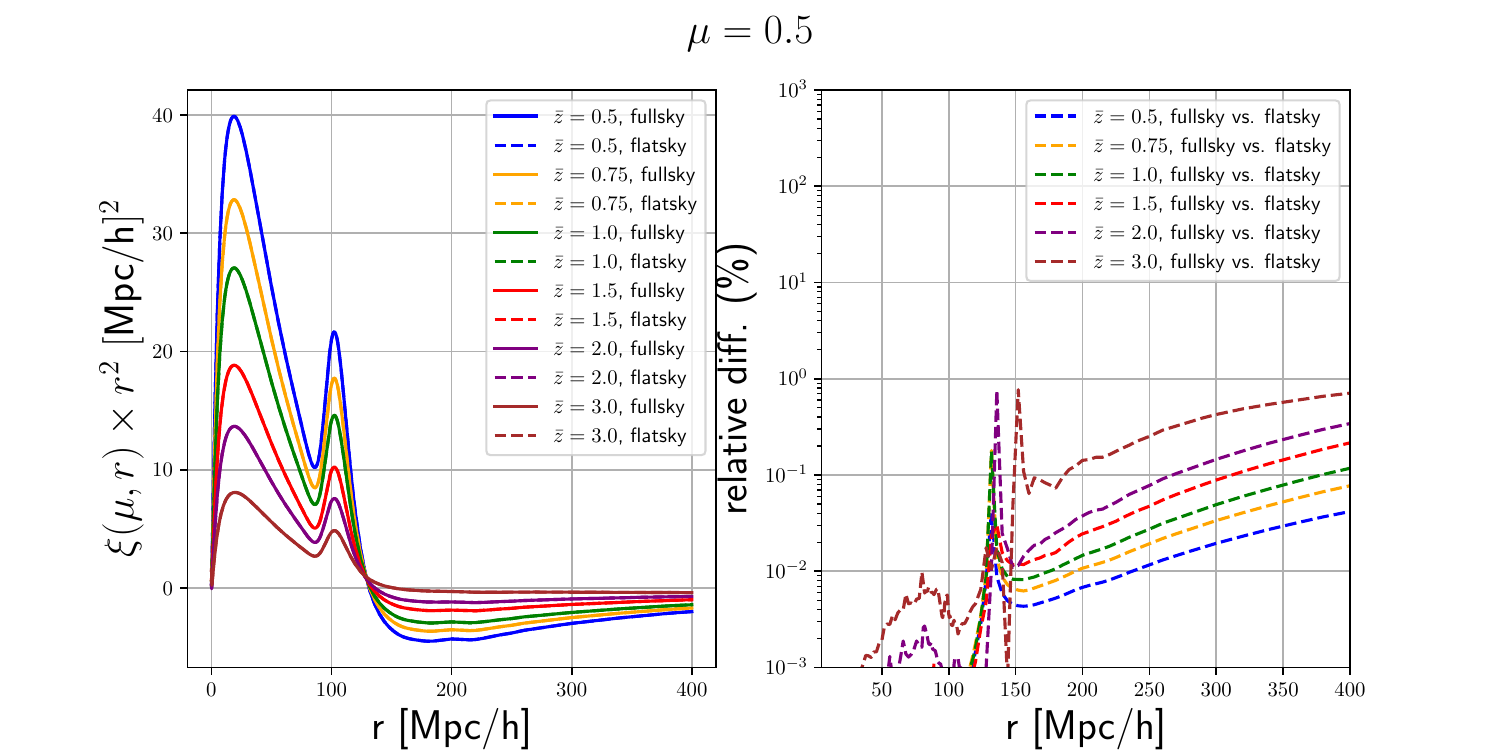}
\label{f:all_all_len_flat_mu1}
\end{subfigure}
\begin{subfigure}[t]{0.49\textwidth}
\includegraphics[trim=20 0 20 2,clip,width=\textwidth]{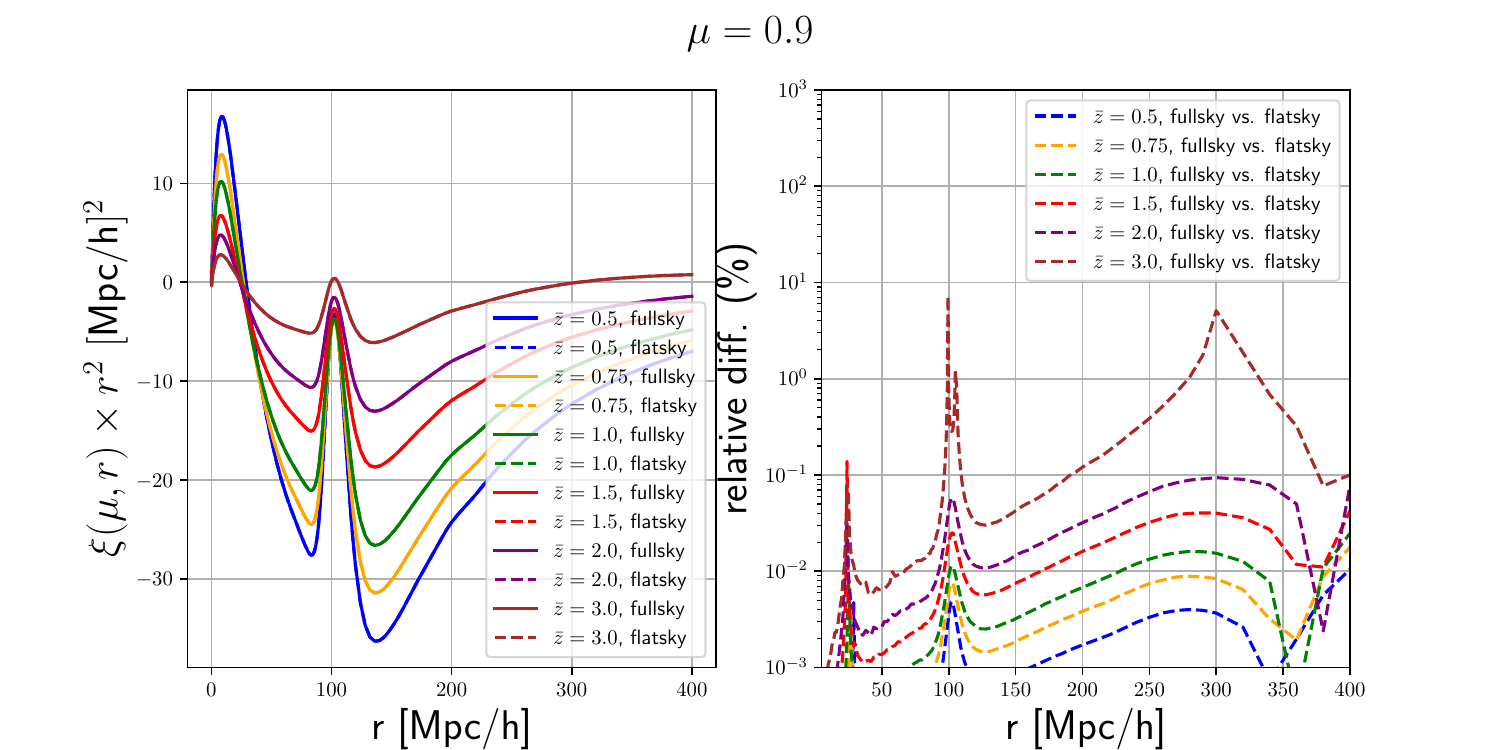}
\end{subfigure}
\label{f:all_all_len_flat_mu2}
\begin{subfigure}[t]{0.49\textwidth}
\includegraphics[trim=20 0 20 2,clip,width=\textwidth]{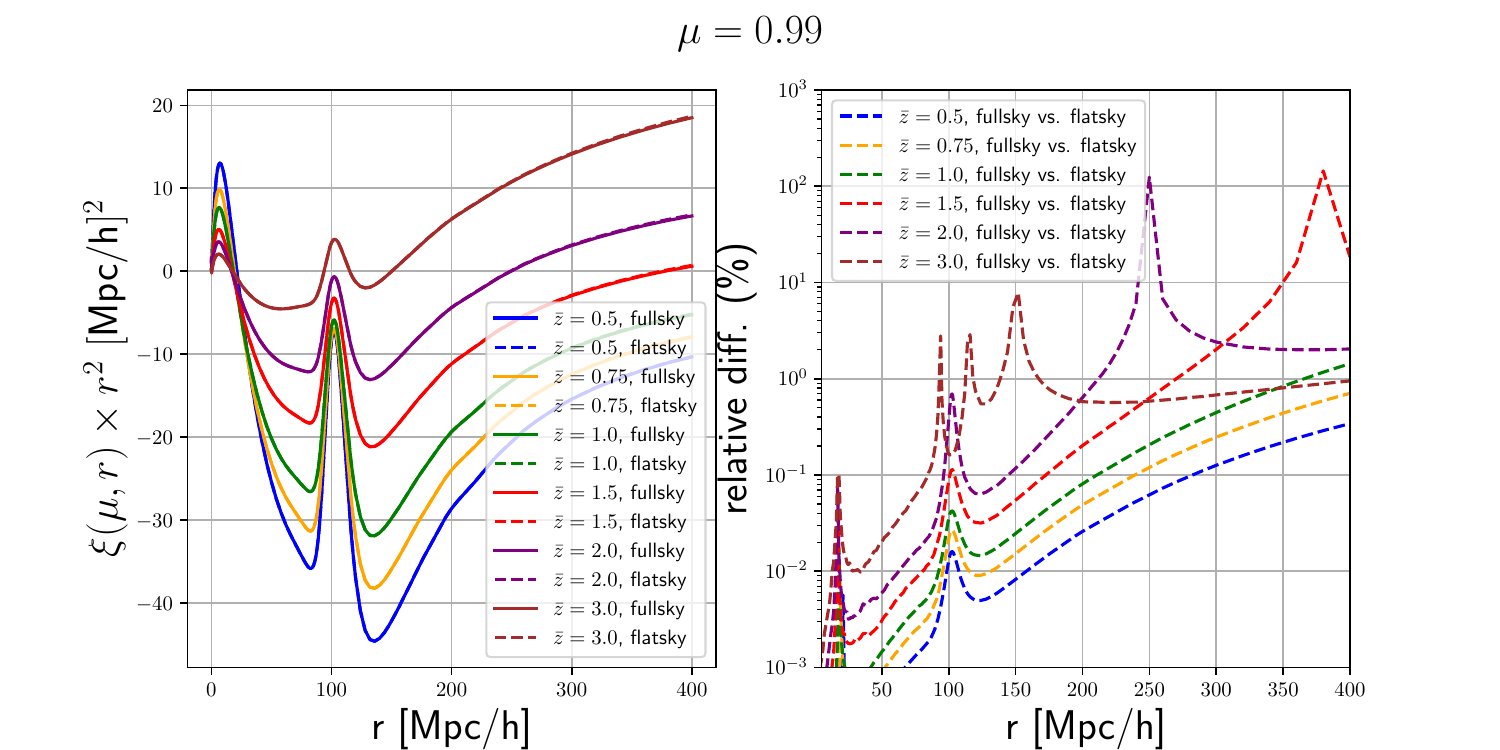}
\label{f:all_all_mu_len_flat3}
\end{subfigure}
\begin{subfigure}[t]{0.49\textwidth}
\includegraphics[trim=20 0 20 2,clip,width=\textwidth]{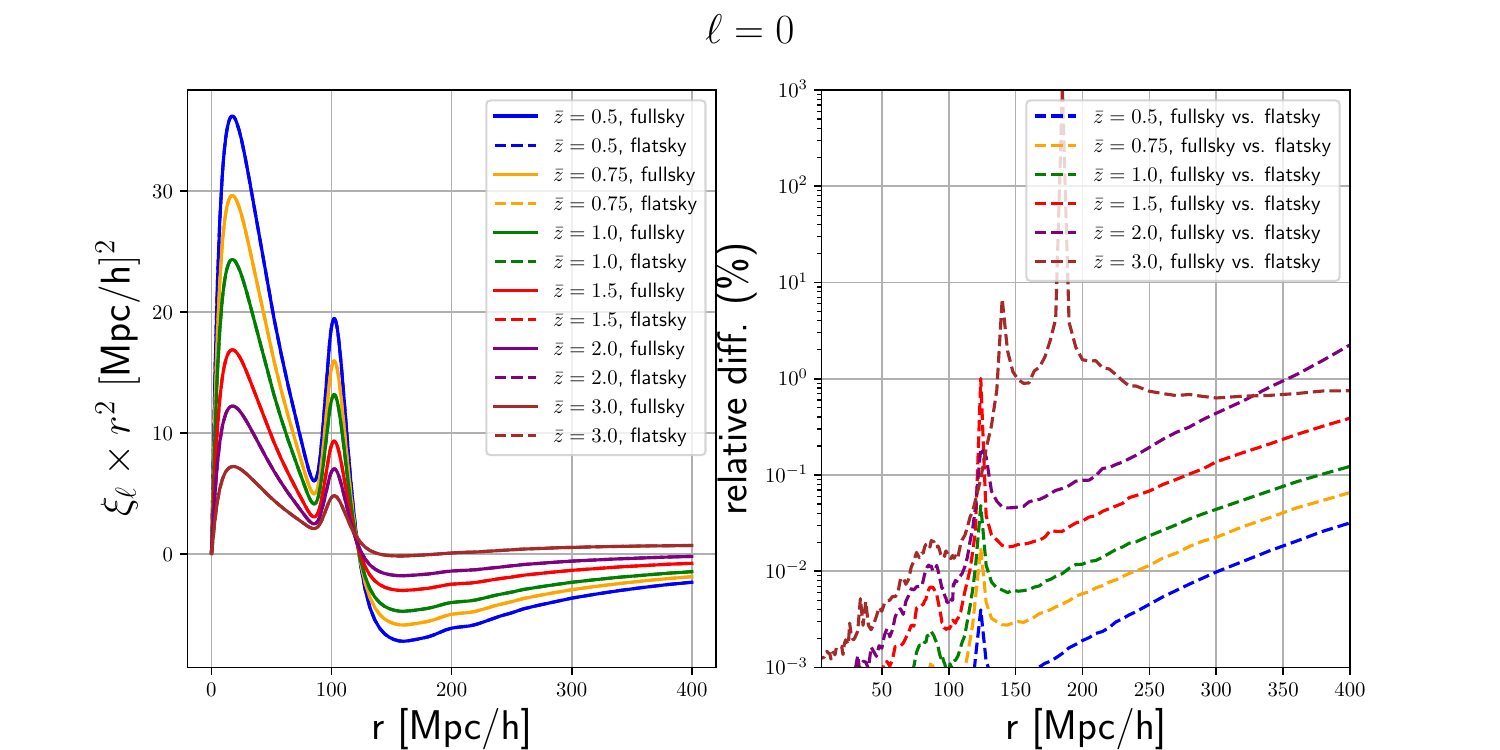}
\label{f:all_all_len_flat0}
\end{subfigure}
\begin{subfigure}[t]{0.49\textwidth}
\includegraphics[trim=20 0 20 2,clip,width=\textwidth]{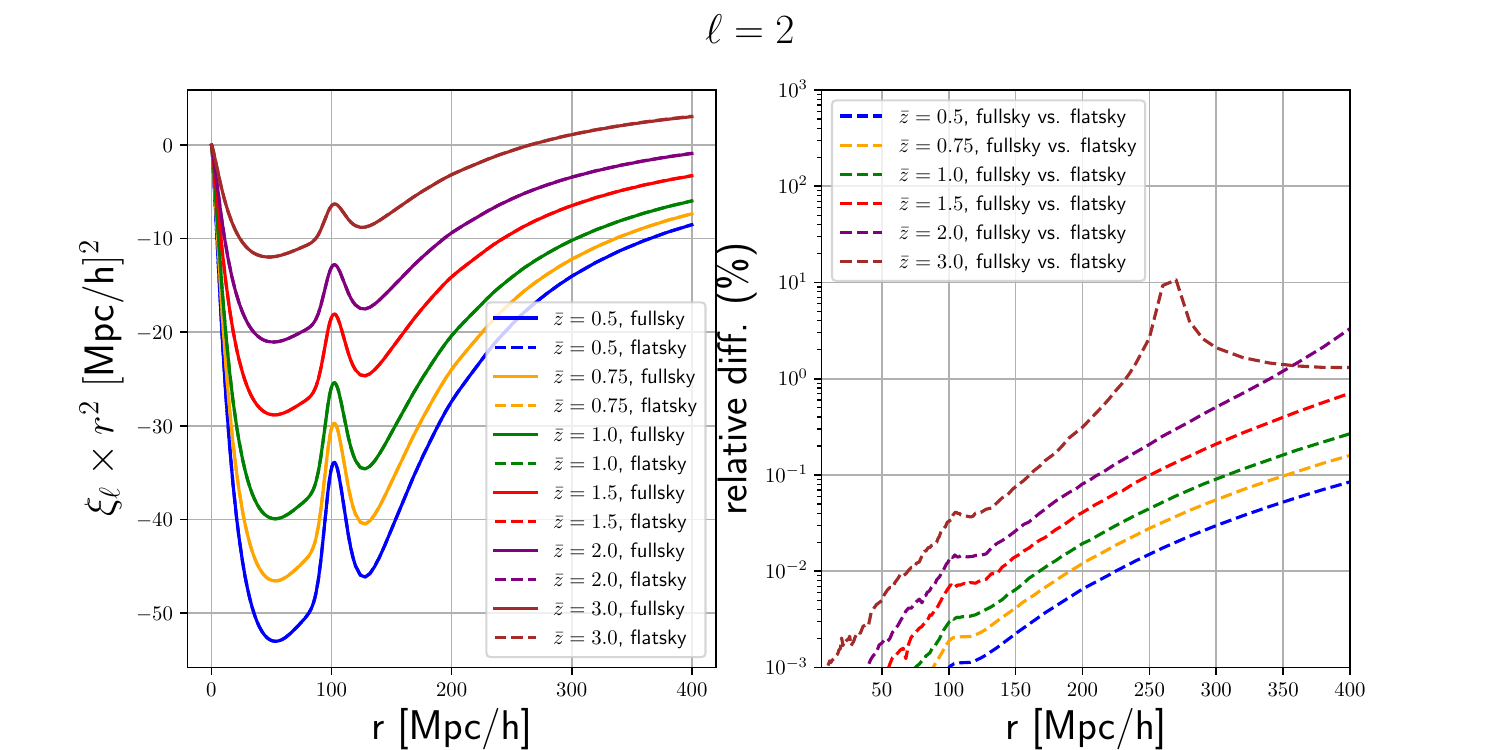}
\label{f:all_all_len_flat2}
\end{subfigure}
\begin{subfigure}[t]{0.6\textwidth}
\includegraphics[trim=20 0 20 2,clip,width=\textwidth]{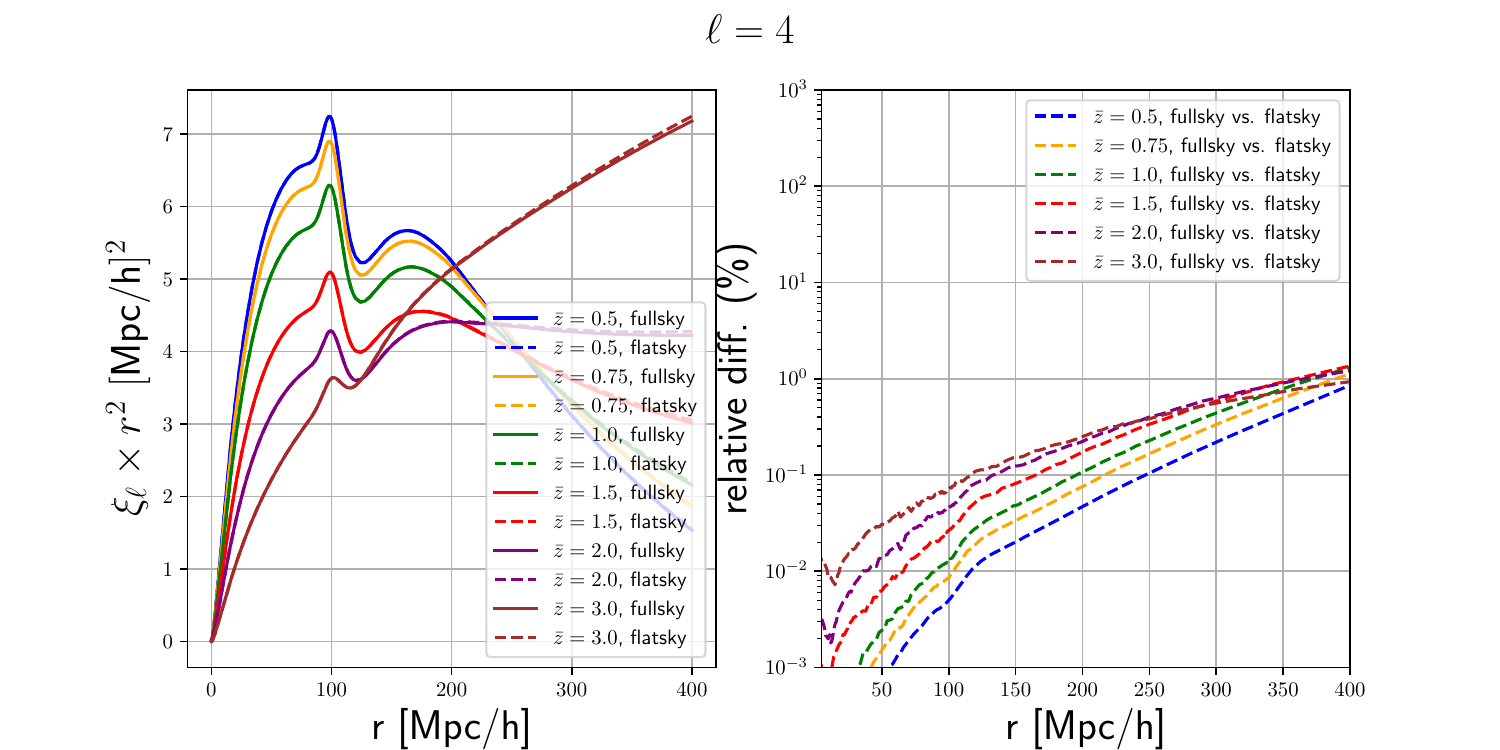}
\label{f:all_all_len_flat4}
\end{subfigure}
\caption{\textit{Left to right, top to bottom}: the 2PCF for $\mu = \{0, 0.5, 0.9, 0.99\}$, and the $\ell = \{0, 2, 4\}$ multipoles, with contributions from density, RSD, Doppler and lensing, computed at various redshifts, in full-sky (solid), and in full-sky, but with flat-sky lensing-lensing (dashed).
The relative difference (in percent) between full-sky and flat-sky is indicated on the right plot of each figure.
}
\label{f:all_all_len_flat_multipoles}
\end{figure}

Since the lensing-lensing contribution is the one that is the most time-consuming to compute, in figure~\ref{f:all_all_len_flat_multipoles} we show results when we only consider the lensing-lensing contribution computed in the flat-sky approximation, while all of the others are computed in full-sky.

We can see that, in this case, the error of the flat-sky approximation with respect to the full-sky result is less than 1\% for all configurations, which suggests a good compromise between the accuracy of the full-sky result, and the performance increase brought by the semi-analytic method developed here.
The various 'glitches' (spikes) are caused by the zero-crossings of the 2PCF and the multipoles.

\subsection{Performance}
\label{s:performance}

In table~\ref{t:speed} we show the results for running COFFE with full-sky, and utilizing the flat-sky Limber approximation for the lensing-lensing contribution, computed for separations $r = 10 \times n\; \mathrm{Mpc}/h$, with $n \in \{1, \ldots, 25\}$, at a mean redshift $\bar z = 1.5$ with $\Delta z = 0.1$.
The 2PCF has been computed for $\mu = 0.9$, while the multipoles of the 2PCF were computed for $\ell \in \{0, 2, 4\}$.
All tests were done on an Intel(R) Core(TM) i5-8350U CPU @ 1.70GHz, using single core performance to rule out any bottlenecks in parallelization.

The results indicate that, using the approximation, the 2PCF can be computed more than 3 orders of magnitude faster, while the multipoles of the 2PCF can be computed more than 4 orders of magnitude faster compared to the full-sky calculation.\footnote{Concerning the seemingly longer run time of the 2PCF than the multipoles, we suspect that this due to the way the computation is implemented in COFFE, while in theory the two should roughly take the same amount of time to compute due to the computation of just one integral (over $\lambda$) in both cases.}
Note that in the table we just report the time it takes to run the COFFE modules which compute either the 2PCF or the multipoles, \textit{not} the entire run-time, which is currently bottlenecked by other modules, notably the on-the-fly computation of the matter power spectrum.

This significant performance increase when utilizing the flat-sky Limber approximation can potentially be used to include lensing when using Markov chain Monte Carlo (MCMC) sampling for cosmological parameter estimation~\cite{Dunkley_2005}, which typically require thousands of evaluations of a given estimator (in our case, $\xi_\ell$); this evaluation would be prohibitively expensive using the full-sky expression.

\begin{table}
\caption{
Speed of the calculation of the full-sky and the flat-sky lensing-lensing in COFFE, measured in seconds, as well as the speed-up of the flat-sky compared to the full-sky calculation.
}
\centering
\begin{tabular}{c c c c}
\toprule
quantity & full-sky & flat-sky & speed-up (full-sky/flat-sky)\\
\midrule
2PCF & 20 s & $1.1 \times 10^{-2}$ s & $\sim$ 1800\\
multipoles & 267 s & $7.5 \times 10^{-3}$ s & $\sim$ 35000\\
\bottomrule
\end{tabular}
\label{t:speed}
\end{table}

\section{Conclusions}
\label{s:conclusions}

We studied the impact of the various contributions to the 2-point correlation function of galaxy number counts in the flat-sky approximation.
The numerical results show that, for density, RSD, and Dopper, i.e. non-integrated terms, the flat-sky approximation is accurate up to 1\% for all separations at redshifts $z \gtrsim 1$, and up to 5\% for lower redshifts.

For the density-lensing term, we found that the flat-sky Limber approximation is not adequate to capture the behavior of the full-sky result, primarily for two reasons:

\begin{enumerate}
\item
The angular correlation function in the approximation vanishes at all redshifts, while it is non-zero in full-sky.
\item
In the $r \rightarrow 0$ limit, the flat-sky Limber correlation function and its multipoles vanish, while they are non-zero in full-sky.
\end{enumerate}

On the other hand, the same approximation for the lensing-lensing contribution is accurate up to 1\% for both the 2PCF and its multipoles at $z \gtrsim 1$, which suggests that cross-correlations between different redshifts, as illustrated in figure~\ref{fig:fullsky_vs_flatsky_tikz}, do not contribute significantly to the result.

Our main result is a semi-analytic method of evaluating the integrated contributions to the multipoles of the 2PCF in the flat-sky Limber approximation, which speeds up the computation by a factor of $\sim 10000$ compared to the full-sky result, which can potentially be used in MCMC sampling for cosmological parameter estimation while including lensing in the signal.
\section*{Acknowledgments}

The author wishes to thank Johannes Trost for his help~\cite{366191} in finding a solution to the density-lensing integral~\eqref{eq:flatsky_density_lensing}, Tamara Radan for her help with the manuscript, Ruth Durrer and William Matthewson for insightful discussions, and the anonymous referee for various comments.
This work is supported by the Swiss National Science Foundation.

\section*{Disclaimer}
This is an author-created, un-copyedited version of an article published in the Journal of Cosmology and Astroparticle Physics (JCAP).
IOP Publishing Ltd is not responsible for any errors or omissions in this version of the manuscript or any version derived from it.
The Version of Record is available online at \url{https://doi.org/10.1088/1475-7516/2021/07/045}.

\appendix

\section{Contributions to the 2-point correlation function}
\label{s:fullsky_expressions}

For completeness, below we state the relevant full-sky contributions for the 2-point correlation function.

We use the following notation, for local terms $A$ and $B$:
\begin{equation}
\xi^{AB}(\bar z, r, \mu)
=
D_1(z_1)\,
D_1(z_2)
\sum_{\ell,n}
\left(
X^n_\ell\big|_A
+
X^n_\ell\big|_{AB}
+
X^n_\ell\big|_{BA}
+
X^n_\ell\big|_B
\right)I_\ell^n(r)
\end{equation}
and the following for the integrated terms:
\begin{equation}
\xi^{AB}(\bar z, r, \mu)
=
Z\big|_A
+
Z\big|_{AB}
+
Z\big|_{BA}
+
Z\big|_B
\end{equation}
\begingroup
\allowdisplaybreaks
\begin{align}
&X_0^0 \big|_\text{den} = b_1 b_2\,,
\label{eq:density}
\\
&X_0^0 \big|_\text{RSD} =  f_1 f_2   \frac{1+2\cos^2\theta}{15 }\,,
\label{eq:rsd1}
\\
&X_2^0 \big|_\text{RSD} = -\frac{f_1 f_2}{21}\left[1+11\cos^2\theta +\frac{18\cos\theta(\cos^2\theta-1)\chi_1\chi_2}{r^2}\right] \,,
\label{eq:rsd2}
\\
&X_4^0 \big|_\text{RSD} = \frac{f_1 f_2}{35r^4} \big\{{4(3\cos^2\theta-1)(\chi_1^4+\chi_2^4)} + {\chi_1\chi_2} (3+\cos^2\theta)\big[ 3 (3+\cos^2\theta)\chi_1\chi_2 \nonumber \\
&\qquad\qquad -8(\chi_1^2+\chi_2^2)\cos\theta \big] \big\} \,,
\label{eq:rsd3}
\\
&X_0^2\big|_\text{d1} = \HH_1\HH_2 f_1 f_2 G_1 G_2  \frac{r^2\cos\theta}{3}\,,
\label{eq:doppler1}
\\
&X_2^2\big|_\text{d1} = - \HH_1\HH_2 f_1 f_2 G_1 G_2 \left((\chi_2-\chi_1\cos\theta)(\chi_1-\chi_2\cos\theta)+\frac{r^2\cos\theta}{3}\right) \,,
\label{eq:doppler2}
\\
&X_0^0 \big|_\text{den-RSD} = \frac{b_1 f_2 }{3} \,,
\label{eq:den-rsd1}
\\
&X_2^0 \big|_\text{den-RSD} = -b_1 f_2  \left(\frac{2}{3} -(1-\cos^2\theta)\frac{\chi_1^2}{r^2}\right) \,,
\label{eq:den-rsd2}
\\
&X_1^1 \big|_\text{den-d1} = -b_1 f_2 \HH_2 G_2 (\chi_1 \cos \theta -\chi_2) \,,
\label{eq:den-doppler1}
\\
&X_1^1 \big|_\text{RSD-d1} = f_1f_2 \HH_2 G_2  \frac{(1+2\cos^2\theta)\chi_2-3 \chi_1\cos\theta}{5} \,,
\label{eq:RSD-doppler1}
\\
&X_3^1 \big|_\text{RSD-d1} = \frac{f_1f_2 \HH_2 G_2}{5r^2} \big[(1-3\cos\theta)\chi_2^3 +\cos\theta(5+\cos^2\theta)\chi_2^2\chi_1-2(2+\cos\theta^2)\chi_2\chi_1^2 \nonumber \\
&\qquad \qquad +2\chi_1^3\cos\theta \big] \,,
\label{eq:RSD-doppler2}
\\
&Z \big|_\text{den-len} = - \frac{3\Omega_m}{2} b_1 \HH_0^2 \frac{2-5s_2}{\chi_2}D_1(z_1) \int\limits_0^{\chi_2} \mathrm{d} \la \frac{\chi_2-\la}{\la}\frac{D_1(\la)}{ a(\la)} \bigg\{ 2\chi_1\la\cos\theta I^1_1(r) \nonumber \\
&\qquad -\frac{\chi_1^2\la^2(1-\cos^2\theta)}{r^2} I^0_2(r)\bigg\} \,,
\label{eq:den-len}
\\
&Z \big|_\text{len}
=
\frac{9 \Omega_m^2}{4}\HH_0^4\frac{(2-5s_1)(2-5s_2)}{\chi_1\chi_2} \int\limits_0^{\chi_1} \!\mathrm{d}\la \int\limits_0^{\chi_2} \!\mathrm{d}\la' \frac{(\chi_1-\la)(\chi_2-\la')}{\la \la'} \frac{D_1(\la)D_1(\la')}{a(\la)a(\la')} \bigg\{
\nonumber
\\
& \quad
\frac{2}{5} (\cos^2\theta-1) \la^2 \la'^2 I^0_0(r)
+
\frac{4 r^2 \cos\theta \la \la'}{3} I^2_0(r)
+
\frac{4 \cos\theta \la \la' (r^2 +6 \cos\theta \la \la')}{15} I^1_1(r)
\nonumber
\\
& \quad
+
\frac{2(\cos^2\theta -1)\la^2\la'^2(2r^4 +3 \cos\theta r^2 \la \la')}{7 r^4} I^0_2(r)
\nonumber
\\
& \quad
+\frac{2 \cos\theta \la \la' \left(2 r^4 +12\cos\theta r^2 \la\la' +15 (\cos^2\theta-1)\la^2\la'^2 \right)}{15 r^2} I^1_3(r)
\nonumber
\\
&\quad +\frac{(\cos^2\theta-1)\la^2\la'^2 \left(6 r^4 +30\cos\theta r^2\la\la' +35 (\cos^2\theta -1)\la^2\la'^2 \right)}{35r^4} I^0_4(r)\bigg\} \,,
\label{eq:len-len}
\end{align}
\endgroup
where
\be
G(z)= \frac{\dot \HH}{\HH^2}+\frac{2-5s}{\chi \HH}+5 s - f_\text{evo} \,.
\ee
Note that inside the integral, $r^2 = \chi_2^2 + \lambda^2 - 2 \chi_2 \lambda \cos\theta$ in the case of density-lensing, and $r^2 = \lambda_1^2 + \lambda_2^2 - 2 \lambda_1 \lambda_2 \cos\theta$ in the case of lensing-lensing, while $\theta$ is the angle at the observer between the two lines of sight.
The result for the other cross-correlations can be obtained by performing the substitution $2 \leftrightarrow 1$.

\section{The Limber approximation for the non-integrated terms}
\label{s:limber_nonint}

Here we sketch a proof of the claim that for non-integrated terms, the Limber approximation yields only terms which are a linear combination of the Heaviside theta and Dirac delta functions.
The relationship between the 2PCF and the angular power spectra for arbitrary contributions $A,B$ is:
\begin{equation}
\xi^{AB}(z_1, z_2, \theta)
=
\frac{1}{4 \pi}
\sum\limits_{\ell = 0}^\infty
(2 \ell + 1) C_\ell(z_1, z_2) P_\ell(\cos \theta)
\end{equation}
with:
\begin{equation}
C^{AB}_\ell(z_1, z_2)
=
\frac{2}{\pi}
\int_0^\infty
\mathrm{d} k\,
k^2
j_{\ell}(k r(z_1))
j_{\ell}(k r(z_2))
T^A(k, z_1)
{T^\star}^B(k, z_2)
P(k)
\label{eq:cell}
\end{equation}
where $T$ denote the transfer functions, $^\star$ denotes complex conjugation, and $P(k)$ is the matter power spectrum at $z=0$.
The (full-sky) Limber approximation amounts to the following substitution for a slowly varying function $f(k)$:
\[
\int_0^\infty
\mathrm{d} k\,
k^2\,
f(k)
j_{\ell}(k r)
j_{\ell}(k r')
\simeq
\frac{\pi}{2 r^2}
\delta(r - r')
f\left[(\ell + 1/2) / r\right]
\]
Inserting this in eq.~\eqref{eq:cell} with $f(k) = T^A {T^\star}^B P(k)$, we obtain:
\begin{equation}
C^{AB}_{\ell,\textrm{Limber}}(z_1, z_2)
=
\frac{1}{r_1^2}
\delta(r_1 - r_2)
{T^{A}}(\nu / r_1, z_1)
{T^{\star B}}(\nu / r_2, z_2)
P(\nu / r_1)
\end{equation}
with $\nu = \ell + 1/2$ and $r_i = r(z_i)$.
After a change of variables, it's easy to show that $\xi^{AB}_\textrm{Limber} (z_1, z_2, \theta) \propto \delta(z_1 - z_2)$.
Note that certain non-integrated terms, such as RSD, contain derivatives of the spherical Bessel functions instead;
these can be reduced to linear combinations of spherical Bessel functions of different order with the help of the following recurrence relations~\cite[Eq. 10.51.1]{NIST:DLMF}:
\[
j'_\ell
=
\frac{1}{2 \ell + 1}
\left[
\ell j_{\ell - 1}
-
(\ell + 1) j_\ell
\right]
\]
As shown in~\cite{maximon1991evaluation}, the evaluation of the integral over the two spherical Bessel functions for $\ell \neq \ell'$ is somewhat more complicated, with the end result being a linear combination of the Heaviside theta and the Dirac delta functions.
The flat-sky results~\eqref{eq:den_flatsky}--\eqref{eq:rsd_d1_flatsky} do not contain such terms, and therefore we do not use the Limber approximation for them.

\section{The flat-sky Limber correlation function for integrated terms}
\label{s:flat_sky_derivation}

This derivation closely follows the derivation in appendix E of~\cite{Tansella:2017rpi}.
For the local terms in linear theory, we may always write in Fourier space $\Delta_\text{effect}(\vec k, z) = f_\text{effect}(k, \nu, z)\, \delta(\vec k)$, where $f_\text{effect}(k, \nu, z)$ is some deterministic function, $\nu = \vec n \cdot \unit k$, where $\bn$ is the unit vector in the direction of the galaxy, and $\delta(\vec k)$ is the overdensity at $z = 0$.
Let us denote $A + B/(k\HH) = \alpha(k, \nu, z)$, where the $A$ is the contribution from density and RSD, and $B$ contains the Doppler term.
We neglect integrated effects in the fourth and fifth lines of eq.~\eqref{eq:number_counts}.
Denoting:
\begin{equation}
\mathbb{F}(\chi \bn, z)
=
\frac{1}{(2\pi)^3}
\int_{-\infty}^\infty
\dd^3 \vec k\,
\mathrm{e}^{-i \vec k \cdot \bn \chi}
\alpha(k, \kcos, z)\,
\delta(\vec k)
\end{equation}
for all of the local terms, and:
\be
\mathbb{I}(\chi \bn, z)
=
\frac{5s - 2}{2\chi(z)}
\int_0^{\chi(z)}
\dd \la\,
\frac{\chi(z)-\la}{\la}
\De_\Om(\Phi + \Psi)
\label{e:Idef}
\ee
for the lensing term, the correlation function of the two contributions above is given by:
\begin{align}
\xi^\text{total}(\bar z, \br)
&=
\langle
\mathbb F(\chi_1\bn_1,z_1)\,
\mathbb F(\chi_2\bn_2,z_2)
\rangle
+
\langle
\mathbb I(\chi_1\bn_1,z_1)\,
\mathbb F(\chi_2\bn_2,z_2)
\rangle
\nonumber
\\
&
+
\langle
\mathbb F(\chi_1\bn_1,z_1)\,
\mathbb I(\chi_2\bn_2,z_2)
\rangle
+
\langle
\mathbb I(\chi_1\bn_1,z_1)\,
\mathbb I(\chi_2\bn_2,z_2)
\rangle
\label{e:corrfctn}
\end{align}
where $\chi_i = \chi(z_i)$, $\br = \chi_2\bn_2-\chi_1\bn_1$, $\bar z = (z_1 + z_2) / 2$, and we assume both $\chi_i\gg r$ and the $z_i$ should not be very different.
The first term is just the local-local contribution, which in flat-sky is given by eqs.~\eqref{eq:den_flatsky}--\eqref{eq:den_d1_flatsky}.
The contribution of the cross term to the correlation function is then given by:
\begin{align}
\xi_{\mathbb I \mathbb F}(\bar z, \br)
=
\langle \mathbb I^{(1)}\, \mathbb F^{(2)} \rangle
=&
\frac{1}{(2\pi)^3}\frac{5s_1 - 2}{2\chi_1}
\int
\dd^3 \vec k\,
k_\perp^2\,
e^{-i\bk\cdot\bn_2\chi_2}\,
\alpha_2(k, \nu, z_2)
\nonumber
\\
&
\times
\int_0^{\chi_1}
\dd\la\,
\la\,
(\chi_1 - \la)\,
P_\text{len-den}(\vec k, z(\la), z_2)\,
e^{i\bk\cdot\bn_1\la}
\label{e:xiif1}
\end{align}
where $X^{(i)} \equiv X(\chi_i \bn_i, z_i)$, and $P_\text{len-den}(\vec k, z(\la), z_2)$ is the (unequal-time) cross-spectrum between lensing and density, defined via:
\begin{equation}
\langle
(\Phi + \Psi)(\vec k, z)\,
\delta(\vec k', z')
\rangle
=
(2 \pi)^3\,
\delta(\vec k - \vec k')\,
P_\text{len-den}(\vec k, z, z')
\end{equation}
In linear theory, we will take $P_\text{len-den}(\vec k, z, z') = \beta (k, z, z')\, P(k)$, where $\beta$ is a deterministic function, and $P(k)$ is the matter power spectrum at $z = 0$.
In the spirit of the flat-sky approximation we now set $\bn_1=\bn_*+\De\bn/2$ and $\bn_2=\bn_*-\De\bn/2$ assuming that $\De\bn$ is very small.
Splitting $\br=\br_\perp +\bn_*r_\pa$ with $\br_\perp=\chi(z)\De\bn$ and $r_\pa=r\cos\al_2$, we then perform the $k$-integral in the direction parallel to $\bn_*$.
In accordance with the Limber approximation, we neglect the dependence of the power spectrum on $k_\pa$, which allows us to use the following result~\cite{kaiser_limber,Bernardeau_2011}:
\begin{equation}
\int_0^\infty
\mathrm{d} k_\parallel\,
f\left(\sqrt{{k}_\perp^2 + k_\parallel^2}\right)
e^{i k_\parallel (r - r')}
\simeq
2\pi f({k}_\perp) \delta(r - r')
\end{equation}
and gives us $2\pi\de(\chi_2-\la) \beta({k}_\perp, z, z') P({k}_\perp)$.
Hence the integral over $\la$ does not contribute if $\chi_2>\chi_1$, otherwise it reduces to the integrand at $\chi_2$:
\begin{align}
\xi_{\mathbb I \mathbb F}(\bar z, \br)
=&
\frac{1}{(2\pi)^2}
\frac{(5s_1 - 2)\Theta(\chi_1-\chi_2)}{2\chi_1}\,
\chi_2\,
(\chi_1 - \chi_2)
\nonumber
\\
&
\times
\int
\dd^2 \vec k_\perp\,
e^{-i\bk_\perp\cdot\br_\perp}\,
k_\perp^2\,
P(k_\perp)\,
\alpha_2(k_\perp, 0, z_2)\,
\beta(k_\perp, z_2)\,
\label{e:xiif2}
\end{align}
where $\Theta$ is the Heaviside theta function.

Using polar coordinates, $\dd^2 \bk_\perp = \dd k_\perp k_\perp \dd \varphi$ we can perform the $\varphi$ integration which yields a Bessel function, $2\pi J_0(k_\perp r_\perp) =2\pi J_0(k_\perp r\sin\al_2)$.
The term $\xi_{\mathbb F \mathbb I}(\bar z, \br)$ contributes in the same way with $z_1$ and $z_2$ exchanged.
Setting $\chi_1-\chi_2 = r_\pa =r\, \mu$ and neglecting the difference between $\chi_1$ and $\chi_2$ ($z_1$ and $z_2$) in all other places, we find for the sum of both mixed terms:
\begin{align}
\xi_{\mathbb I \mathbb F + \mathbb F\mathbb I}(\bar z, \br)
=&
\frac{r \mu}{4\pi}
\left[
(5s_1 - 2)\, b_2\, \Theta(\chi_1-\chi_2)\,
-
(5s_2 - 2)\, b_1\, \Theta(\chi_2-\chi_1)
\right]
\nonumber
\\
&
\times
D_1(\bar z)\,
\int
\dd k_\perp\,
k_\perp^3\,
P(k_\perp)\,
J_0(k_\perp r\sqrt{1-\mu^2})
\beta(k_\perp, \bar z)
\label{e:xiiffi}
\end{align}
where we've taken into account that $\alpha_i(k_\perp, 0, z) = b_i(z)\, D_1(z)$.
Here we have also neglected the difference between $\cos\al_2$ and $\mu$, since in the flat-sky approximation all these angles are equal.
Note also that, since $k_\pa = 0$ in the flat-sky Limber limit, the integrated term is not correlated with redshift space distortions, nor the Doppler term.

Finally, we can use the following simplification:
\begin{equation}
a\, x\, \Theta(x) - b\, x\, \Theta(-x)
=
\frac{1}{2}
\left[
(a - b)x
+
(a + b)|x|
\right]
\label{eq:heaviside_theta}
\end{equation}
which allows us to write:
\begin{align}
\xi_{\mathbb I \mathbb F + \mathbb F\mathbb I}(\bar z, \br)
=&
\frac{r}{8\pi}
\left\{
\left[
(5s_1 - 2)\, b_2
-
(5s_2 - 2)\, b_1
\right]\mu
+
\left[
(5s_1 - 2)\, b_2
+
(5s_2 - 2)\, b_1
\right]|\mu|
\right\}
\nonumber
\\
&
\times
D_1(\bar z)\,
\int
\dd k_\perp\,
k_\perp^3\,
P(k_\perp)\,
J_0(k_\perp r\sqrt{1-\mu^2})
\beta(k_\perp, \bar z)
\label{eq:xiiffi}
\end{align}

If we consider only one population of galaxies, the first term in the curly brackets vanishes.
Due to the Poisson equation, in general relativity we have that:
\begin{equation}
\beta(k, z)
=
-3\,
\frac{
D_1(z)\, H_0^2\, (1 + z)\, \Omega_m
}
{
k^2
}
\end{equation}
which gives us exactly eq.~\eqref{eq:den_len_flat_2pcf}.

Let us finally compute the double integrated term:
\begin{align}
\xi_{\mathbb I \mathbb I}(z, \br)
=&
\frac{(2 - 5s_1)(2 - 5s_2)}{(2\pi)^3 4\chi^2}
\int
\dd^3 \vec k\,
P(k)\,
\int_0^{\chi_1}
\dd\la\,
\int_0^{\chi_2}
\dd\la'\,
\la(\chi_1-\la)k_\perp^{2}
\nonumber
\\
&\hspace{-0.1cm}
\times
\la'(\chi_2-\la')k_\perp^{2}\,
D_1(z(\la))\,
(1+z(\la))\,
D_1(z(\la'))\,
(1+z(\la'))\,
e^{-i\bk \cdot (\bn_1\la-\bn_2\la')}
\end{align}

Via the same procedure as above, the integration over $k_\pa$ leads to $2\pi\de(\la-\la')$ and we find:
\begin{align}
\xi_{\mathbb I \mathbb I}(\br,z)
=&
\frac{(3\Om_mH_0^2)^2(2 - 5s_1)(2 - 5s_2)}
{(2\pi)^2 4\chi^2}
\int
\frac{\dd^2 k_\perp}{k_\perp^4}\,
P(k_\perp)
\nonumber
\\
&
\times
\int_0^{\chi}
\dd\la\,
[\la(\chi-\la)k_\perp^{2}]^2\,
D_1^2(z(\la))\,
(1+z(\la))^2\,
e^{-i\bk_\perp \cdot \br_\perp(\la/\chi)}
\end{align}

We now perform a change of variables, $\bk_\perp\mapsto (\la/\chi)\bk_\perp$.
In terms of this new variable, the integral contribution to the correlation function becomes:
\begin{align}
\xi_{\mathbb I \mathbb I}(\br,z)
=&
\frac{(3\Om_mH_0^2)^2(2 - 5 s_1)(2 - 5 s_2)}{(2\pi)^24\chi^2}
\int_0^{\chi}\dd\la\,
\int\frac{\dd^2 k_\perp}{k_\perp^4}P_\de(k_\perp\chi/\la)e^{-i\bk_\perp \cdot \br_\perp}
\nonumber
\\
&
\times
\left(\frac{\la}{\chi}\right)^2\left[\frac{(\chi-\la)\chi^2}{\la}k_\perp^{2}\right]^2D^2(z(\la))(1+z(\la))^2
\end{align}

Again, performing the $\varphi$ integration we end up with:
\begin{align}
\xi_{\mathbb I \mathbb I}(\br,z)
=&
\frac{(3\Om_mH_0^2)^2(2-5 s_1)(2 - 5 s_2)}{8\pi\chi^2}
\int_0^{\chi}\dd\la\,
\int \dd k_\perp k_\perp P_\de(k_\perp\chi/\la)J_0(k_\perp r\sqrt{1-\mu^2})
\nonumber
\\
&
\times
\left(\frac{\la}{\chi}\right)^2\left[\frac{(\chi-\la)\chi^2}{\la}\right]^2D^2(z(\la))(1+z(\la))^2
\end{align}
which gives us eq.~\eqref{eq:lens_lens_2pcf}.

\section{Analytic integrals}
\label{s:integrals}

Here we outline how we obtained the results eqs.~\eqref{eq:density_lensing_multipoles} and~\eqref{eq:integral_bessel}.

\subsection{Density-lensing integral}
\label{s:den_len_appendix}

The integral we are looking for is of the form:
\begin{equation}
\mathcal{J}_\ell(\alpha)
\equiv
\int_{-1}^{1}
\dd x\,
|x|\,
P_\ell(x)\,
J_0(\alpha \sqrt{1 - x^2})
\end{equation}

As a first step, we do an expansion of the Legendre polynomials of the form (see for instance~\cite[Eq. 12.8]{arfken1999mathematical}):
\begin{equation}
P_\ell(z)
=
\frac{1}{2^\ell}
\sum\limits_{k=0}^{\left\lfloor \frac{\ell}{2}\right\rfloor}
(-1)^k
\begin{pmatrix} \ell \\ k \end{pmatrix}
\begin{pmatrix} 2\ell - 2k \\ \ell \end{pmatrix}
z^{\ell - 2k}
\label{eq:legendre_expansion}
\end{equation}
where $\left\lfloor A \right\rfloor$ denotes the floor of the real number $A$.

We will also take advantage of the following identity~\cite[Eq. 10.22.19]{NIST:DLMF}:
\begin{equation}
\int_{0}^{\frac{1}{2}\pi}J_{\mu}\left(z\sin\theta\right)(\sin\theta)^{\mu+1}(%
\cos\theta)^{2\nu+1}\mathrm{d}\theta=2^{\nu}\Gamma\left(\nu+1\right)z^{-\nu-1}%
J_{\mu+\nu+1}\left(z\right)
\label{eq:bessel_trig_integral}
\end{equation}
where the only requirement is that $\mathrm{Re}(\mu) > -1$ and $\mathrm{Re}(\nu) > -1$.

The calculation can now be performed as follows:
\allowdisplaybreaks
\begin{align*}
\mathcal{J}_\ell(\alpha)
&=
\int_{-1}^1 \dd x\, |x|\, J_0(\alpha\sqrt{1 - x^2}) P_\ell(x)\\
&=
[(-1)^\ell + 1]
\int_0^1 \dd x\, x\, J_0(\alpha \sqrt{1 - x^2}) P_\ell(x)\\
&
|\mathrm{substitution}\;x = \cos \phi|\\
&=
[(-1)^\ell + 1]
\int_0^\frac{\pi}{2} \dd \phi\, \sin \phi\, \cos \phi\, P_\ell(\cos \phi)\, J_0 (\alpha \sin \phi)\\
&
|\mathrm{expansion\;of}\;P_\ell|\\
&=
[(-1)^\ell + 1]
\sum\limits_{k=0}^{\left\lfloor \frac{\ell}{2}\right\rfloor}
(-1)^k
\begin{pmatrix} \ell \\ k \end{pmatrix}
\begin{pmatrix} 2\ell - 2k \\ \ell \end{pmatrix}
\int_0^\frac{\pi}{2} \dd \phi\, \sin \phi\, \cos \phi\, (\cos \phi)^{\ell - 2k} J_0 (\alpha \sin \phi)\\
&=
[(-1)^\ell + 1]
\sum\limits_{k=0}^{\left\lfloor \frac{\ell}{2}\right\rfloor}
(-1)^k
\begin{pmatrix} \ell \\ k \end{pmatrix}
\begin{pmatrix} 2\ell - 2k \\ \ell \end{pmatrix}
\int_0^\frac{\pi}{2} \dd \phi\, \sin \phi\, (\cos \phi)^{\ell - 2k + 1} J_0 (\alpha \sin \phi)
\end{align*}
The integral in the above has the same form as the Bessel identity~\eqref{eq:bessel_trig_integral}, with $\mu = 0$ and $\nu = \ell / 2 - k$, so that our final result is exactly eq.~\eqref{eq:density_lensing_multipoles}, completing the proof.

Note that there are two generalizations of the above result which we will exploit in appendix~\ref{s:bias_full}, which are:
\begin{align}
\mathcal{M}(n, \ell; \alpha)
=&
\int_{-1}^{1}
\dd x\,
x^n\,
P_\ell(x)\,
J_0(\alpha \sqrt{1 - x^2})
\\
\mathcal{N}(n, \ell; \alpha)
=&
\int_{-1}^{1}
\dd x\,
|x|\,
x^n\,
P_\ell(x)\,
J_0(\alpha \sqrt{1 - x^2})
\end{align}

Proceeding exactly as above, we obtain the results:
\begin{align}
\mathcal{M}(n, \ell; \alpha)
=&
[(-1)^{n + \ell} + 1]
\Omega(n, \ell; \alpha)
\label{eq:omega_n}
\\
\mathcal{N}(n, \ell; \alpha)
=&
[(-1)^{n + \ell} + 1]
\Omega(n + 1, \ell; \alpha)
\label{eq:omega_abs_n}
\end{align}
where we define $\Omega(n, \ell; \alpha)$ as:
\begin{equation}
\Omega(n, \ell; \alpha)
\equiv
\frac{1}{2^\ell}
\sum\limits_{k = 0}^{\lfloor \frac{\ell}{2}\rfloor}
(-1)^k
\begin{pmatrix}\ell \\ k \end{pmatrix}
\begin{pmatrix}2 \ell - 2 k \\ \ell \end{pmatrix}
2^\nu
\Gamma(\nu + 1)
\alpha^{-\nu - 1}
J_{\nu + 1}(\alpha)
\label{eq:omega_def}
\end{equation}
with $\nu = n / 2 + \ell / 2 - k - 1 / 2$.

\subsection{Lensing-lensing integral}
\label{s:len_len_appendix}

The integral we are looking for is of the form:
\begin{equation}
\mathcal{I}_\ell(\alpha)
\equiv
\int_{-1}^{1}
\dd x\,
P_\ell(x)\,
J_0(\alpha \sqrt{1 - x^2})
\end{equation}

Note that, since the integrand is odd if $\ell$ is odd, and the integration limits are symmetric around zero, the integral vanishes, so the result can only be non-zero if $\ell$ is even.

Now, we will take advantage of the standard orthogonality property of the Legendre polynomials:
\begin{equation}
\int_{-1}^1
\dd x\,
P_\ell(x)\,
P_n(x)
=
\frac{2}{2l+1} \delta_{\ell, n}
\end{equation}
as well as the following expansion of the Bessel function~\cite[Eq. 10.60.10]{NIST:DLMF}:
\begin{equation}
J_{0}\left(z\sin\alpha\right)
=
\sum_{n=0}^{\infty}(4n+1)\frac{(2n)!}{2^{2n}(n!)%
^{2}}j_{2n}\left(z\right)P_{2n}\left(\cos\alpha\right)
\label{eq:bessel_sum}
\end{equation}

The calculation is then straightforward:
\begin{align*}
\mathcal{I}_\ell(\alpha)
&\equiv
\int_{-1}^{1}
\dd x\,
P_\ell(x)\,
J_0(\alpha \sqrt{1 - x^2})
\\
&=
\sum_{n=0}^{\infty}(4n+1)\frac{(2n)!}{2^{2n}(n!)%
^{2}}
j_{2n}(\alpha)
\int_{-1}^{1}
\dd x\,
P_\ell(x)\,
P_{2n}(x)
\\
&=
\sum_{n=0}^{\infty}(4n+1)\frac{(2n)!}{2^{2n}(n!)%
^{2}}
j_{2n}(\alpha)
\frac{2}{2\ell + 1}
\delta_{\ell, 2n}
\\
&=
\frac{\ell!}{2^{\ell - 1}\left[(\ell / 2)!\right]^{2}}
j_{\ell}(\alpha)
\\
&=
C(\ell)
j_\ell(\alpha)
\end{align*}
which is identical to eq.~\eqref{eq:integral_bessel} with the coefficients $C(\ell)$ given by eq.~\eqref{eq:coefficients}.

Note that we could have obtained the same result by using eq.~\eqref{eq:omega_n} with $n = 0$.

\section{Multipoles for a redshift-dependent bias}
\label{s:bias_full}

Here we derive eq.~\eqref{eq:bias_analytic}.
\begingroup
\allowdisplaybreaks
\begin{align}
\xi_{\ell}
\propto&
\int_{-1}^1
\dd \mu\,
P_\ell(\mu)\,
Q_i\,
Q_j\,
J_0(\alpha\, \sqrt{1 - \mu^2})
\nonumber
\\
=&
\sum_{n_1,n_2}
a_{n_1}^{(i)}\,
b_{n_2}^{(j)}\,
\int_{-1}^1
\dd \mu\,
P_\ell(\mu)\,
\chi_1^{n_1}\,
\chi_2^{n_2}\,
J_0(\alpha\, \sqrt{1 - \mu^2})
\nonumber
\\
=&
\sum_{n_1,n_2}
a_{n_1}^{(i)}\,
b_{n_2}^{(j)}\,
\int_{-1}^1
\dd \mu\,
P_\ell(\mu)\,
\left[\bar\chi - \frac{1}{2}\mu\, r\right]^{n_1}
\left[\bar\chi + \frac{1}{2}\mu\, r\right]^{n_2}
J_0(\alpha\, \sqrt{1 - \mu^2})
\nonumber
\\
=&
\sum_{n_1,n_2}
a_{n_1}^{(i)}\,
b_{n_2}^{(j)}\,
\sum_{k_1 = 0, k_2 = 0}^{n_1, n_2}
\factorial{n_1}{k_1}\,
\factorial{n_2}{k_2}\,
\bar\chi^{n_1 + n_2 - k_1 - k_2}
\left(
\frac{r}{2}
\right)^{k_1 + k_2}
(-1)^{k_1}
\nonumber
\\
&\times
\int_{-1}^1
\dd \mu\,
\mu^{k_1 + k_2}\,
P_\ell(\mu)\,
J_0(\alpha\, \sqrt{1 - \mu^2})
\nonumber
\\
=&
\sum_{n_1,n_2}
a_{n_1}^{(i)}\,
b_{n_2}^{(j)}\,
\sum_{k_1 = 0, k_2 = 0}^{n_1, n_2}
\factorial{n_1}{k_1}\,
\factorial{n_2}{k_2}\,
\bar\chi^{n_1 + n_2 - k_1 - k_2}
\left(
\frac{r}{2}
\right)^{k_1 + k_2}
(-1)^{k_1}
\nonumber
\\
&\times
\left[
(-1)^{k_1 + k_2 + \ell} + 1
\right]
\Omega\left(k_1 + k_2, \ell; \alpha \right)
\label{eq:bias_analytic_full}
\end{align}
\endgroup
where $\Omega(n, \ell, \alpha)$ is given by eq.~\eqref{eq:omega_def}.

Note that we can use expansion~\cite[Eq. 15]{weisstein} to also obtain the odd multipoles for density-lensing, eq.~\eqref{eq:den_len_flat_2pcf}, with the final result being a linear combination of integrals of the form~\eqref{eq:integral_bessel}.
Furthermore, note that eq.~\eqref{eq:bias_analytic_full} contains roughly $\mathcal{O}(n^4)$ terms, where $n$ is the number of terms in the expansion~\eqref{eq:bias_taylor};
this is seemingly suboptimal, as for $n=10$ we need $\sim 10^{4}$ terms, which would greatly slow down the computation.
However, we can use the method to construct a Taylor polynomial for the bias in \textit{each} redshift bin instead:
\begin{equation*}
s(\chi)
\simeq
\sum\limits_{k = 0}^n
\frac{1}{k!}
\frac{\partial^k s}{\partial \chi^k}\bigg|_{\chi(\bar z)}
[\chi - \chi(\bar z)]^k
\label{eq}
\end{equation*}
Taking $n = 1$, i.e. just a constant plus a linear term, this approximation differs by less than 1\% for the magnification bias~\eqref{eq:bias}, and is much better than the naive case $s(z)=\textrm{const}.$
In this case, we only need a handful of terms in eq.~\eqref{eq:bias_analytic_full}, instead of potentially thousands, to capture the redshift dependence of a given bias.%

\bibliographystyle{JHEP}
\bibliography{refs}

\end{document}